\documentclass[aip,jcp,reprint,twocolumn,superscriptaddress,showpacs,floatfix]{revtex4-2}
\usepackage{amsmath}
\usepackage{graphicx}% Include figure files
\usepackage{dcolumn}% Align table columns on decimal point
\usepackage{bm}% bold math
\usepackage{verbatim}
\usepackage[utf8]{inputenc}
\usepackage[T1]{fontenc}
\usepackage{mathptmx}
\usepackage{etoolbox}
\usepackage{subcaption} % For subfigure support
\usepackage{float}
\usepackage{bm}% bold math
\usepackage{mathrsfs}
\usepackage{multirow}
\usepackage{ragged2e}
\DeclareUnicodeCharacter{2212}{-}
\usepackage{hyperref}
\hypersetup{
    colorlinks=true,    % Enables colored links
    linkcolor=blue,     % Color of internal links
    citecolor=magenta,      % Color of citations
    pdfborder={0 0 0}   % Removes the red boxes around links
}

\makeatletter
\def\@email#1#2{%
 \endgroup
 \patchcmd{\titleblock@produce}
  {\frontmatter@RRAPformat}
  {\frontmatter@RRAPformat{\produce@RRAP{*#1\href{mailto:#2}{#2}}}\frontmatter@RRAPformat}
  {}{}
}%
\makeatother

\begin{document}
\preprint{AIP/123-QED}
\title{Reduced-cost Relativistic Equation-of-Motion Coupled Cluster Method based on Frozen Natural Spinors:  A State-Specific Approach}

\author{Tamoghna Mukhopadhyay}
\affiliation{\small Department of Chemistry, Indian Institute of Technology Bombay, Powai, Mumbai 400076, India}
\author{Mrinal Thapa}
\affiliation{\small Department of Chemistry, Indian Institute of Technology Bombay, Powai, Mumbai 400076, India}
\author{Somesh Chamoli}
\affiliation{\small Department of Chemistry, Indian Institute of Technology Bombay, Powai, Mumbai 400076, India}
\author{Xubo Wang}
\affiliation{\small Department of Chemistry, The Johns Hopkins University, Baltimore, Maryland 21218, United States}
\author{Chaoqun Zhang}
\affiliation{\small Department of Chemistry, Yale University, New Haven, Connecticut 06520, United States}
\author{Malaya K. Nayak}
\affiliation{\small Theoretical Chemistry Section, Bhabha Atomic Research Centre, Trombay,
Mumbai 400085, India}
\author{Achintya Kumar Dutta}
\thanks{Corresponding author}
\email[e-mail: ]{achintya@chem.iitb.ac.in}
\affiliation{\small Department of Chemistry, 
Indian Institute of Technology Bombay, Powai, Mumbai 400076, India}
\affiliation{ \small Department of Inorganic Chemistry, Faculty of Natural Sciences, Comenius University, Ilkovičova 6, Mlynská dolina 84215 Bratislava, Slovakia \\
}%
\email[e-mail: ]{achintya.kumar.dutta@uniba.sk}
%\maketitle

\begin{abstract}

We present the theoretical framework, implementation, and benchmark results for a reduced-cost relativistic equation-of-motion coupled cluster singles and doubles (EOM-CCSD) method based on state-specific frozen natural spinors (SS-FNS). In this approach, the state-specific frozen natural spinors are derived from the second-order algebraic diagrammatic construction (ADC(2)) method, providing a compact virtual space for excited-state calculations. The excitation energies computed with the SS-FNS-EE-EOM-CCSD method exhibit smooth convergence with respect to the size of the virtual space and demonstrate significant improvements over those obtained using the conventional MP2-based FNS approach. We have implemented the relativistic SS-FNS-EE-EOM-CCSD method using both the four-component Dirac–Coulomb and the exact two-component atomic mean-field (X2CAMF) Hamiltonians for excitation energies and transition properties. The X2CAMF-based relativistic EOM-CCSD method emerges as a promising approach for large-scale excited-state calculations, achieving excellent agreement with the standard relativistic EOM-CCSD method based on the untruncated canonical spinor basis, but at a significantly reduced computational cost.
\end{abstract}
\maketitle

%\pagebreak
%\clearpage
%\newpage~\newpage~\newpage
%%% \setstretch{1.5}

%%%%%%%%%%%%%%%%%%%%%%%%%%%%%%%%%%%%%%%%%%%%%%%% Introduction %%%%%%%%%%%%%%%%%%%%%%%%%%%%%%%%%%%%%%%%%%%%%%%%%%%%%%%%
\section{Introduction}
\label{sec1}
Quantum chemistry has made substantial strides in precisely determining atomic and molecular energies and properties. Various advances in theoretical methodologies and computational capabilities over several decades have significantly improved the precision and reliability of predicting these values. In recent years, there has been a pronounced interest in excited-state energies and properties due to their extensive applications in disciplines such as spectroscopy and photochemistry\cite{gonzalezQuantumChemistryDynamics2020}. A large variety of quantum chemical methods exist in the literature for the prediction of excitation energies and properties. Each of them has its own set of merits and demerits. Among the various single-reference electron correlation methods available, the coupled cluster (CC) method\cite{cizekCorrelationProblemAtomic1966, paldusCorrelationProblemsAtomic1972, shavittManyBodyMethodsChemistry2009} is considered among the most accurate for its size-extensive treatment of energy and systematically improvable nature. The ground state coupled cluster method can be extended to excited states using the equation-of-motion (EOM)\cite{roweEquationsofMotionMethodExtended1968, stantonEquationMotionCoupledcluster1993, nooijenEquationMotionCoupled1995, krylovEquationofMotionCoupledClusterMethods2008, kowalski2000active} framework. Similar results can be obtained using the linear response coupled cluster\cite{kochCoupledClusterResponse1990, monkhorstCalculationPropertiesCoupledcluster1977, mukherjeeResponsefunctionApproachDirect1979} approach. An alternative methodology is the $\Delta$-based approach, in which the ground and excited states are independently calculated using separate SCF\cite{bagusDirectNearHartreeFockCalculations1971, bagusSelfConsistentFieldWaveFunctions1965} and correlation calculations\cite{leeExcitedStatesCoupled2019, zhengPerformanceDeltaCoupledClusterMethods2019, south4ComponentRelativisticCalculations2016}. The individual calculations for ground and excited states in $\Delta$-based approaches may present several technical challenges, such as convergence issues and symmetry-breaking.  The direct difference of energy-based methods has the additional advantage of predicting transition properties, which is essential for simulating experiments. A significant drawback of the CC method is the sheer high computational cost. The coupled cluster method in singles and doubles (CCSD) approximation has a formal computational scaling of $O(N^6)\approx O(N_{occ}^2N_{vir}^4)$ for ground-state calculations, where $N$ is the total number of basis functions used in the calculation and $N_{occ}$ and $N_{vir}$ are the total number of occupied and virtual spinors, respectively.  The storage requirement, on the other hand, scales as  $O(N^4)$. The EOM-based excited-state calculations in singles doubles approximation (EOM-CCSD)\cite{shavittManyBodyMethodsChemistry2009} also have similar scaling and storage requirements. Inclusion of higher-order excitations, such as triples (EOM-CCSDT)\cite{wattsInclusionConnectedTriple1994, kucharski2001coupled} or quadruples (EOM-CCSDTQ)\cite{hirataHigherorderEquationofmotionCoupledcluster2004}  further increases the computational costs, making them practically infeasible beyond small molecules on standard computers.

In systems containing heavy elements, including relativistic effects, is paramount for getting both qualitative and quantitative accuracy. One of the most straightforward approaches to include relativistic effects for electronic structure calculations of multi-electronic systems is to use a four-component Dirac-Coulomb (4c-DC) Hamiltonian\cite{dyallIntroductionRelativisticQuantum2007}. A relativistic CCSD calculation using the 4c-DC Hamiltonian is at least 32 times more expensive than its non-relativistic analog\cite{dyallIntroductionRelativisticQuantum2007}. The high computational cost of excited-state energy\cite{sheeEquationofmotionCoupledclusterTheory2018}  and property\cite{mukhopadhyayAnalyticCalculationTransition2025} calculations within the relativistic EOM-CC framework limits its applicability to atoms and small molecules, typically only with modest basis sets. Relativistic multireference Fock-space coupled cluster (FSCC)\cite{EliavFockSpace2021}, although highly accurate and size-extensive, suffers the same problem.
Various strategies have been described in the literature to reduce the computational cost of relativistic coupled cluster calculations. One approach is to use two-component relativistic theories.\cite{hessRelativisticElectronicstructureCalculations1986,vanlentheRelativisticRegularTwocomponent1996,dyallInterfacingRelativisticNonrelativistic1997,nakajimaNewRelativisticTheory1999,baryszTwocomponentMethodsRelativistic2001,liuExactTwocomponentHamiltonians2009,saueRelativisticHamiltoniansChemistry2011,dyallIntroductionRelativisticQuantum2007} Among various two-component approaches available, the exact two-component (X2C) theory based on atomic mean field (AMF) spin-orbit integrals (X2CAMF)  has emerged out to be one the most promising one.\cite{hessMeanfieldSpinorbitMethod1996,liuAtomicMeanfieldSpinorbit2018,zhangAtomicMeanFieldApproach2022,knechtExactTwocomponentHamiltonians2022} The X2CAMF based CC method provides a balance between computational efficiency and accuracy making it an optimal method for correlation calculations of heavy element containing systems\cite{zhangAtomicMeanFieldApproach2022}. In addition, one can approximate the two-electron integrals by using resolution of identity\cite{kelleyLargescaleDiracFock2013,batesFullyRelativisticComplete2015} or Cholesky decomposition,\cite{helmich-parisRelativisticCholeskydecomposedDensity2019,banerjeeRelativisticResolutionoftheidentityCholesky2023,uhlirovaCholeskyDecompositionSpinFree2024,zhangCholeskyDecompositionBasedImplementation2024} 
 which can drastically reduce the storage and memory requirements. One can also use the massively parallel programs, which are designed to scale across multiple compute nodes, to accelerate the floating-point operations.\cite{pototschnigImplementationRelativisticCoupled2021,deprinceCoupledClusterTheory2011} The other alternative is to use truncated natural spinors\cite{chamoliReducedCostFourcomponent2022} to reduce the floating-point operations involved in the calculations. Analogous to their non-relativistic counterparts \cite{lowdinQuantumTheoryManyParticle1955}, the natural spinors are the eigenfunctions of the spin-coupled correlated one-body reduced density matrix. The efficacy of frozen natural spinors (FNS) to reduce the computational cost of the standard relativistic four-component ground state coupled cluster method\cite{chamoliReducedCostFourcomponent2022} and unitary coupled cluster method\cite{majeeReducedCostFourcomponent2024} has already been demonstrated. Furthermore, this framework has also been extended to core and valence ionization problems\cite{surjuseLowcostFourcomponentRelativistic2022}. Gomes and co-workers have reported a similar natural spinor-based approach for the ground state coupled cluster method within the two-component framework\cite{majeeReducedCostFourcomponent2024}. The natural spinor framework can be combined with the Cholesky decomposition-based X2CAMF-CC method to obtain further reduction in the computational cost\cite{chamoli2025frozen}. However, Gomes and co-workers\cite{yuanFrequencyDependentQuadraticResponse2023} have shown that, unlike ground-state energy calculations, the standard MP2-based natural spinors do not provide a consistent description of various kinds of excited states. The standard natural spinors generated from the MP2 wavefunction do not contain information about the excited state electronic distribution. To get an optimal virtual space for correlated excited state calculations, one needs to use a set of natural spinors, which is generated from a good first-order description of the excited state wave function. 

In this manuscript, we present a cost-efficient implementation of the state-specific frozen natural spinor (SS-FNS)-based relativistic equation-of-motion coupled cluster method for the excited states.

%%%%%%%%%%%%%%%%%%%%%%%%%%%%%%%%%%%%%%%%%%%%%%%%%%%%%% Theory %%%%%%%%%%%%%%%%%%%%%%%%%%%%%%%%%%%%%%%%%%%%%%%%%%%%
\section{Theory}
\label{sec2}
\subsection{Relativistic coupled cluster theory}
\label{sec2.1}
The relativistic coupled cluster theory\cite{cizekCorrelationProblemAtomic1966, paldusCorrelationProblemsAtomic1972, lindrothNumericalSolutionRelativistic1988, visscherFormulationImplementationRelativistic1996, visscherFormulationImplementationRelativistic2001} uses an exponential ansatz on a Dirac-Hartree-Fock reference determinant $\left| {{\phi }_{0}} \right\rangle $ to generate the exact wave function $\left| {{\psi }_{CC}} \right\rangle$
\begin{equation}
\label{eqn1}
    \left| {{\psi }_{\text{CC}}} \right\rangle = {{e}^{{\hat{T}}}}\left| {{\phi }_{0}} \right\rangle
\end{equation}
The cluster operator $\hat{T}$ in Eq. (\ref{eqn1}) consists of a set of excitation operators ${{\hat{T}}}_{n}$ up to $n^{th}$ excitation manifold and can be represented as 
\begin{equation}
\label{eqn2}
    \hat{T}=\sum\limits_{n}{{{{\hat{T}}}_{n}}}
\end{equation}
The operator ${\hat{T}}_{n}$ can be written as
\begin{equation}
\label{eqn3}
    {{\hat{T}}_{n}}=\sum\limits_{\begin{smallmatrix} 
     i<j... \\ 
     a<b... 
    \end{smallmatrix}}{t_{ij...}^{ab...}\left\{ \hat{a}_{a}^{\dagger }\hat{a}_{b}^{\dagger }...\ {{{\hat{a}}}_{j}}{{{\hat{a}}}_{i}}...\right\}}
\end{equation}
Where $t_{ij...}^{ab...}$ are cluster amplitudes, $\hat{a}^{\dagger }$  and $\hat{a}$ denote second quantization creation and annihilation operators, indices $(i, j, ...)$ denote occupied spinors, and $(a, b, ...)$ represent virtual spinors.
In the coupled cluster singles and doubles (CCSD) approximation, the excitation manifold is considered up to doubles $\left( n=2 \right)$. Thus,
\begin{equation}
\label{eqn4}
    \hat{T}={{\hat{T}}_{1}}+{{\hat{T}}_{2}}
\end{equation}
The cluster amplitudes in the CCSD approximation ($t_i^a$, and $t_{ij}^{ab}$) can be obtained by solving a set of coupled nonlinear equations.
\begin{equation}
\label{eqn5}
    \left\langle  \phi _{i}^{a} \right|\bar{H}\left| {{\phi }_{0}} \right\rangle=0
\end{equation}
\begin{equation}
\label{eqn6}
    \left\langle  \phi _{ij}^{ab} \right|\bar{H}\left| {{\phi }_{0}} \right\rangle=0
\end{equation}
The relativistic ground state CCSD energy\cite{lindrothNumericalSolutionRelativistic1988, visscherFormulationImplementationRelativistic1996, visscherFormulationImplementationRelativistic2001} can be obtained as
\begin{equation}
\label{eqn7}
    \left\langle  {{\phi }_{o}} \right|\bar{H}\left| {{\phi }_{0}} \right\rangle ={{E}_{\text{CC}}}
\end{equation}
In Eq. (\ref{eqn5}-\ref{eqn6}), $\left| {\phi }_i^a \right\rangle$ and $\left| {\phi }_{ij}^{ab}
\right\rangle$  are the excited determinants and $\bar{H}$ is the similarity transformed Hamiltonian.

In the four-component formalism, $\bar{H}$ is  defined as
\begin{equation}
\label{eqn8}
    \bar{H}=e^{-\hat{T}}\hat{H}^{\text{4c}}e^{\hat{T}}
\end{equation}
\begin{comment}
where $\hat{H}^{\text{4c}}$ is the four-component Dirac-Coulomb (DC) Hamiltonian expressed as:
\begin{equation}
\label{eqn9}
    {{\hat{H}}^{\text{4c}}}=\sum\limits_{i}^{N}{\left[ c{{{\vec{\alpha }}}_{i}}.{{{\vec{p}}}_{i}}+{{\beta }_{i}}{{m}_{0}}{{c}^{2}}+\sum\limits_{A}^{{{N}_{\text{nuc}}}}{{{{\hat{V}}}_{iA}}} \right]}+\sum\limits_{i<j}^{N}{\frac{1}{{{r}_{ij}}}{{{\hat{I}}}_{4}}}
\end{equation}
with $\alpha$ and $\beta$ as Dirac matrices, $\sigma$ denoting the Pauli spin matrices and $\hat{p}$ as the momentum operator. $c$ and ${{m}_{0}}$ are the speed of light and the rest mass of the electron, respectively. $\hat{V}_{iA}$ represents the nuclear potential. ${{\hat{I}}_{4}}$ is a $4\times 4$ identity matrix and ${{r}_{ij}}$ is the distance between ${{i}^{th}}$ and ${{j}^{th}}$ electrons.
\end{comment}
where $\hat{H}^{\text{4c}}$ is the four-component Dirac-Coulomb (DC) Hamiltonian, which in second-quantized form is expressed as:
\begin{equation}
\label{eqn9}
    \hat{H}^{\text{4c}} = \sum_{pq}{h^{\text{4c}}_{pq}\hat{a}_p^{\dagger}\hat{a}_q} 
    + \sum_{pqrs}{\frac{1}{4}g^{\text{4c}}_{pqrs}\hat{a}_p^{\dagger}\hat{a}_q^{\dagger}\hat{a}_s\hat{a}_r}
\end{equation}
As we are only interested in the positive energy spectrum, the no-pair approximation\cite{sucherFoundationsRelativisticTheory1980, reiherRelativisticQuantumChemistry2015, dyallIntroductionRelativisticQuantum2007} is used for the correlation calculations.

\subsection{Equation-of-motion approach for excited states}
\label{sec2.2}
In the equation-of-motion formalism\cite{roweEquationsofMotionMethodExtended1968}, we consider two states simultaneously\cite{shavittManyBodyMethodsChemistry2009}, initial state $\left| {{\psi }_{0}} \right\rangle$ and target state $\left| {{\psi }_{k}} \right\rangle$.
\begin{equation}
\label{eqn10}
    \hat{H}^{\text{4c}}\left| {{\psi }_{0}} \right\rangle ={{E}_{0}}\left| {{\psi }_{0}} \right\rangle 
\end{equation}
\begin{equation}
\label{eqn11}
    \hat{H}^{\text{4c}}\left| {{\psi }_{k}} \right\rangle ={{E}_{k}}\left| {{\psi }_{k}} \right\rangle 
\end{equation}
The target state $\left| {{\psi }_{k}} \right\rangle $ can be generated from the initial state $\left| {{\psi }_{0}} \right\rangle $ by operating a linear operator ${{\hat{R}}_{k}}$ on it
\begin{equation}
\label{eqn12}
    \left| {{\psi }_{k}} \right\rangle ={{\hat{R}}_{k}}\left| {{\psi }_{0}} \right\rangle 
\end{equation}
For the excited states, the operator ${{\hat{R}}_{k}}$ has the following form
\begin{equation}
\label{eqn13}
    \hat{R}_{k} = r_{0}
    + \sum_{i,a} r_{i}^{a} \left\{ \hat{a}_{a}^{\dagger} \hat{a}_{i} \right\}
    + \sum_{\substack{i<j \\ a<b}} r_{ij}^{ab} \left\{ \hat{a}_{a}^{\dagger} \hat{a}_{i} \hat{a}_{b}^{\dagger} \hat{a}_{j} \right\}
    + \dots
\end{equation}

\begin{comment}
By using the normal-ordered Hamiltonian in equations (\ref{eqn10}) and (\ref{eqn11}), one can obtain
\begin{equation}
\label{eqn14}
    {{\hat{H}}^{\text{4c}}_{N}}\left| {{\psi }_{0}} \right\rangle =\Delta {{E}_{0}}\left| {{\psi }_{0}} \right\rangle 
\end{equation}
\begin{equation}
\label{eqn15}
    {{\hat{H}}_{N}^{\text{4c}}}\left| {{\psi }_{k}} \right\rangle =\Delta {{E}_{k}}\left| {{\psi }_{k}} \right\rangle
\end{equation}
Where $\Delta E_{0}=E_{0}-E_{\text{DHF}}$  and $\Delta E_{k}=E_{k}-E_{\text{DHF}}$ . \\
\end{comment}

When the coupled cluster ground state $\left| {{\psi }_{CC}} \right\rangle $ is taken to be the initial state, then equation (\ref{eqn12}) modifies to
\begin{equation}
\label{eqn14}
    \left| {{\psi }_{k}} \right\rangle ={{\hat{R}}_{k}}{{e}^{{\hat{T}}}}\left| {{\phi }_{0}} \right\rangle
\end{equation}
From equation (\ref{eqn11})
\begin{equation}
\label{eqn15}
    	{{\hat{H}}^{\text{4c}}_{}}{{\hat{R}}_{k}}{{e}^{{\hat{T}}}}\left| {{\phi }_{0}} \right\rangle = {{E}_{k}}{{\hat{R}}_{k}}{{e}^{{\hat{T}}}}\left| {{\phi }_{0}} \right\rangle
\end{equation}
One can directly calculate the excitation energy by using the commutator form of equation (\ref{eqn15})
\begin{equation}
\label{eqn16}
    \left[ {{{\bar{H}}}^{\text{}}_{}},{{{\hat{R}}}_{k}} \right]\left| {{\phi }_{0}} \right\rangle =\left({{E}_{k}}- {{E}_{0}} \right){{\hat{R}}_{k}}\left| {{\phi }_{0}} \right\rangle ={{\omega }_{k}}{{\hat{R}}_{k}}\left| {{\phi }_{0}} \right\rangle 
\end{equation}
where ${{\omega }_{k}}$ is the excitation energy for the ${{k}^{th}}$ excited state
\begin{equation}
\label{eqn17}
    {{\omega }_{k}}={{E}_{k}}-{{E}_{0}}
\end{equation}
Due to the non-Hermitian nature of
${\bar{H}}^{\text{}}_{}$, a corresponding left eigenvector also exists, which is bi-orthogonal to the right eigenvector
\begin{equation}
\label{eqn18}
  \left\langle  {{\phi }_{o}} \right|
  \hat{L}_{i} \hat{R}_{j} \left| {{\phi }_{0}} \right\rangle =\delta_{ij}
 \end{equation}
For excited states $\hat{L}_{k}$ operator is defined as
\begin{equation}
\label{eqn19}
\hat{L}_k =  \sum_{ia} l_a^i 
\left\{ \hat{a}_i^\dagger \hat{a}_a \right\}
+ \sum_{\substack{i<j \\ a<b}} l_{ab}^{ij} 
\left\{ \hat{a}_i^\dagger \hat{a}_a \hat{a}_j^\dagger \hat{a}_b \right\} 
+ \dots
\end{equation}
To calculate any first-order property corresponding to the operator $\Theta$ in the EOM-CC framework, one needs to calculate the expectation value\cite{mukhopadhyayAnalyticCalculationTransition2025}
 \begin{equation}
 \label{eqn20}
\langle \Theta \rangle = \langle \psi | \hat{\Theta} | \psi \rangle 
  \end{equation}
For example, the square of the ground to excited transition dipole moment can be calculated as 
\begin{equation}
\label{eqn21}
\left| \mu_{o \to k} \right|^2 =
\langle \phi_0 | (1 + \hat{\Lambda}) \bar{\mu} \hat{R}_k | \phi_0 \rangle
\langle \phi_0 | \hat{L}_k \bar{\mu} | \phi_0 \rangle
\end{equation}
Where $\hat{\Lambda}$ is the coupled cluster de-excitation operator represented as
\begin{equation}
\label{eqn22}
\hat{\Lambda} = \hat{\Lambda}_1 + \hat{\Lambda}_2 + \hat{\Lambda}_3 + \dots 
\end{equation}
with
\begin{comment}
\begin{equation}
\label{eqn23}
\begin{gathered}
    \hat{\Lambda}_1 = \sum_{ia} \lambda^a_i 
    \left\{ \hat{a}^\dagger_i \hat{a}_a \right\}, \\
    \hat{\Lambda}_2 = \frac{1}{4} \sum_{i<j a<b} \lambda^{ab}_{ij} 
    \left\{ \hat{a}^\dagger_i \hat{a}_a \hat{a}^\dagger_j \hat{a}_b \right\},
\end{gathered}
\end{equation}
and 
\end{comment}
\begin{equation}
\label{eqn23}
\hat{\Lambda}_n = \left(\frac{1}{n!} \right)^2
\sum_{\substack{i<j\cdots \\ a<b\cdots}} 
\lambda^{ab\cdots}_{ij\cdots} 
\left\{ \hat{a}^\dagger_i \hat{a}_a... \hat{a}^\dagger_j \hat{a}_b \cdots \right\}
\end{equation}
The experimentally observable property, oscillator strength, is related to the square of the transition dipole moment.
\begin{equation}
\label{eqn24}
f_{0 \to k} = \frac{2}{3} \Delta \omega_{k} \left| \mu_{0 \to k} \right|^2.
\end{equation}
In the EOM-CCSD method, the $\hat{T}$, $\hat{R}$, $\hat{L}$ and $\hat{\Lambda}$ are truncated at singles and doubles excitation level and the most expensive step in the calculation scales as $O(N_{occ}^2N_{vir}^4)$ power of the basis set, where $N_{occ}$ and $N_{vir}$ are no of occupied and virtual spinor respectively.
For the calculation of excitation energy, it is sufficient to solve for either left or right eigenvectors, in addition to the ground state coupled cluster amplitudes. For the calculation of properties, one needs to calculate both left and right eigenvectors, as well as ground state $t$ and $\lambda$ amplitudes. Therefore, the transition property calculation in the relativistic EOM-CCSD method is at least twice as costly as the energy calculations.  One can consider natural spinor-based techniques to truncate the virtual space.

\subsection{State-specific frozen natural spinors}
\label{sec2.3}
Natural spinors\cite{chamoliReducedCostFourcomponent2022, yuanAssessingMP2Frozen2022, chamoli2025frozen, majeeReducedCostFourcomponent2024} are the relativistic counterparts of the natural orbitals introduced by Löwdin\cite{lowdinQuantumTheoryManyParticle1955}. They are the eigenfunctions of the correlated spin-coupled one-body reduced density matrix\cite{chamoliRelativisticReducedDensity2024}. The ground-state natural spinors are typically constructed from the first-order MP2 wave function. In the FNS framework, the natural spinors are generated for the virtual space, whereas the occupied space is kept at their DHF description. After the SCF calculation, a partial AO to MO integral transformation is performed to generate integrals with two external indices, and one can calculate the virtual-virtual block of the one-particle reduced density matrix at the MP2 level
  \begin{equation}
    \label{eqn25}
    D_{ab}=\frac{1}{2}\sum_{cij}^{} {\frac{\langle ac||ij \rangle \hspace{0.1cm}\langle ij||bc \rangle}{\varepsilon_{ij}^{ac} \hspace{0.2cm}\varepsilon_{ij}^{bc}}}
    \end{equation}

The virtual-virtual block of this reduced density matrix $\left( {{D}_{ab}}\right)$ is then diagonalized to obtain virtual natural spinors (VNS) as eigenfunctions $V$ and the corresponding occupancies as eigenvalues $\eta$.
\begin{equation}
\label{eqn26}
    D_{ab}V=V\eta
\end{equation}
A pre-determined threshold ${{\eta }_{crit}}$ can be used to drop off the virtual natural spinors with lesser occupancies, significantly reducing the computational cost for further calculations.
\begin{equation}
\label{eqn27}
    \tilde{V}=VT
\end{equation}
where ${{T}_{ij}}={{\delta }_{ij}}$ if ${{\eta }_{i}}\ge {{\eta }_{crit}}$ and ${{T}_{ij}}=0$ otherwise.
The virtual-virtual block of the Fock matrix is transformed to this truncated natural spinor basis.
\begin{equation}
\label{eqn28}
    \tilde{F}={{\tilde{V}}^{\dagger }}F\tilde{V}
\end{equation}
The transformed fock matrix $\left(\tilde{F}\right)$ is diagonalized to semi-canonicalize the basis,	
\begin{equation}
\label{eqn29}
    \tilde{F}\tilde{Z}=\tilde{Z}\tilde{\epsilon }
\end{equation}
The set of eigenvectors $\tilde{Z}$ and the retained virtual natural spinors $\tilde{V}$ together form the transformation matrix $\left( B \right)$ from the canonical to the truncated FNS basis.
\begin{equation}
\label{eqn30}
    B=\tilde{V}\tilde{Z}
\end{equation}
\begin{comment}
If the occupied and virtual spinors are obtained from atomic integrals using the transformation matrix ${{U}_{occ}}$ and ${{U}_{vir}}$ respectively, then the direct transformation to the truncated FNS basis from the atomic orbital basis can be obtained using the following transformation matrices respectively
\begin{equation}
\label{eqn38}
    {{\tilde{U}}_{occ}}={{U}_{occ}}
\end{equation}
\begin{equation}
\label{eqn39}
    {{\tilde{U}}_{vir}}={{U}_{vir}}B
\end{equation}
\end{comment}

In subsequent discussions, this scheme of generating frozen natural spinors from MP2 density will be called standard FNS. The natural spinors\cite{chamoliReducedCostFourcomponent2022, yuanAssessingMP2Frozen2022, chamoli2025frozen, majeeReducedCostFourcomponent2024} obtained in the aforementioned way are optimized for recovering the ground state correlation energy, but are not suitable for the excited states. Now, there is no common consensus on how to generate an appropriate set of natural orbitals for the excited states, even in the non-relativistic case. Hättig and co-workers\cite{helmich2011local} have used state-specific pair natural orbitals (PNOs) obtained from the CIS(D) method to generate an appropriate virtual space for the excited states. Kállay and co-workers\cite{mester2017reduced}, on the other hand, used state-specific excited state natural orbitals generated from CIS(D) density to reduce the computational cost of CC2 calculations. Valeev and coworkers \cite{peng2018state} have used state-averaged PNOs for excited states. Neese and co-workers\cite{dutta2018exploring} have taken an alternate strategy to combine the technique of similarity transformation with the ground state PNOs to arrive at a low-cost implementation of the coupled cluster-based excited state method. We have recently shown that the ADC(2) method gives an accurate first-order description of the excited state wave function, and the ADC(2) natural orbital-based EOM-CCSD method\cite{manna2025reducedcostequationmotion} gives uniform accuracy for valence, Rydberg, and charge transfer states.

Following the same philosophy, the state-specific one-particle reduced density matrix calculated at the relativistic ADC(2) level of theory\cite{pernpointnerRelativisticPolarizationPropagator2014, pernpointnerFourComponentPolarizationPropagator2018, chakraborty2025relativistic} can be used to generate the state-specific natural spinors for the relativistic EOM-CCSD method. In the present case, the zeroth-order intermediate state representation\cite{schirmer2004intermediate} has been used to calculate the excited state density.  For the $k^{th}$ excited state, the virtual-virtual block of the state-specific reduced density matrix can be obtained as,
\begin{equation}
\label{eqn31}
    D_{ab}^{\text{SS}}(k)=D_{ab}^{\text{MP2}}+D_{ab}^{\text{EE-ADC(2)}}(k)
\end{equation}
Here, $D_{ab}^{\text{MP2}}$ are the virtual-virtual block of the one-particle reduced density matrix calculated at the MP2 level and $D_{ab}^{\text{EE-ADC(2)}}(k)$ is the difference density matrix for the k$^{th}$ excited state calculated at EE-ADC(2) level of theory, respectively.  The explicit expressions for these reduced density matrices are provided in the Appendix.  The programmable expression for the relativistic ADC(2) sigma vectors can be found in Ref. \onlinecite{chakraborty2025relativistic}. The state-specific reduced density matrix obtained from Eq. (\ref{eqn31}) can be put into Eq. (\ref{eqn26}) to obtain the transformation matrix for the state-specific frozen natural spinor basis using Eqs. (\ref{eqn26}-\ref{eqn30}).  The subsequent coupled cluster and EOM-CCSD calculation is performed in the state-specific frozen natural spinor (SS-FNS) basis.
Now, the ground-state coupled cluster amplitudes need to be solved for each excited state separately.  This incurs an additional cost for each excited state, but the computational gain from the reduced virtual space size offsets this cost. Moreover,  this makes the wave function for each excited state bi-orthogonal to the ground state wave function and gives a way to calculate the ground-to-excited-state transition property. 

One can make a perturbative correction for the truncated natural spinors by taking the difference between the EE-ADC(2) energies in the canonical basis and in the truncated state-specific natural spinor basis. This correction in energy can be added to the uncorrected SS-FNS-EE-EOM-CCSD as a perturbative correction for the excited states.

{\small
\begin{equation}
\label{eqn32}
        \omega_{\text{SS-FNS-EOM}}^{\text{corrected}}(k) = \omega_{\text{SS-FNS-EOM}}^{\text{uncorrected}}(k) + 
        \omega_{\text{EE-ADC(2)}}^{\text{canonical}}(k)-\omega_{\text{EE-ADC(2)}}^{\text{SS-FNS}}(k)
\end{equation}}
Now, to ensure that each excited state is solved in the SS-FNS basis designed for it, one needs to solve for each root separately using a root-wise modified Davidson solver\cite{hiraoGeneralizationDavidsonsMethod1982}, where at each iteration, the overlap of the new vectors is checked with the guess vectors to ensure correct root homing.
The canonical ADC(2) eigen vectors are transformed to the corresponding natural spinor basis and used as the initial guess for the subsequent  EOM-CCSD and ADC(2) calculation in the FNS basis.

In order to solve the relativistic ADC(2)\cite{pernpointnerRelativisticPolarizationPropagator2014, pernpointnerFourComponentPolarizationPropagator2018, chakraborty2025relativistic} problem for excited states, one needs to generate the integrals with one, two, and three external indices in the canonical basis, in addition to the two-external integral necessary for standard FNS calculations. 
The generation of three-indices external integrals in the 4c-DC framework can become the bottleneck for calculations in large basis sets. Moreover, the full integral transformation, even in the truncated natural spinor basis, must be repeated for each excited state and can be extremely costly. Thus, unlike the case of a ground-state coupled cluster method, one cannot have a practical implementation of a natural spinor-based low-cost EOM-CCSD method based on a four-component Dirac-Coulomb Hamiltonian without taking additional approximations. As a potential solution, one can switch to two-component methods.

\subsection{The X2CAMF approximation }
\label{sec2.4}
\begin{comment}
The four-component DC Hamiltonian $\left(\hat{H}^{\text{4c}}\right)$ in the second quantized form can be written as:
\begin{equation}
\label{eqn37}
    \hat{H}^{\text{4c}} = \sum_{pq}{h^{\text{4c}}_{pq}\hat{a}_p\hat{a}_q} 
    + \sum_{pqrs}{\frac{1}{4}g^{\text{4c}}_{pqrs}\hat{a}_p\hat{a}_q\hat{a}_s\hat{a}_r}
\end{equation}
In the exact two-component (X2C) theory, the above four-component DC Hamiltonian can be transformed using the X2C decoupling scheme.\cite{dyallInterfacingRelativisticNonrelativistic1997} 
\end{comment}
Under the spin-separation scheme,\cite{dyallExactSeparationSpinfree1994}, the two-electronic part of the four-component DC Hamiltonian ($\hat{H}^{\text{4c}}$) can be separated into spin-free (SF) and spin-dependent (SD) parts.
\begin{equation}
\label{eqn33}
    \hat{H}^{\text{4c}} = \sum_{pq}{h^{\text{4c}}_{pq}\hat{a}_p^{\dagger}\hat{a}_q}
    + \frac{1}{4}\sum_{pqrs}{g^{\text{4c,SF}}_{pqrs}\hat{a}_p^{\dagger}\hat{a}_q^{\dagger}\hat{a}_s\hat{a}_r}
    + \frac{1}{4}\sum_{pqrs}{g^{\text{4c,SD}}_{pqrs}\hat{a}_p^{\dagger}\hat{a}_q^{\dagger}\hat{a}_s\hat{a}_r}
\end{equation}
%\label{eqn11}
%    \hat{H}^{4c} = \sum_{pq}{h^{4c}_{pq}\hat{a}_p\hat{a}_q}
%    &+ \frac{1}{4}\sum_{pqrs}{g^{4c,SF}_{pqrs}\hat{a}_p\hat{a}_q\hat{a}_s\hat{a}_r} \\
%%    &+ \frac{1}{4}\sum_{pqrs}{g^{4c,SD}_{pqrs}\hat{a}_p\hat{a}_q\hat{a}_s\hat{a}_r} \\
%\end{align}

The spin-dependent part of the above Hamiltonian can be treated within the atomic mean field (AMF) approximation\cite{hessMeanfieldSpinorbitMethod1996,liuAtomicMeanfieldSpinorbit2018,knechtExactTwocomponentHamiltonians2022, zhangAtomicMeanFieldApproach2022}, exploiting the localized nature of the spin-orbit interactions.
\begin{equation}
\label{eqn34}
    \frac{1}{4}\sum_{pqrs}{g^{\text{4c,SD}}_{pqrs}\hat{a}_p^{\dagger}\hat{a}_q^{\dagger}\hat{a}_s\hat{a}_r}
    \approx \sum_{pq}{g^{\text{4c,AMF}}_{pq}\hat{a}_p^{\dagger}\hat{a}_q}=\sum_{pqiA}{n_{iA}}g^{\text{4c,SD}}_{p_{iA}q_{iA}}\hat{a}_p^{\dagger}\hat{a}_q
\end{equation}
Where $A$ represents the distinct atoms in the molecule, $i$ is occupied spinors for atom $A$, and $n_{iA}$ denotes their corresponding occupation numbers. Substituting Eq. (\ref{eqn34}) in Eq. (\ref{eqn33}), we get,
\begin{equation}
\label{eqn35}
    %\begin{align}
    \hat{H}^{\text{4c}} = \sum_{pq}{h^{\text{4c}}_{pq}\hat{a}_p^{\dagger}\hat{a}_q}
    + \frac{1}{4}\sum_{pqrs}{g^{\text{4c,SF}}_{pqrs}\hat{a}_p^{\dagger}\hat{a}_q^{\dagger}\hat{a}_s\hat{a}_r}
    + \sum_{pq}{g^{\text{4c,AMF}}_{pq}\hat{a}_p^{\dagger}\hat{a}_q}
    %\end{align}
\end{equation}
The above four-component Hamiltonian can be transformed into a two-component picture by utilizing the relation between coefficients of the large and small component wavefunctions through the  $X$ matrix and a relation between coefficients of the large component and two-component wavefunction through the $R$ matrix.
\begin{equation}
\label{eqn36}
    C^S=XC^L
\end{equation}
\begin{equation}
\label{eqn37}
    C^L=RC^{\text{2c}}
\end{equation}
The spin-free contribution of the Coulomb interaction $g^{\text{4c,SF}}$ is reduced to the non-relativistic two-electron integrals $g^{\text{NR}}$ when scalar two-electron picture-change (2e-pc) contributions are neglected.
%\begin{equation}
%\label{eqn43}
%    g^{\text{4c,SF}}_{pqrs} \approx g^{\text{NR}}_{pqrs}
%\end{equation}
The four-component Hamiltonian, after transformation, becomes the two-component X2CAMF Hamiltonian
\begin{equation}
\label{eqn38}
    %\begin{align}
    \hat{H}^{\text{X2CAMF}} = \sum_{pq}{h^{\text{X2C}}_{pq}\hat{a}_p^{\dagger}\hat{a}_q}
    + \frac{1}{4}\sum_{pqrs}{g^{\text{NR}}_{pqrs}\hat{a}_p^{\dagger}\hat{a}_q^{\dagger}\hat{a}_s\hat{a}_r}
    + \sum_{pq}{g^{\text{2c,AMF}}_{pq}\hat{a}_p^{\dagger}\hat{a}_q}
    %\end{align}
\end{equation}
The above Hamiltonian can be written in terms of an effective one-electron operator and a non-relativistic two-electron operator as:
\begin{equation}
\label{eqn39}
    \hat{H}^{\text{X2CAMF}} = \sum_{pq}{h^{\text{X2CAMF}}_{pq}\hat{a}_p^{\dagger}\hat{a}_q}
    + \frac{1}{4}\sum_{pqrs}{g^{\text{NR}}_{pqrs}\hat{a}_p^{\dagger}\hat{a}_q^{\dagger}\hat{a}_s\hat{a}_r}
\end{equation}
Where $h^{\text{X2CAMF}} = h^{\text{X2C}}+g^{\text{2c,AMF}}$ is the effective one-electron operator. The main advantage of using the above Hamiltonian is that it completely avoids the construction of the relativistic two-electron integrals. In X2CAMF-based CC methods, the similarity transformed Hamiltonian can be written as
\begin{equation}
\label{eqn40}
    \bar{H}=e^{-\hat{T}}\hat{H}^{\text{X2CAMF}}e^{\hat{T}}
\end{equation}

\subsection{Cholesky decomposition}
\label{sec2.5} 
%We have used the Cholesky decomposition (CD) technique for the two-electron integrals as it %does not require a pre-optimized auxiliary basis set as compared to the closely related %density fitting (DF) or resolution of identity (RI) %techniques\cite{hattigStructureOptimizationsExcited2005,hohensteinDensityFittingCholesky2010}.
Under Cholesky decomposition (CD) approximation, electron repulsion integrals (ERIs) can be approximated as
\begin{equation}
\label{eqn41}
    \left(\mu\nu|\kappa\lambda\right) \approx \sum_{P}^{n_{\text{CD}}}{L_{\mu\nu}^{P}L_{\kappa\lambda}^{P}}
\end{equation}
Where  $\mu,  \nu,  \kappa, \lambda$  are atomic spinor indices, $L_{\mu\nu}^{P}$  denotes the Cholesky vectors and $n_{\text{CD}}$ is the dimension of Cholesky vectors. The present implementation follows a one-step algorithm where the Cholesky vectors are generated using an iterative procedure by identifying the largest diagonal elements of the ERI  matrix, and the process goes on till this largest diagonal element comes below the predefined Cholesky threshold ($\tau$). Afterwards, the Cholesky vectors are transformed to the molecular spinor basis as follows:
\begin{equation}
\label{eqn42}
    L_{pq}^{P} = \sum_{\mu\nu}C_{\mu p}^* L_{\mu\nu}^{P}C_{\nu q}
\end{equation}
The molecular basis (MO) Cholesky vectors can be used to form anti-symmetrized two-electron integrals
\begin{equation}
\label{eqn43}
    \left\langle pq||rs\right\rangle = \sum_{P}^{n_{\text{CD}}} \left ( L_{pr}^{P}L_{qs}^{P} - L_{ps}^{P}L_{qr}^{P} \right )
\end{equation}
In the current implementation, formation or storage of integrals with four external indices  $\left( \left\langle ab||cd\right\rangle \right)$ and three external indices $\left( \left\langle ab||ci\right\rangle \right)$  is completely avoided. 
These are generated on the fly wherever needed. Other integrals are explicitly constructed and stored in the FNS basis. More details on the FNS and Cholesky decomposed X2CAMF-based implementation of the relativistic coupled cluster method can be found in Ref. \onlinecite{chamoli2025frozen}.

%%%%%%%%%%%%%%%%%%%%%%%%%%%%%%%%%%%%%%%%%%%% Computational Details %%%%%%%%%%%%%%%%%%%%%%%%%%%%%%%%%%%%%%%%%%%%%%%%%%%%
\section{Computational Details}
\label{sec3}
% BAGH information
The SS-FNS-EE-EOM-CCSD method based on both 4c-DC and X2CAMF Hamiltonian has been implemented in our in-house software package BAGH. \cite{duttaBAGHQuantumChemistry2025}.It is primarily written in Python, with performance-critical sections written in Cython and FORTRAN. It is interfaced with PySCF\cite{sunLibcintEfficientGeneral2015, sunPySCFPythonbasedSimulations2018, sunRecentDevelopmentsSCF2020}, socutils\cite{wangXubwaSocutils2025}, DIRAC\cite{bastDIRAC232023} and GAMESS-US\cite{barcaRecentDevelopmentsGeneral2020}. 
% Algorithm
The 4c-DC-SS-FNS-EE-EOM-CCSD and X2CAMF-SS-FNS-EE-EOM-CCSD methods follow slightly different algorithms. In the 4c-DC-SS-FNS-EE-EOM-CCSD method, after the DHF calculation, $\left\langle  ai \right|\left| jk \right\rangle $, $\left\langle  ij \right|\left| ab \right\rangle$ , $\left\langle  ia \right|\left| jb \right\rangle $, and  $\left\langle  ia \right|\left| bc \right\rangle $ integrals are generated in the canonical basis, and the canonical ADC(2) calculation is performed to generate state-specific natural spinors.   The  $\left\langle  ab\right|\left| cd \right\rangle$ and $\left\langle  ij \right|\left| kl \right\rangle$ integrals are only generated in the truncated SS-FNS basis, and the rest of the integrals are transformed to the truncated FNS basis for each root.  Whereas in the case of the X2CAMF-SS-FNS-EE-EOM-CCSD method, none of the integrals are generated in the canonical basis. The steps involved in the X2CAMF-SS-FNS-EE-EOM-CCSD method are as follows. 
\begin{enumerate}
    \item Solve the X2CAMF-HF equations and construct the ground state reference wave function.
    \item Generate  three-centered two-electron integrals - $L^P_{ij}$, $L^P_{ia}$ and $L^P_{ab}$ in the canonical natural spinor basis using Cholesky decomposition. 
    \item Perform relativistic ADC(2) calculation, save the converged eigenvectors, and generate the truncated SS-FNS basis for each state.
    \item Loop over each excited state.
    \item Transform the  ADC(2) eigenvector and three-centered two-electron integrals to the truncated SS-FNS basis of the corresponding state.
    \item Generate four-centered two-electron integrals up to two external indices from the transformed three-centered two-electron integrals. 
    \item Perform ADC(2) calculation in the truncated SS-FNS basis.
    \item Perform the EE-EOM-CCSD calculation in the truncated SS-FNS basis.
    \item Add the perturbative correction.
\end{enumerate}

\begin{figure}[htbp]
    \centering
    \includegraphics[width=0.45\textwidth]{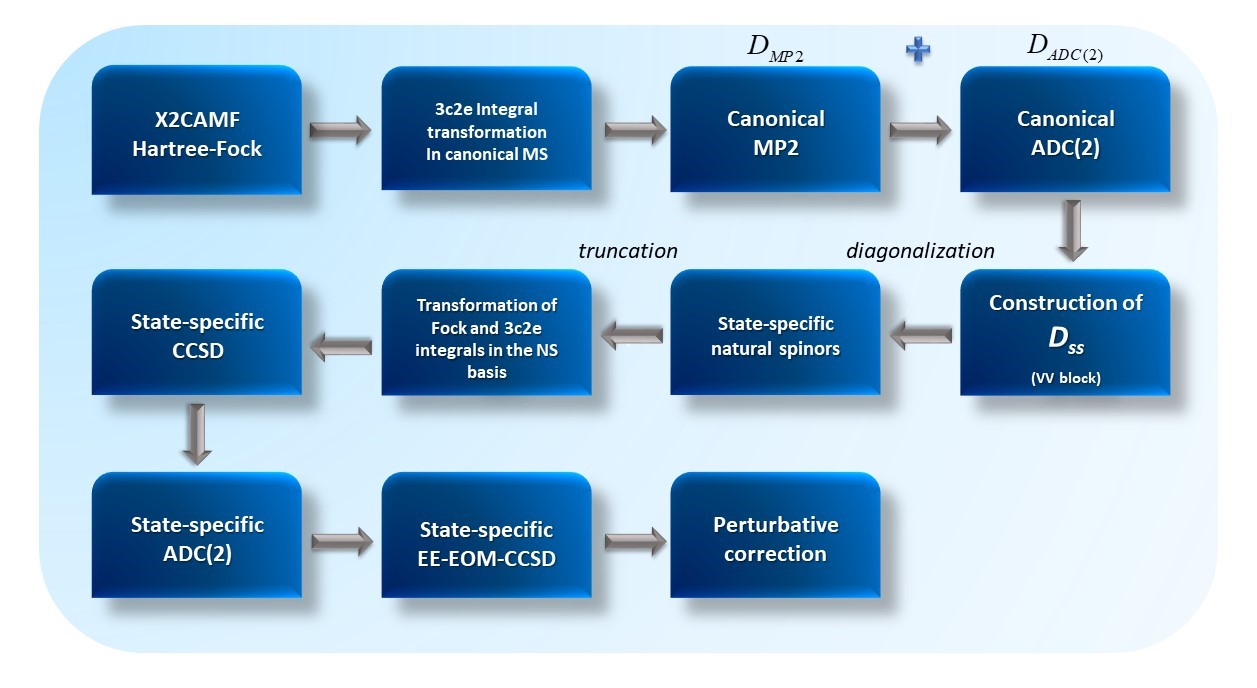}
    \caption{\justifying{The schematic diagram of the algorithm of the SS-FNS-EE-EOM-CCSD method. A CD threshold of $10^{-3}$ has been used for all CD-X2CAMF calculations.}}
    \label{fig:algorithm}
\end{figure}

FIG. \ref{fig:algorithm}  presents a schematic description of the X2CAMF-SS-FNS-EE-EOM-CCSD algorithm.

% FIGURE: algorithm

% Basis set information

%For majority of the calculations involve Dyall's basis sets.\cite{dyallRelativisticNonrelativisticFinite2002, dyallRelativisticDoublezetaTriplezeta2007, dyallRelativisticDoubleZetaTripleZeta2009, dyallRelativisticDoublezetaTriplezeta2004, dyallRelativisticDoublezetaTriplezeta2011, dyallRelativisticDoublezetaTriplezeta2012, dyallRelativisticDoublezetaTriplezeta2007a, dyallRelativisticDoublezetaTriplezeta2016, dyallCoreCorrelatingBasis2012} For all calculations involving Zn atom, dyall.v2z basis set has been used. For AuH molecule calculations, singly augmented dyall.v2z basis set has been used for Au atom and uncontracted aug-cc-pVDZ basis set has been used for H atom. The uncontracted triply augmented cc-pVTZ basis set has been used for calculations involving Ga$^+$ ion and uncontracted triply augmented dyall.v3z basis set has been used for the calculations involving In$^+$ and Tl$^+$ ions. For triiodide ion (I$_3^-$), dyall.av3z basis set has been employed. Xe atom excitation energies and transition dipole moments are calculated using triply augmented dyall.ae3z basis set.

%%%%%%%%%%%%%%%%%%%%%%%%%%%%%%%%%%%%%%%%%%%% Results and Discussion %%%%%%%%%%%%%%%%%%%%%%%%%%%%%%%%%%%%%%%%%%%%%%%%
\section{Results and Discussion}
\label{sec4}
%\textcolor{blue}{We have benchmarked the energies and transition properties of the SS-FNS-EE-%EOM-CCSD method for both atoms and molecules.} 

\subsection{Determining an appropriate threshold }
\label{sec4.1}

% FIGURE: Zn_combined_pct
\begin{figure*}[ht!]
\centering
    % First subplot
    \begin{subfigure}{0.45\textwidth} % Adjust width as needed
        \includegraphics[width=\linewidth]{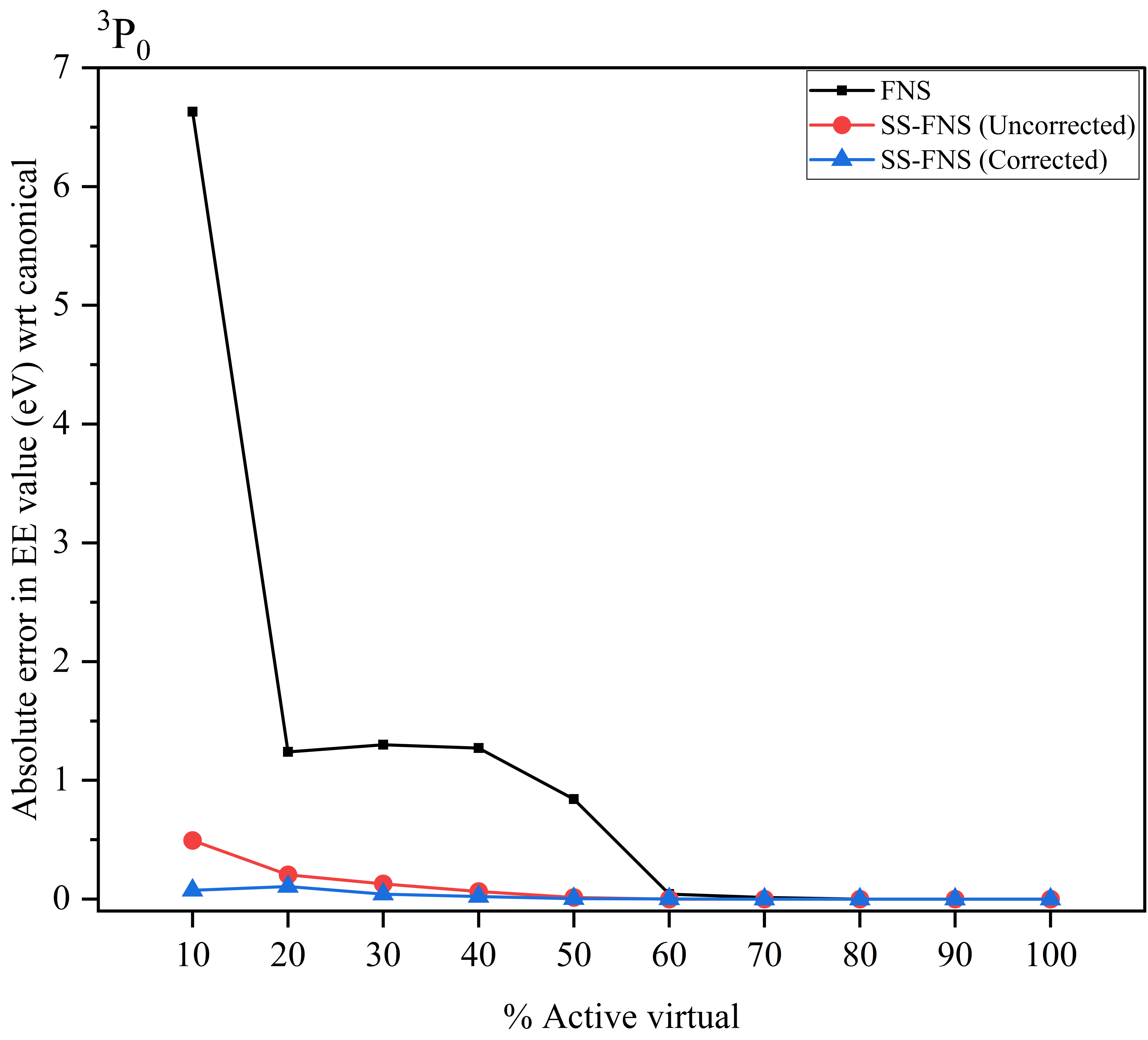} % Replace with your file
        \caption{}
        \label{fig:Zn_combined_pct:sub1}
    \end{subfigure}
    \begin{subfigure}{0.45\textwidth}
        \includegraphics[width=\linewidth]{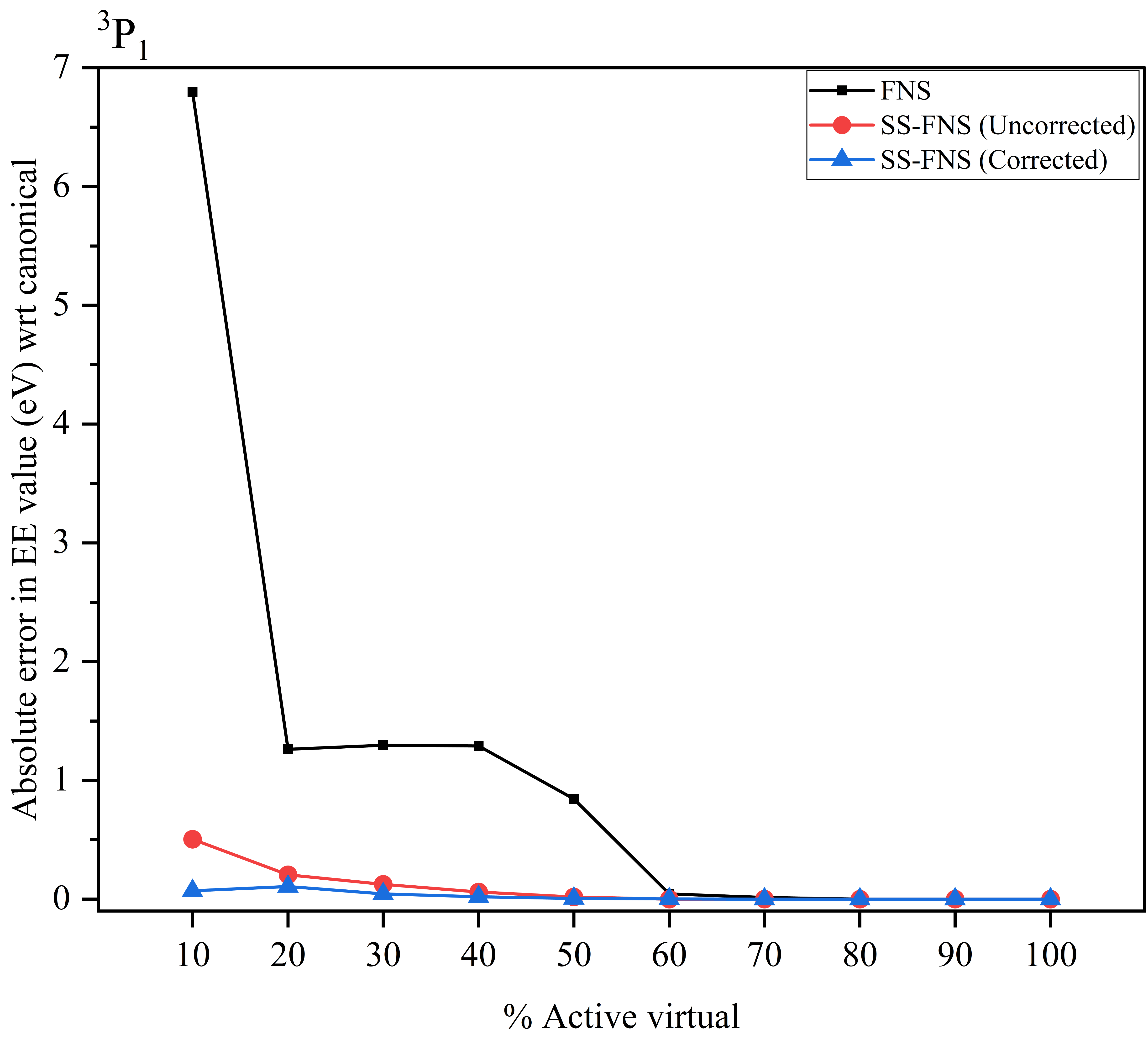}
        \caption{}
        \label{fig:Zn_combined_pct:sub2}
    \end{subfigure}

    % Second column (three figures)
    \begin{subfigure}{0.45\textwidth}
        \includegraphics[width=\linewidth]{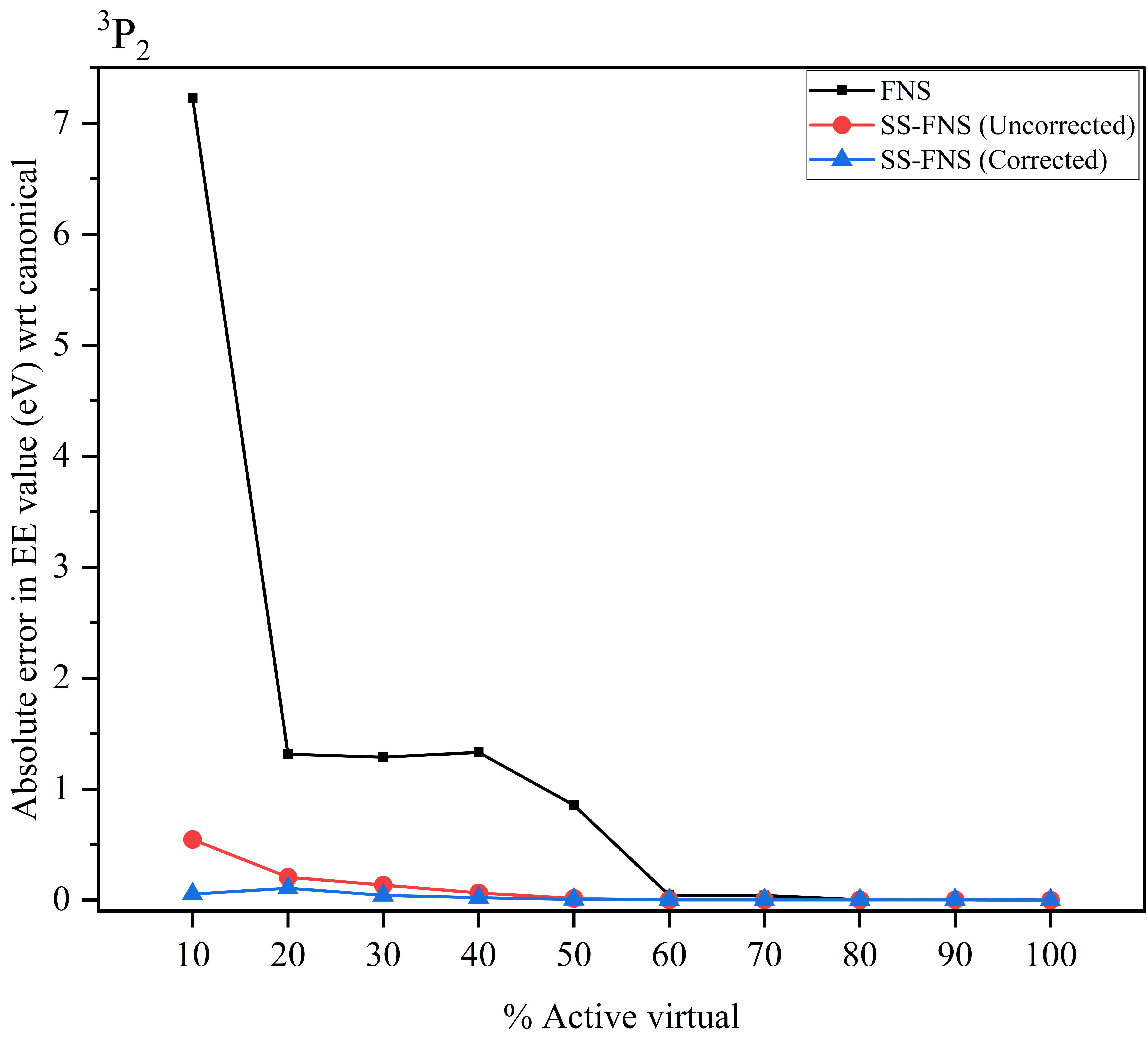}
        \caption{}
        \label{fig:Zn_combined_pct:sub3}
    \end{subfigure}
    \begin{subfigure}{0.45\textwidth}
        \includegraphics[width=\linewidth]{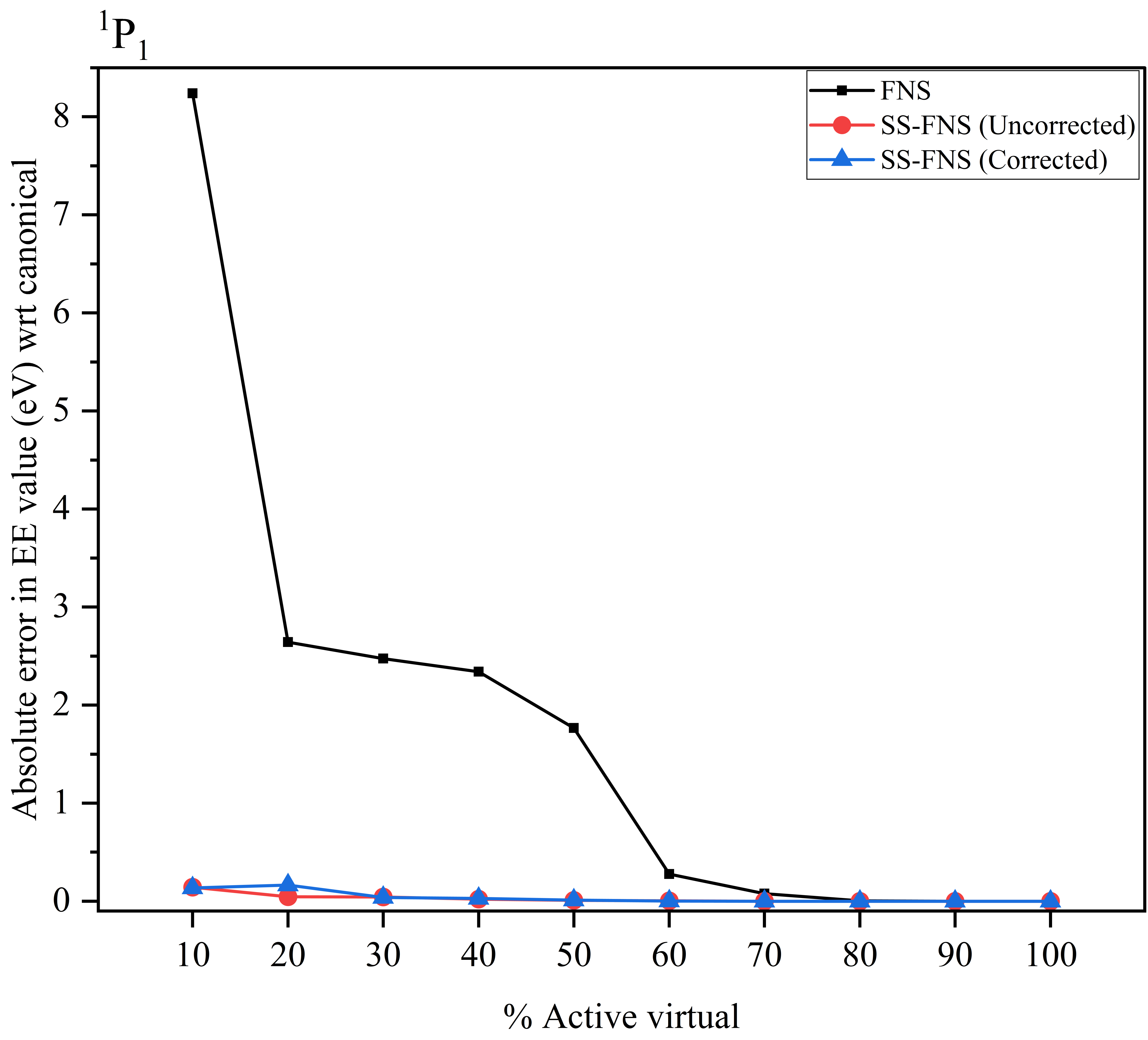}
        \caption{}
        \label{fig:Zn_combined_pct:sub4}
    \end{subfigure}

    % Main caption for the figure
    \caption{
    \label{fig:Zn_combined_pct}
    \justifying{ The comparison of the absolute error in excitation energies (in eV) vs percentages of active virtual spinors retained in FNS and SS-FNS based 4c-DC-EE-EOM-CCSD for ((\ref{fig:Zn_combined_pct:sub1}) $^3P_0$, (\ref{fig:Zn_combined_pct:sub2}) $^3P_1$, (\ref{fig:Zn_combined_pct:sub3}) $^3P_2$ and (\ref{fig:Zn_combined_pct:sub4}) $^1P_1$ states) of Zn atom. The dyall.v2z basis set has been used, and excitation energies in the untruncated canonical basis have been used as reference.}}
\end{figure*}
A key requirement of any state-specific scheme for excited states is its ability to treat each excited state equitably. The accuracy of the SS-FNS-EE-EOM-CCSD method depends on the truncation threshold.
Since Zn is a potential candidate for atomic clocks\cite{chamoli2024relativistic}, it would be a good test case to benchmark the threshold. 
We have analyzed the convergence of excitation energies with respect to the virtual space size for the first four excited states of Zn atom using the 4c-DC-SS-FNS-EE-EOM-CCSD method with dyall.v2z basis set.
%To determine an appropriate threshold, we have analyzed the convergence of excitation %energies for the first four excited states with respect to the virtual space size in the Zn %atom using the 4c-DC-SS-FNS-EE-EOM-CCSD method with a dyall.v2z basis set. The Zn atom is %chosen because of its potential as a candidate for an optical atomic clock
%\cite{chamoli2024relativistic}.
For comparison, we have also plotted the EE-EOM-CCSD results in the standard FNS basis. The states for which the comparison is shown are $^3P_0$ (FIG. \ref{fig:Zn_combined_pct:sub1}), $^3P_1$ (FIG. \ref{fig:Zn_combined_pct:sub2}), $^3P_2$ (FIG. \ref{fig:Zn_combined_pct:sub3}) and $^1P_1$ (FIG. \ref{fig:Zn_combined_pct:sub4}) states of Zn atom. It can be seen that in the standard FNS basis, the excitation energy achieves convergence with around 70$\%$ of the virtual space.  The convergence is much faster in the SS-FNS basis.  The excitation energies for all four states converge at around 40$\%$ 
of the virtual space.   The convergence behavior improves with the inclusion of the perturbative correction, and the excitation energies converge with 30 $\%$ of the virtual space. 
The occupation threshold is generally a more appropriate criterion for truncation than the size of the virtual space.  FIG. S1 illustrates the convergence of absolute error in excitation energies (in eV) for both the standard and state-specific FNS schemes, evaluated at different truncation thresholds for the Zn atom. It is observed that out of the two truncation schemes, the SS-FNS approximation gives better results in all truncation thresholds for all four states of the Zn atom. The error values in excitation energies calculated using the FNS-EE-EOM-CCSD approximation converge at the $10^{-7}$ threshold for all states. The SS-FNS truncation scheme significantly improves accuracy and converges at $10^{-4.5}$ threshold. The behavioral pattern of both truncation schemes is similar for the triplet states. However, the singlet state shows a slightly different pattern for the SS-FNS scheme, which converges at a lower threshold $10^{-3}$. The size of the virtual space selected
is almost similar between the FNS and SS-FNS schemes.
Around 57\% of the virtual natural spinors are retained in both truncation schemes at the threshold of $10^{−4.5}$, while in the
SS-FNS scheme, for all four states of the Zn
atom  (see TABLE S1). Yet the huge difference in accuracy shows that the use of  ADC(2) one-particle reduced-density matrix in the SS-FNS scheme can capture the correct electronic distribution for various kinds of excited states. The inclusion of perturbative corrections improves the SS-FNS results, at least at a lower truncation threshold, and the result converges at $10^{-4}$ threshold.

Gomes and co-workers\cite{yuanFrequencyDependentQuadraticResponse2023} have studied the performance of MP2-based FNS for the excitation energy and two-photon transition $^1S_0\rightarrow\;^1S_0$ in Ga$^+$ ion. It has been observed that MP2-based standard FNS gives poor performance for the excitation energy, and one needs to manually include spinors, which are important for the excitation to get good accuracy.  TABLE \ref{tab:Ga+} presents the excitation energy of the Ga$^+$ ion, calculated both in canonical and truncated natural spinor basis. The calculations on Ga$^+$ are performed in the uncontracted triply augmented cc-pVTZ basis set. The truncation threshold used is 10$^{-4.5}$ for both standard FNS and SS-FNS. It can be seen that the use of standard FNS shows a huge error of 9.474 eV at the 10$^{-4.5}$ threshold with respect to the canonical value. The SS-FNS approach, on the other hand, shows a very good agreement with the canonical value, with an error of 0.009 eV.  The inclusion of perturbative corrections has a small effect and shows an error of -0.007 eV. 

% TABLE: Ga+ table.
\begin{table*}[htbp]
\caption{
\label{tab:Ga+}
The comparison of excitation energies (in eV) of Ga$^+$ ion at truncation threshold of $10^{-4.5}$ calculated using uncontracted triply augmented cc-pVTZ basis set with respect to the canonical and experimental results
}
\begin{ruledtabular}
\begin{tabular}{ l c c c c }
Method &Uncorrected &Corrected &Canonical &Expt.\cite{AtomicSpectraDatabase2009} \\
\hline
4c-DC-FNS-EE-EOM-CCSD&22.586 &-      &\multirow{2}{*}{13.112} &\multirow{2}{*}{13.225} \\
4c-DC-SS-FNS-EE-EOM-CCSD&13.121 &13.105 &         &                        \\
%X2CAMF-FNS-EE-EOM-CCSD    &23.144 &-                          &                        \\
%X2CAMF-SS-FNS-EE-EOM-CCSD &13.117 &13.102 &         &                        \\
\end{tabular}
\end{ruledtabular}
\end{table*}
% FIGURE: AuH_thresh
\begin{figure}[htbp]
    \centering
    \includegraphics[width=0.45\textwidth]{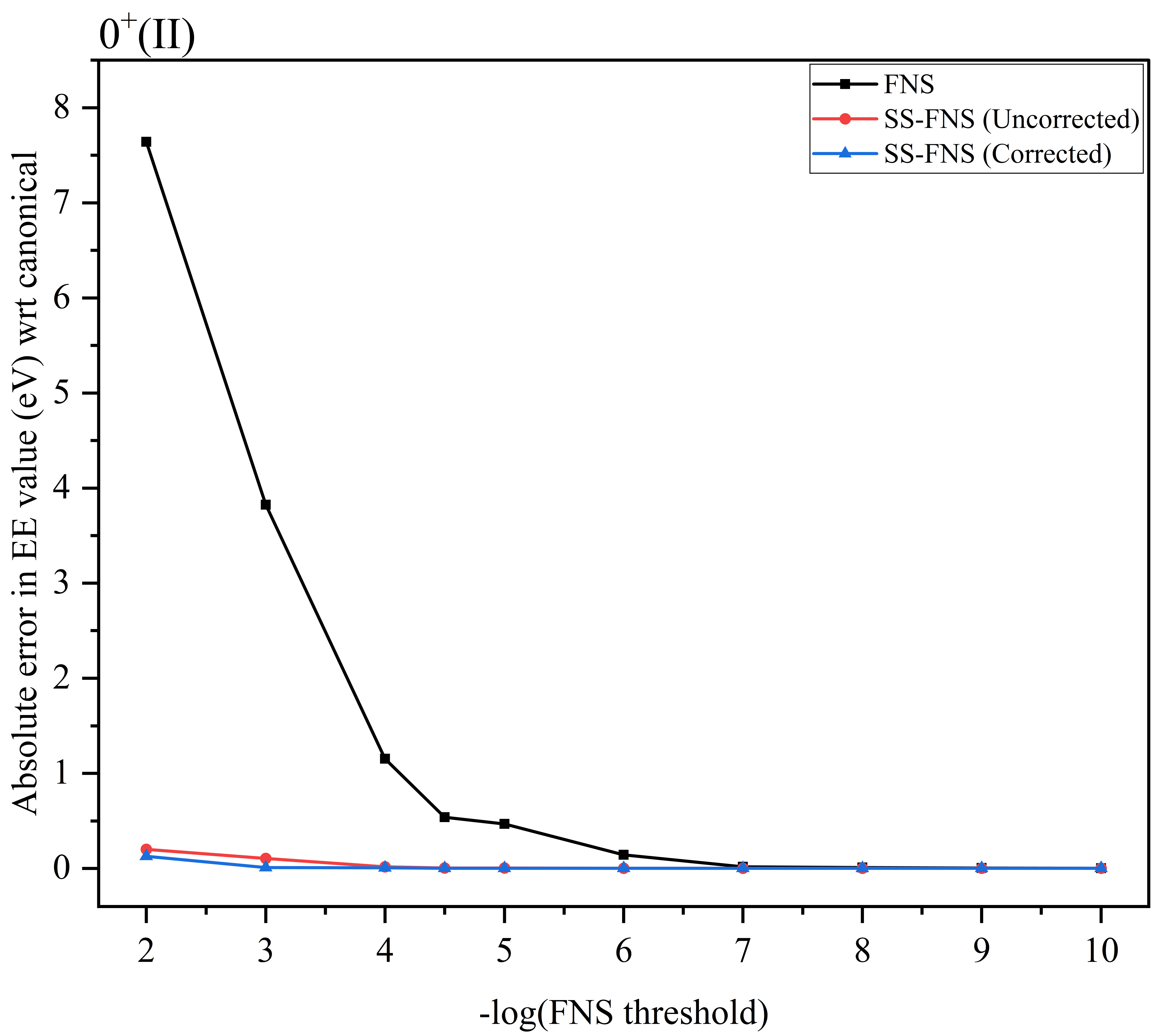}
    \caption{\justifying{
    The absolute error in excitation energies (in eV) vs occupation threshold for FNS and SS-FNS truncation schemes of 4c-DC-EE-EOM-CCSD for AuH molecule ($0^+$(II) state). Singly augmented dyall.v2z basis set is used for Au, and uncontracted aug-cc-pVDZ basis set is used for H.Untruncated canonical value has been used as the reference.}}
    \label{fig:AuH_thresh}
\end{figure}

One also needs to benchmark the truncation thresholds for molecules. FIG. \ref{fig:AuH_thresh}  shows the comparison of absolute error in excitation energy (in eV) for 0$^+$(II) state of the AuH molecule with respect to the canonical result at different truncation thresholds using both FNS and SS-FNS truncation schemes. The singly augmented dyall.v2z basis set is used for Au, and the uncontracted version of the aug-cc-pVDZ basis set is used for H. From FIG. \ref{fig:AuH_thresh}, it can be seen that 4c-DC-FNS-EE-EOM-CCSD met convergence much slower compared to the corresponding SS-FNS analogue. At the threshold of $10^{-7}$, the error in the FNS approximation becomes negligible and appears to be converged. The 4c-DC-SS-FNS-EE-EOM-CCSD method shows a faster convergence behaviour, and the error converges at a threshold of  $10^{-4}$. The inclusion of perturbative corrections improves the convergence behaviour, and the excitation energy converges at the threshold $10^{-3}$. At the threshold of $10^{-4.5}$, the  SS-FNS scheme selects 132 virtual spinors out of a total of 382 virtual spinors (see TABLE S2). While the standard FNS method in the same threshold selects 126 virtual spinors with a much larger error of 0.535 eV. The significant increase in accuracy in the SS-FNS scheme, in spite of having a comparable number of active virtual spinors, can be attributed to a proper description of the excited state wave function by state-specific ADC(2) natural spinors. Thus, the SS-FNS truncation scheme can be applied to larger systems containing heavy systems. For both atoms and molecules, we will use the conservative threshold of $10^{-4.5}$ for further calculations.

\subsection{Comparison between 4c-DC and X2CAMF-SS-FNS-EE-EOM-CCSD results}
\label{sec4.2}
TABLE \ref{tab:SI_CD} compares the 4c-DC and X2CAMF-SS-FNS-EE-EOM-CCSD method for the above-mentioned states of Zn atom, Ga$^+$ ion, and AuH molecule. The canonical and experimental results\cite{AtomicSpectraDatabase2009} are also presented for reference. It can be seen that for the Zn atom, the two methods show excellent agreement with each other. The maximum deviation between the 4c-DC and X2CAMF versions is 0.003 eV for $^3P_1$ state (uncorrected). Thus, a two-component description using the X2CAMF Hamiltonian is sufficient for the calculation of excited states of heavy elements.

% SI table
% TABLE: SI_CD
\begin{table}[h]
\caption{
\label{tab:SI_CD}
The comparison of excitation energies (in eV) of Zn, Ga$^{+}$, and AuH calculated in four-component (4c-DC) and X2CAMF versions of SS-FNS-EE-EOM-CCSD at FNS threshold of $10^{-4.5}$ with canonical 4c-DC-EE-EOM-CCSD and experimental result
}
\begin{ruledtabular}
\begin{tabular}{ c c c c c c c c }
System &States &\multicolumn{2}{c}{4c-DC} &\multicolumn{2}{c}{X2CAMF} &Can &Expt.\cite{AtomicSpectraDatabase2009, huberConstantsDiatomicMolecules1979} \\
\cline{3-4}
\cline{5-6}
                    &          &Unc &Corr &Unc &Corr & & \\
\hline
\multirow{4}{*}{\footnotemark[1] Zn}&$^3P_0$   &3.902 &3.905 &3.900 &3.903 &3.903 &4.006 \\
                    &$^3P_1$   &3.925 &3.928 &3.922 &3.926 &3.926 &4.030 \\
                    &$^3P_2$   &3.971 &3.975 &3.969 &3.973 &3.973 &4.078 \\
                    &$^1P_1$   &6.093 &6.092 &6.091 &6.090 &6.090 &5.796 \\
                    &          &     &     &     &     &     &     \\
\footnotemark[2] Ga$^{+}$&$^1S_0$  & 13.121   & 13.105  & 13.117  & 13.102    & 13.112    &  13.225    \\
 &          &     &     &     &     &     &     \\
\footnotemark[3] AuH                 &$0^+$(II) &3.456 &3.452 &3.456 &3.452 &3.453 &3.430 \\
\end{tabular}
\end{ruledtabular}
\footnotetext[1]{\mbox{dyall.v2z basis set is used.}}
\footnotetext[2]{\mbox{uncontracted triply augmented cc-pVTZ basis set is used.}}
\footnotetext[3]{\parbox[t]{9 cm}{singly augmented dyall.v2z basis set is used for Au 
and uncontracted version of aug-cc-pVDZ basis set is used for H.}}
\end{table}
A similar trend is observed for Ga$^+$. The 4c-DC and X2CAMF variants give identical results up to two significant digits and show a negligible error of 0.01 eV with respect to the canonical 4c-DC-EE-EOM-CCSD values.
For the 0$^+$(II) state of AuH, both 4c-DC and X2CAMF results match exactly up to three significant digits, and the perturbative correction results in a small improvement. The error with respect to the canonical 4c-DC-EE-EOM-CCSD  is only 0.001 eV. This shows that the CD-based X2CAMF scheme can indeed be used as an affordable alternative to the 4c-DC Hamiltonian for the SS-FNS-EE-EOM-CCSD method, especially for calculations with a large number of basis functions. For all the subsequent discussions in this work, SS-FNS-EE-EOM-CCSD uses the CD-X2CAMF scheme by default.

\subsection{Fine structure splitting: Ga$^+$, In$^+$, Tl$^+$ ions}
\label{sec4.3}
% TABLE: Splitting table.
\begin{table*}[htbp]
\caption{
\label{tab:splitting}
The comparison of fine structure splitting (in eV) and excitation energies from $^1S_0$ ground state (in eV) of Ga$^+$, In$^+$, and Tl$^+$ ions with experimental results using X2CAMF version of SS-FNS-EE-EOM-CCSD.
}
\begin{ruledtabular}
\begin{tabular}{ c l c c c c c c c c c }
Atom    &           &\multicolumn{2}{c}{SS-FNS-EE-EOM-CCSD} &Canonical EE-EOM-CCSD &Expt.\cite{AtomicSpectraDatabase2009} \\
\cline{3-4}
        &                                   &Uncorrected &Corrected  &               &      \\
\hline
Ga$^+$  &$\;^1S_0 \rightarrow \;^3P_0$      &5.751  &5.754  &5.745  &5.874  \\
        &$\;^1S_0 \rightarrow \;^3P_1$      &5.804  &5.808  &5.798  &5.928  \\
        &$\;^1S_0 \rightarrow \;^3P_2$      &5.916  &5.920  &5.910  &6.044  \\
        &$\;^1S_0 \rightarrow \;^1P_1$      &8.775  &8.771  &8.768  &8.766  \\
        &$\;^1S_0 \rightarrow \;^3S_1$      &12.612 &12.615 &12.612 &12.764 \\
        &$\;^1S_0 \rightarrow \;^1S_0$      &13.117 &13.102 &13.108 &13.225 \\
        &\;\;\;Splitting                    &       &       &       &       \\
        &$\;^3P_0 \rightarrow \;^3P_1$      &0.053  &0.053  &0.053  &0.054  \\
        &$\;^3P_1 \rightarrow \;^3P_2$      &0.112  &0.112  &0.112  &0.116  \\
        &                                   &       &       &       &       \\
In$^+$  &$\;^1S_0 \rightarrow \;^3P_0$      &4.896  &4.899  &4.891  &5.242  \\
        &$\;^1S_0 \rightarrow \;^3P_1$      &5.026  &5.029  &5.021  &5.375  \\
        &$\;^1S_0 \rightarrow \;^3P_2$      &5.316  &5.319  &5.311  &5.682  \\
        &$\;^1S_0 \rightarrow \;^1P_1$      &7.796  &7.784  &7.787  &7.816  \\
        &$\;^1S_0 \rightarrow \;^3S_1$      &11.248 &11.249 &11.246 &11.645 \\
        &$\;^1S_0 \rightarrow \;^1S_0$      &11.641 &11.633 &11.631 &12.030 \\
        &\;\;\;Splitting                    &       &       &       &       \\
        &$\;^3P_0 \rightarrow \;^3P_1$      &0.130  &0.130  &0.130  &0.133  \\
        &$\;^3P_1 \rightarrow \;^3P_2$      &0.290  &0.290  &0.290  &0.307  \\
        &                                   &       &       &       &       \\
Tl$^+$  &$\;^1S_0 \rightarrow \;^3P_0$      &6.170  &6.170  &6.161 &6.131   \\
        &$\;^1S_0 \rightarrow \;^3P_1$      &6.537  &6.535  &6.526 &6.496   \\
        &$\;^1S_0 \rightarrow \;^3P_2$      &7.608  &7.615  &7.606 &7.653   \\
        &$\;^1S_0 \rightarrow \;^1P_1$      &9.503  &9.497  &9.497 &9.381   \\
        &$\;^1S_0 \rightarrow \;^3S_1$      &12.909 &12.913 &12.905 &13.047 \\
        &$\;^1S_0 \rightarrow \;^1S_0$      &13.287 &13.280 &13.279 &13.390 \\
        &\;\;\;Splitting                    &       &       &       &       \\
        &$\;^3P_0 \rightarrow \;^3P_1$      &0.367  &0.365  &0.365 &0.365   \\
        &$\;^3P_1 \rightarrow \;^3P_2$      &1.071  &1.080  &1.080 &1.157   \\
        &                                   &       &       &      &        \\
\end{tabular}
\end{ruledtabular}
\end{table*}

\begin{figure}[htbp]
    \centering
    \includegraphics[width=0.45\textwidth]{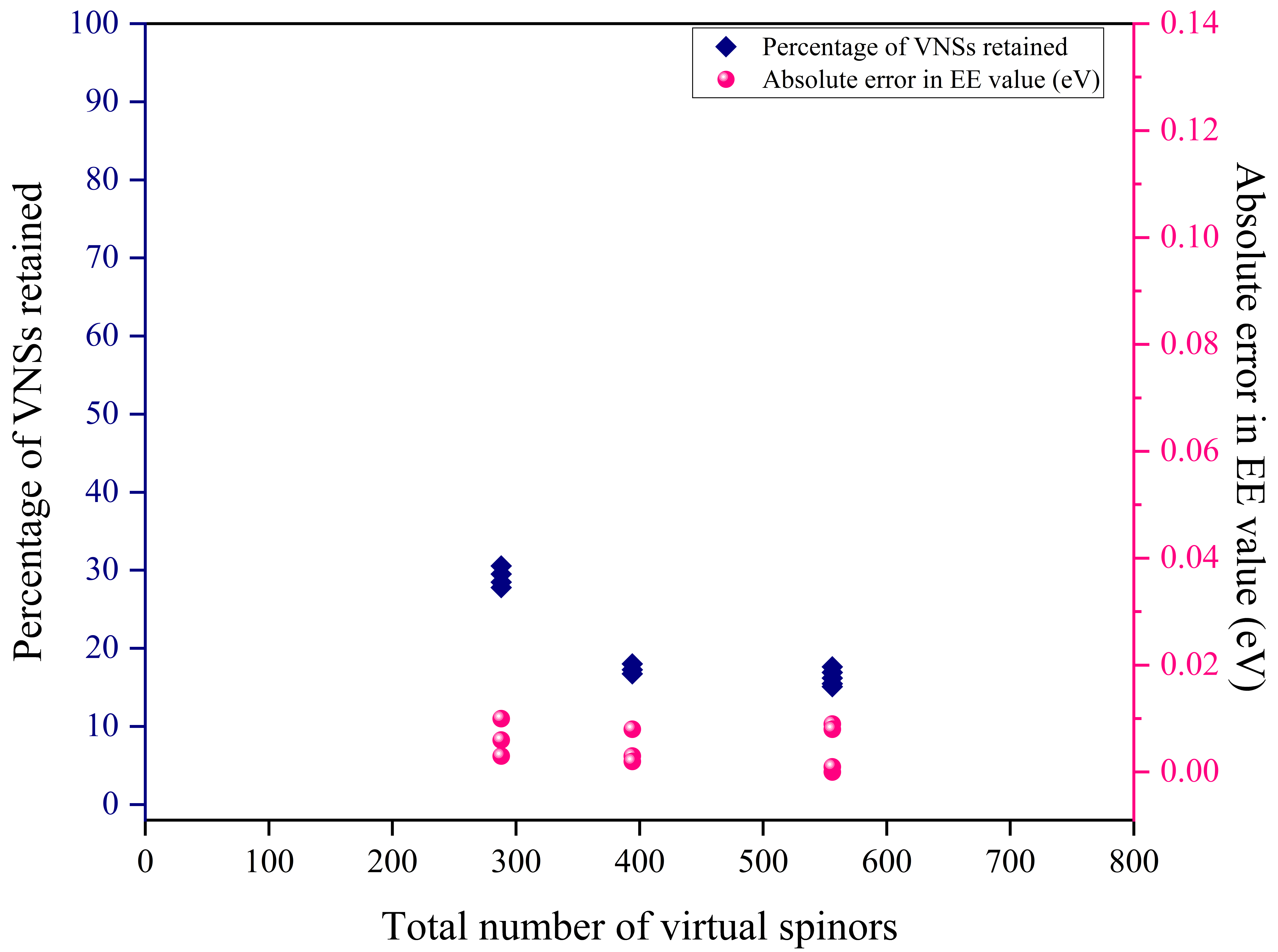}
    \caption{\justifying{The percentage of SS-FNSs retained and absolute error in excitation energy (eV) (with respect to canonical results) as a function of the size of the virtual space for Ga$^+$, In$^+$, and Tl$^+$ ions.}}
    \label{fig:dot_13}
\end{figure}

Apart from excitation energies, fine structure splittings are important observables and serve as an important criterion for assessing a method's accuracy. 
As energy differences,\cite{BorschevskyAtomicTransitionEnergies2006} fine structure splittings offer deeper insights into whether different excited states are treated on an equal footing.
Moreover, the fine structure splitting in the triplet state is a consequence of the spin-orbit coupling, which can only be accounted for in a relativistic framework. TABLE \ref{tab:splitting}  presents the excitation energies and the fine structure splittings of group 13 cations (Ga$^+$, In$^+$, and Tl$^+$). The uncontracted triply augmented cc-pVTZ basis set has been used for Ga$^+$ and triply augmented dyall.v3z basis set has been used for In$^+$ and Tl$^+$.  The excitation energies are shown for $^3P_0$, $^3P_1$, $^3P_2$, $^1P_1$, $^3S_1$ and $^1S_0$ states of all three ions. All the transitions are considered from the ground state $^1S_0$. For all three ions, the SS-FNS-EE-EOM-CCSD excitation energies are in excellent agreement with the corresponding canonical value, and the errors are within 0.01 eV. The effect of perturbation correction is negligible for all the cases. Figure \ref{fig:dot_13}  illustrates the relationship between the size of the virtual space and the error observed in excitation energy for the group 13 cations. It can be seen that even with the inclusion of less than 40 $\%$ of the virtual space, one can get $\le$ 0.01 eV accuracy with respect to the canonical EOM-CCSD results.
%{Revise the sentence below.}
However, the performance of the EOM-CCSD method with respect to the experimentally measured excitation energy is not consistent.
For  Ga$^+$ and Tl$^+$, the excitation energy values are within 0.2 eV of the experimental values, which is generally considered as the error bar of the standard EOM-CCSD method\cite{dutta2018exploring, schreiber2008benchmarks}. A slightly larger error is obtained for In$^+$, where the errors can be as high as 0.4 eV.  One presumably needs to include three or higher-body operators in the calculation to get a quantitative agreement with experiments.

% FIGURE: dot_13

The splitting, on the other hand, is in very good agreement with the experimental results. One can see that the experimental trend of increase in the value of a particular fine structure splitting down the group from Ga$^+$ to Tl$^+$ is retained in the SS-FNS results. Both with and without perturbative correction, the results follow the same trend, indicating the proper handling of the spin-orbit coupling. For all the cases, the perturbative correction has either kept unchanged or improved the splitting results. A maximum absolute error of 0.077 eV relative to the experimental value is observed for the $^3P_1\rightarrow\;^3P_2$ fine structure splitting of the Tl$^+$ ion. Thus, the SS-FNS-EE-EOM-CCSD method can be used to simulate the fine-structure splittings in a range of systems.

\subsection{Benchmarking beyond diatomics: I$^-_3$ ion}
\label{sec4.4}
The triiodide anion (I$_3^-$) acts as a popular benchmark for evaluating relativistic electron correlation methods for excited states. Numerous theoretical studies on I$_3^-$ have been reported in the literature, employing methods such as Intermediate Hamiltonian Fock-Space Coupled Cluster (IHFS-CC),\cite{visscherFormulationImplementationRelativistic2001, landauMixedsectorIntermediateHamiltonian2004} EOM-CCSD,\cite{sheeEquationofmotionCoupledclusterTheory2018} CASPT2,\cite{anderssonSecondorderPerturbationTheory1990, roosCompleteActiveSpace1980} MRCI,\cite{valaInitioDiatomicsMolecule2001, fleigGeneralizedActiveSpace2003, fleigGeneralizedActiveSpace2006}, ADC(3)\cite{chakraborty2025relativistic} and TD-DFT\cite{grossTimeDependentDensityFunctionalTheory1990} within a relativistic framework. Wang and co-workers have reported EOM-CCSD calculations for the same molecule, incorporating spin-orbit coupling (SOC) effects through a perturbative approach\cite{wangEquationofMotionCoupledClusterTheory2014}. Following the work of Gomes and co-workers\cite{sheeEquationofmotionCoupledclusterTheory2018}, the calculations were performed using a bond length of 2.93 \AA and dyall.av3z basis set.
Out of the total occupied spinors, 138 spinors were kept frozen in the correlation calculations. Two different schemes were used for the virtual spinors. In the first scheme, canonical virtuals up to energies \mbox{$\leq12.0E_{h}$} were correlated following Ref. \onlinecite{sheeEquationofmotionCoupledclusterTheory2018}. It leads to a total of 332 virtual spinors in the correlation space, which are subsequently truncated using the SS-FNS approach. In the second scheme, the full canonical virtual space of 974 spinors was considered and subsequently truncated using the SS-FNS method.
%, but unlike their work, 138 occupied spinors were frozen, leaving 22 electrons for the correlation calculation.
The results obtained using the SS-FNS-EE-EOM-CCSD method are summarized in TABLE \ref{tab:I3-}.  The canonical 4c-DC-EE-EOM-CCSD results from Gomes and co-workers\cite{sheeEquationofmotionCoupledclusterTheory2018} were used as a reference. It can be seen that in both cases, the number of active virtual spinors in the two schemes is almost the same in the truncated FNS basis. The size of the virtual space in the FNS basis for all the states is 226, except for the 18th state, where it selects 230 and 232 virtual spinors in the former and latter schemes, respectively. Statistical parameters are only calculated for the former scheme, as the canonical 4c-DC-EE-EOM-CCSD \cite{sheeEquationofmotionCoupledclusterTheory2018} are only available in the truncated basis.

% TABLE: I3- table.
\begin{table*}[ht!]
\caption{
\label{tab:I3-}
The comparison of excitation energies (in eV) of I$_3^-$ ion calculated using X2CAMF-SS-FNS-EE-EOM-CCSD method and dyall.av3z basis set and is compared with the relativistic 4c-DC-EE-EOM-CCSD canonical results.
}
\begin{ruledtabular}
\begin{tabular}{ c c c c c c c c c }
States &$\Omega$  &4c-DC-EE-EOM-CCSD\cite{sheeEquationofmotionCoupledclusterTheory2018} &\multicolumn{3}{c}{SS-FNS-EE-EOM-CCSD\footnotemark[1]} &\multicolumn{2}{c}{SS-FNS-EE-EOM-CCSD\footnotemark[2]} \\
\cline{4-6}
\cline{7-9}
        &          &     &Uncorrected &Corrected &$N_{vir}$ &Uncorrected &Corrected &$N_{vir}$ \\
\hline
1		&$2_g  $   &2.24 &2.271	&2.263	&226	&2.271	&2.263 &226 \\
2		&$1_g  $   &2.37 &2.405	&2.400	&226	&2.405	&2.400 &226 \\
3		&$0_u^-$   &2.37 &2.412	&2.407	&226	&2.411	&2.407 &225 \\
4		&$1_u  $   &2.38 &2.402	&2.395	&226	&2.403	&2.394 &226 \\
5		&$0_g^-$   &2.84 &2.864	&2.858	&226	&2.865	&2.857 &226 \\
6		&$0_g^+$   &2.89 &2.917	&2.910	&226	&2.918	&2.910 &226 \\
7		&$1_g  $   &3.07 &3.089	&3.081	&226	&3.089	&3.081 &226 \\
8		&$2_u  $   &3.32 &3.351	&3.342	&226	&3.352	&3.341 &226 \\
9		&$1_u  $   &3.41 &3.434	&3.424	&226	&3.433	&3.423 &226 \\
10		&$0_u^+$   &3.66 &3.679	&3.669	&226	&3.679	&3.669 &226 \\
11		&$2_g  $   &4.09 &4.125	&4.116	&226	&4.125	&4.116 &226 \\
12		&$0_u^-$   &4.08 &4.102	&4.093	&226	&4.102	&4.093 &226 \\
13		&$1_u  $   &4.18 &4.203	&4.193	&226	&4.203	&4.193 &226 \\
14		&$1_g  $   &4.21 &4.243	&4.234	&226	&4.243	&4.233 &226 \\
15		&$0_u^+$   &4.49 &4.496	&4.485	&226	&4.496	&4.485 &226 \\
16		&$0_g^-$   &4.69 &4.721	&4.713	&226	&4.722	&4.713 &226 \\
17		&$0_g^+$   &4.70 &4.732	&4.724	&226	&4.733	&4.723 &226 \\
18		&$1_g  $   &4.90 &4.940	&4.928	&230	&4.940	&4.928 &232 \\
\hline
MAD     &          &     &0.042	&0.037	&	    &	    &     &    \\
ME      &          &     &0.028	&0.019	&	    &	    &     &    \\
MAE     &          &     &0.028	&0.019	&	    &	    &     &    \\
STD     &          &     &0.008	&0.009	&	    &	    &     &    \\
RMSD    &          &     &0.029	&0.021	&	    &	    &     &    \\
\end{tabular}
\end{ruledtabular}
\footnotetext[1]{\mbox{Only 332 virtual spinors are considered in the canonical basis.}}
\footnotetext[2]{\mbox{All virtual spinors are considered in the canonical basis.}}
\end{table*}

From TABLE \ref{tab:I3-}, it can be seen that the excitation energies for all states exhibit a blue shift because of the truncation. This is also reflected in the value of ME (Mean Error) and MAE (Mean Absolute Error). The uncorrected SS-FNS-EE-EOM-CCSD results are very good for almost all of the states. However, the inclusion of perturbative corrections leads to improved results in cases. The truncation of virtual spinors at the canonical level (considering 332 virtual spinors in the correlation space) has very little or insignificant effect on the SS-FNS-EE-EOM-CCSD result. This also reaffirms the correct choice of the energy cut-off for the virtual spinors by Gomes and coworkers\cite{sheeEquationofmotionCoupledclusterTheory2018} for the correlation calculations on a canonical basis.  The MAD (Maximum Absolute Deviation) observed among all the states is 0.04 eV,  and it is observed for four of the states {$1_g$(2),$0_u^-$(3),$2_g$(11),$1_g$(18)}. The inclusion of perturbative corrections improved the results for all of them. The perturbative correction has improved the MAE value from 0.028 to 0.019. The STD (Standard Deviation) value is 0.008 and 0.009 for uncorrected and corrected results, respectively, signifying a high accuracy of the SS-FNS-EE-EOM-CCSD method, even in the uncorrected version. The RMSD (Root Mean Square Deviation) value of the uncorrected SS-FNS-EE-EOM-CCSD is 0.029. RMSD value decreases from 0.029 to 0.021 upon adding the perturbative correction, reflecting an improvement in the results.

\subsection{Benchmarking of transition property: Xe atom}
\label{sec4.5}
%\subsubsection*{Xe atom}

% TABLE: Xe table.
\begin{table*}[ht!]
\caption{
\label{tab:Xe}
The comparison of excitation energies (in eV) and transition dipole moments (in a.u.) of the Xe atom calculated using X2CAMF version of SS-FNS-EE-EOM-CCSD, triply augmented dyall.ae3z basis set with 4c-DC-EE-EOM-CCSD.
}
\begin{ruledtabular}
\begin{tabular}{ c c c c c c c c c }
States &\multicolumn{2}{c}{4c-DC-EE-EOM-CCSD\cite{mukhopadhyayAnalyticCalculationTransition2025}} &\multicolumn{4}{c}{SS-FNS-EE-EOM-CCSD} &\multicolumn{2}{c}{Expt.\cite{zaitsevskiiFiniteFieldCalculationsTransition2020, sansonettiHandbookBasicAtomic2005}} \\
\cline{2-3}
\cline{4-7}
\cline{8-9}
 &EE    &TDM   &EE (Uncorrected) &EE (Corrected) &TDM &$N_{vir}$\footnotemark[1] &EE &TDM\\
\hline
$5p^5\left(^2P_{3/2}\right)6s^2[3/2]^o_1$ &8.43	 &0.637	&8.446	&8.434	&0.657	&158 &8.437  &0.654 $\pm$ 0.0042 \\
$5p^5\left(^2P_{1/2}\right)6s^2[1/2]^o_1$ &9.56	 &0.510	&9.571	&9.559	&0.536	&160 &9.570  &0.521 $\pm$ 0.0034 \\
$5p^5\left(^2P_{3/2}\right)5d^2[1/2]^o_1$ &9.92	 &0.124	&9.938	&9.924	&0.114	&162 &9.917  &0.120 $\pm$ 0.0028 \\
$5p^5\left(^2P_{3/2}\right)5d^2[3/2]^o_1$ &10.43 &0.706	&10.442	&10.432	&0.695	&162 &10.401 &0.704 $\pm$ 0.0174 \\
\end{tabular}
\end{ruledtabular}
\footnotetext[1]{\mbox{No of active virtuals.}}
\end{table*}

To benchmark the SS-FNS approach for transition properties, we have computed the transition dipole moments (TDMs) for the first four bright Rydberg states of the Xe atom. These values were previously calculated within the Fock-space relativistic coupled-cluster (FSCC) framework by Eliav and co-workers.\cite{zaitsevskiiFiniteFieldCalculationsTransition2020} In our recent works, we reported TDMs of these excited states of Xe using 4c-DC-EE-EOM-CCSD\cite{mukhopadhyayAnalyticCalculationTransition2025} and 4c-DC-EE-ADC(3)\cite{chakraborty2025relativistic} with various basis sets. 
In the current work, we have selected the triply augmented dyall.v3z basis set, as it has been shown\cite{mukhopadhyayAnalyticCalculationTransition2025} to give good agreement with experimental data. The canonical basis contained 562 basis functions. As shown in TABLE \ref{tab:Xe}, the number of retained virtual spinors in the SS-FNS formalism for each state is significantly lower than in the canonical basis. Despite truncating approximately 70\% of the virtual spinors in nearly all states, the excitation energies remain in close agreement with their canonical counterparts. The deviation in excitation energy is within 0.02 eV of the canonical result, which falls well within the EE-EOM-CCSD error bar\cite{schreiber2008benchmarks}. Similarly, the TDM values for all four states show very good agreement with the canonical data. As shown in TABLE \ref{tab:Xe}, $5p^5\left(^2P_{3/2}\right)5d^2[3/2]^o_1$ is the brightest while $5p^5\left(^2P_{3/2}\right)5d^2[1/2]^o_1$, is the least bright among the four states in both cases (canonical and SS-FNS).  The SS-FNS framework slightly overestimates the excitation energy, but the inclusion of ADC(2) correction eliminates this overestimation, ensuring agreement up to two significant digits. No perturbative correction is considered for the transition properties.

\subsection{Computational Efficiency}
\label{sec4.6}
To show the applicability of the SS-FNS-EE-EOM-CCSD method beyond small molecules, the first excited state of [I$_3$(H$_2$O)$_6]^-$ complex has been calculated. The solvated triiodide ion has been studied\cite{margulis2001monte} extensively because it provides a convenient system for femtosecond spectroscopy. The geometry of the hexa-aqua triiodide complex has been obtained by adding water molecules around I$_3^-$ molecule by the solvation module of ORCA\cite{ORCA5}, followed by a geometry optimization at BP86/X2C-TZVPall level of theory. The spin-free-X2C1e approximation has been used to include the scalar relativistic effect in the geometry optimization. The structure is shown in FIG. \ref{fig:complex} and the geometry of the complex have been provided in the Supplementary Material.
The SS-FNS-EE-EOM-CCSD calculation is performed on a dedicated workstation with two Intel(R) Xeon(R) Gold 5315Y CPUs @ 3.20 GHz and 2 TB of total RAM. The dyall.av3z basis set has been used for the I atoms, and the uncontracted version of the cc-pVDZ basis set has been used for the O and H atoms. The hexa-aqua triiodide complex consists of 21 atoms with 220 electrons. The aforementioned basis set uses 1394 virtual spinors on a canonical basis. Cholesky threshold has been set to 10$^{-3}$, resulting in 2580 Cholesky vectors. The frozen core approximation has been considered, and 150 electrons have been frozen for the correlation calculations. The SS-FNS approximation has reduced the number of virtual spinors to 517 at the threshold of 10$^{-4.5}$. The total time taken by the calculation is 5 days, 1 hour, 41 minutes, out of which the SCF calculation took 53 minutes. In the canonical basis, the  Cholesky vectors are formed in 18 minutes, the transformation of Cholesky vectors took 5 minutes, and the generation of integrals took 24 minutes. The time taken by MP2, CIS, and ADC(2) methods in the canonical basis is 23 minutes, 8 hours 9 minutes, and 22 hours 29 minutes, respectively. The time taken to generate the SS-FNS basis for that state is 23 minutes. The total time spent in the SS-FNS-EE-EOM-CCSD calculation is 3 days, 16 hours, 32 minutes.
The excitation energy for the $2_g$ state of [I$_3$(H$_2$O)$_6]^-$ complex obtained from this calculation is 2.279 eV.

% FIGURE: complex
\begin{figure}[htbp]
    \centering
    \includegraphics[width=0.45\textwidth]{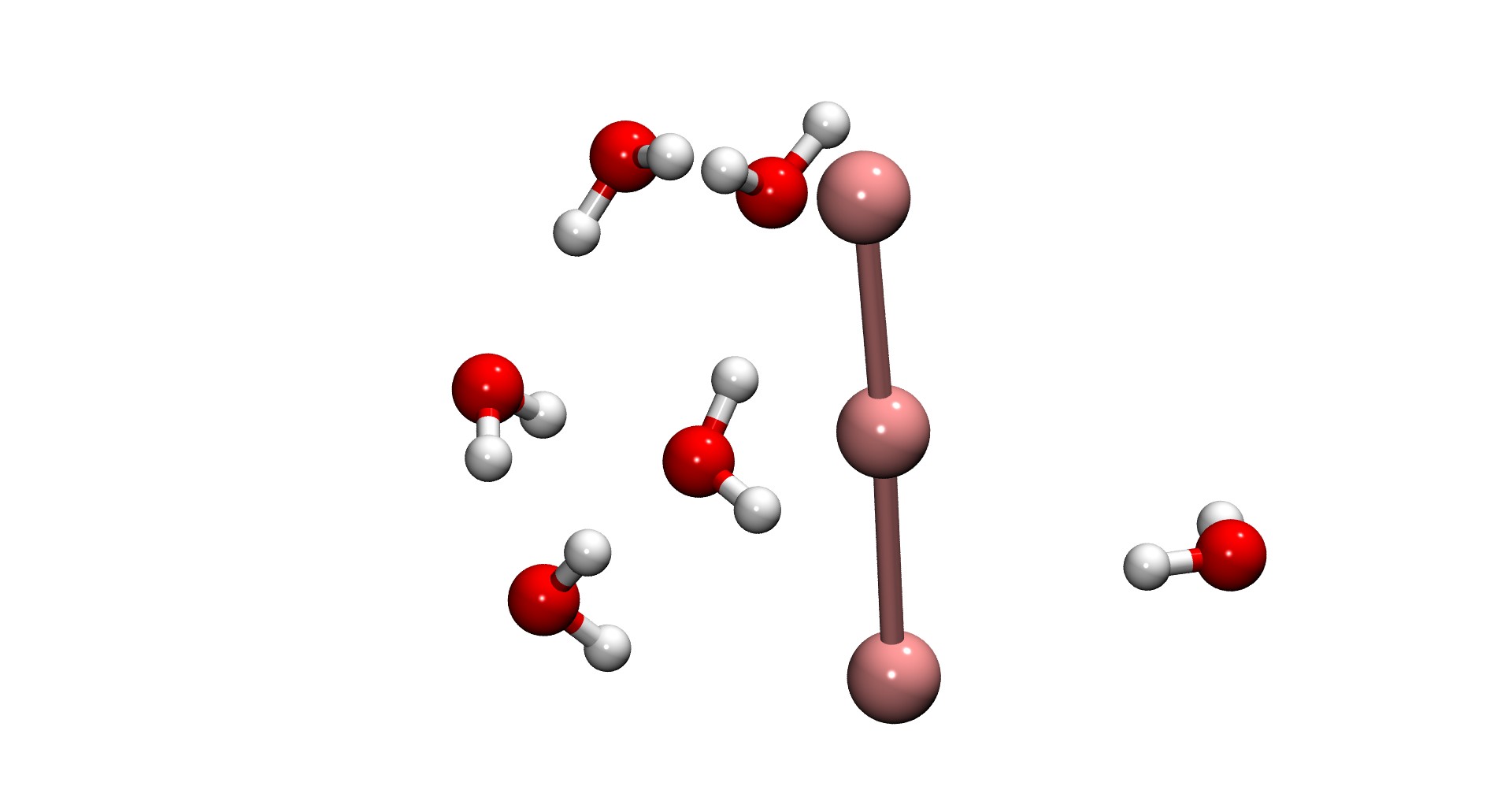}
    \caption{\justifying{The molecular structure of the [I$_3$(H$_2$O)$_6]^-$ complex.}}
    \label{fig:complex}
\end{figure}

\section{Conclusion}
\label{sec5}
We present the theory, implementation, and benchmark of a low-cost relativistic EOM-CCSD method based on state-specific frozen natural spinors. The use of state-specific FNS generated from the relativistic ADC(2) method ensures smooth convergence of the excitation energy with respect to the size of the virtual space and performs much better than the standard FNS generated from ground-state MP2 calculation.  The relativistic SS-FNS-EE-EOM-CCSD method gives excellent agreement with the canonical EOM-CCSD method for excitation energy, fine structure splitting, and transition properties.  The SS-FNS-EE-EOM-CCSD requires separate integral transformations for each excited state, and one needs to use an efficient approximation to generate the two-electron integrals. The SS-FNS-EE-EOM-CCSD method, based on X2CAMF Hamiltonian and Cholesky decomposition treatment of the two-electron integrals, appears as an attractive option and gives excellent agreement with the SS-FNS-EE-EOM-CCSD method based on 4c-DC Hamiltonian. 
The SS-FNS approach is general and can be extended to triples and higher body excitation. Work towards this direction is in progress and will be reported in future studies.

\section{Supplementary Material}
\label{sec6}
The Supplementary Material contains the geometry of triiodide hexaaqua complex, the variation of absolute error in excitation energies of Zn atom ($^3P_0$, $^3P_1$, $^3P_2$ and $^1P_1$ states) across different truncation thresholds and the excitation energies of Zn atom (same four states) and AuH molecule ($0^+$(II) state) at different truncation thresholds using FNS and SS-FNS versions of 4c-DC-EE-EOM-CCSD method.

\begin{acknowledgments}
AKD, TM, MT, and SC  acknowledge the support from IIT Bombay, CRG (Project No. CRG/2022/005672) and MATRICS (Project No. MTR/2021/000420) project of DST-SERB, CSIR-India (Project No. 01(3035)/21/EMR-II), and ISRO for financial support, IIT Bombay super computational facility, and C-DAC. Supercomputing resources (Param Smriti, Param Bramha, and Param Rudra) for computational time.
AKD acknowledges the research fellowship funded by the EU NextGenerationEU through the Recovery and Resilience Plan for Slovakia under project No. 09I03-03-V04-00117.
The author acknowledges Prof. Balasubramanian Sundaram, JNCASR, India, for stimulating insights on the solvated (I$_3^-$) example. 
SC acknowledges the Prime Minister's Research Fellowship (PMRF).
MT acknowledges DST for the Fellowship.
\end{acknowledgments}

\appendix
\section{Programmable expressions for state-specific reduced density matrix for closed-shell systems in canonical basis}
% Redefine equation numbering for the appendix
\renewcommand{\theequation}{A\arabic{equation}}
\setcounter{equation}{0} % Reset counter to 0
\begin{equation}
\label{eqn51}
    D^{SS}_{ab} = D^{MP2}_{ab} + D^{EE-ADC(2)}_{ab}
\end{equation}
\begin{equation}
\label{eqn52}
    D^{MP2}_{ab} = \frac{1}{2}\left(D_{vv}+D_{vv}^{\dagger}\right)    
\end{equation}
\begin{equation}
\label{eqn53}
    D_{vv} = \frac{1}{2}t_{in}^{ea*(1)}t_{in}^{eb(1)}
\end{equation}
\begin{equation}
\label{eqn54}
    D^{EE-ADC(2)}_{ab} = r_{i}^{a*}r_{i}^{b} + \frac{1}{2}r_{ij}^{ca*}r_{ij}^{cb}
\end{equation}
In 4c-DC:
\begin{equation}
\label{eqn55}
    t_{ij}^{ab(1)} = \frac{\langle ij||ab \rangle} {f_{ii}+f_{jj}-f_{aa}-f_{bb}}
\end{equation}
In CD-X2CAMF:
\begin{equation}
\label{eqn56}
    t_{ij}^{ab(1)} = \frac{\left(L^{P}_{ia}L^{P}_{jb} - L^{P}_{ib}L^{P}_{ja}\right)}{f_{ii}+f_{jj}-f_{aa}-f_{bb}}
\end{equation}
where $L^{P}_{ia}$ are half-transformed three-centered two-electron Cholesky vectors.  
% ADC(2) matvec
%\textbf{ADC(2) matrix-vector equations}

%\textit{1h1p-1h1p block}
%\begin{equation}
 %   \begin{align}
 %       \sigma_{i}^{a} &= r_i^b I^{[1]}_{ab} + r_j^a I^{[2]}_{ij} - (P|ij)(P|ab) r_j^b + (P|bj)(P|ia) r_j^b \\
 %                  &+ \frac{1}{2}t_{ik}^{ac} \left[ (P|ck)(P|bj)r_j^b - (P|cj)(P|bk)r_j^b \right] \\
%                   &+ \frac{1}{2}r_{j}^{b}t_{jk}^{bc} \left[ (P|kc)(P|ia) - (P|ic)(P|ka) \right] \\
%    \end{align}
%\end{equation}
%\textit{1h1p-2h2p block}
%\begin{equation}
 %       \sigma_{i}^{a} = \langle jk||ib\rangle r_{jk}^{ab} + (P|ab)\left[r_{ij}^{bc}(P|jb)-r_{ij}^{cb}(P|jb)\right]
%\end{equation}
%\textit{2h2p-1h1p block}
%\begin{equation}
  %  \begin{align}
   %     \sigma_{ij}^{ab} &= \frac{1}{2}\mathscr{P}(ij) r_{i}^{c} \left[(P|cb)(P|ja)-(P|ca)(P|jb)\right] \\
  %      &- \frac{1}{2}\mathscr{P}(ab)\langle ij||ka\rangle r_{k}^{b}
  %  \end{align}
%\end{equation}
%\textit{2h2p-2h2p block}
%\begin{equation}
%        \sigma_{ij}^{ab} = \mathscr{P}(ab)f_{bc} r_{ij}^{ac} - \mathscr{P}(ij)f_{ik} r_{kj}^{ab}
%\end{equation}

%\textit{ADC(2) Intermediates}
%\begin{equation}
%    I_{ab}^{[1]} = f_{ab} + \frac{1}{4}\mathscr{P}_+(ab) (P|jb) \left[t_{ij}^{ac} (P|ci) - t_{ji}^{ac(1)} (P|cj) \right]
%\end{equation}
%\begin{equation}
%    I_{ij}^{[2]} = -f_{ij} + \frac{1}{4}\mathscr{P}_+(ij) (P|jb) \left[t_{ik}^{ab} (P|ak) - t_{ik}^{ba(1)} (P|ak) \right]
%\end{equation}

%\pagebreak
%\clearpage
%\newpage
\renewcommand{\refname}{References} % For article class
\bibliographystyle{aipnum4-1}
%\bibliography{paper}

\begin{thebibliography}{91}%
\makeatletter
\providecommand \@ifxundefined [1]{%
 \@ifx{#1\undefined}
}%
\providecommand \@ifnum [1]{%
 \ifnum #1\expandafter \@firstoftwo
 \else \expandafter \@secondoftwo
 \fi
}%
\providecommand \@ifx [1]{%
 \ifx #1\expandafter \@firstoftwo
 \else \expandafter \@secondoftwo
 \fi
}%
\providecommand \natexlab [1]{#1}%
\providecommand \enquote  [1]{``#1''}%
\providecommand \bibnamefont  [1]{#1}%
\providecommand \bibfnamefont [1]{#1}%
\providecommand \citenamefont [1]{#1}%
\providecommand \href@noop [0]{\@secondoftwo}%
\providecommand \href [0]{\begingroup \@sanitize@url \@href}%
\providecommand \@href[1]{\@@startlink{#1}\@@href}%
\providecommand \@@href[1]{\endgroup#1\@@endlink}%
\providecommand \@sanitize@url [0]{\catcode `\\12\catcode `\$12\catcode `\&12\catcode `\#12\catcode `\^12\catcode `\_12\catcode `\%12\relax}%
\providecommand \@@startlink[1]{}%
\providecommand \@@endlink[0]{}%
\providecommand \url  [0]{\begingroup\@sanitize@url \@url }%
\providecommand \@url [1]{\endgroup\@href {#1}{\urlprefix }}%
\providecommand \urlprefix  [0]{URL }%
\providecommand \Eprint [0]{\href }%
\providecommand \doibase [0]{http://dx.doi.org/}%
\providecommand \selectlanguage [0]{\@gobble}%
\providecommand \bibinfo  [0]{\@secondoftwo}%
\providecommand \bibfield  [0]{\@secondoftwo}%
\providecommand \translation [1]{[#1]}%
\providecommand \BibitemOpen [0]{}%
\providecommand \bibitemStop [0]{}%
\providecommand \bibitemNoStop [0]{.\EOS\space}%
\providecommand \EOS [0]{\spacefactor3000\relax}%
\providecommand \BibitemShut  [1]{\csname bibitem#1\endcsname}%
\let\auto@bib@innerbib\@empty
%</preamble>
\bibitem [{\citenamefont {Gonz{\'a}lez}\ and\ \citenamefont {Lindh}(2020)}]{gonzalezQuantumChemistryDynamics2020}%
  \BibitemOpen
  \bibfield  {author} {\bibinfo {author} {\bibfnamefont {L.}~\bibnamefont {Gonz{\'a}lez}}\ and\ \bibinfo {author} {\bibfnamefont {R.}~\bibnamefont {Lindh}},\ }\href {\doibase 10.1002/9781119417774} {\emph {\bibinfo {title} {Quantum {{Chemistry}} and {{Dynamics}} of {{Excited States}}: {{Methods}} and {{Applications}}}}}\ (\bibinfo  {publisher} {John Wiley \& Sons, Ltd},\ \bibinfo {year} {2020})\BibitemShut {NoStop}%
\bibitem [{\citenamefont {{\v C}{\'i}{\v z}ek}(1966)}]{cizekCorrelationProblemAtomic1966}%
  \BibitemOpen
  \bibfield  {author} {\bibinfo {author} {\bibfnamefont {J.}~\bibnamefont {{\v C}{\'i}{\v z}ek}},\ }\href {\doibase 10.1063/1.1727484} {\bibfield  {journal} {\bibinfo  {journal} {J. Chem. Phys.}\ }\textbf {\bibinfo {volume} {45}},\ \bibinfo {pages} {4256} (\bibinfo {year} {1966})}\BibitemShut {NoStop}%
\bibitem [{\citenamefont {Paldus}, \citenamefont {{\v C}{\'i}{\v z}ek},\ and\ \citenamefont {Shavitt}(1972)}]{paldusCorrelationProblemsAtomic1972}%
  \BibitemOpen
  \bibfield  {author} {\bibinfo {author} {\bibfnamefont {J.}~\bibnamefont {Paldus}}, \bibinfo {author} {\bibfnamefont {J.}~\bibnamefont {{\v C}{\'i}{\v z}ek}}, \ and\ \bibinfo {author} {\bibfnamefont {I.}~\bibnamefont {Shavitt}},\ }\href {\doibase 10.1103/PhysRevA.5.50} {\bibfield  {journal} {\bibinfo  {journal} {Phys. Rev. A}\ }\textbf {\bibinfo {volume} {5}},\ \bibinfo {pages} {50} (\bibinfo {year} {1972})}\BibitemShut {NoStop}%
\bibitem [{\citenamefont {Shavitt}\ and\ \citenamefont {Bartlett}(2009)}]{shavittManyBodyMethodsChemistry2009}%
  \BibitemOpen
  \bibfield  {author} {\bibinfo {author} {\bibfnamefont {I.}~\bibnamefont {Shavitt}}\ and\ \bibinfo {author} {\bibfnamefont {R.~J.}\ \bibnamefont {Bartlett}},\ }\href {\doibase 10.1017/CBO9780511596834} {\emph {\bibinfo {title} {Many-{{Body Methods}} in {{Chemistry}} and {{Physics}}: {{MBPT}} and {{Coupled-Cluster Theory}}}}},\ Cambridge {{Molecular Science}}\ (\bibinfo  {publisher} {Cambridge University Press},\ \bibinfo {address} {Cambridge},\ \bibinfo {year} {2009})\BibitemShut {NoStop}%
\bibitem [{\citenamefont {ROWE}(1968)}]{roweEquationsofMotionMethodExtended1968}%
  \BibitemOpen
  \bibfield  {author} {\bibinfo {author} {\bibfnamefont {D.~J.}\ \bibnamefont {ROWE}},\ }\href {\doibase 10.1103/RevModPhys.40.153} {\bibfield  {journal} {\bibinfo  {journal} {Rev. Mod. Phys.}\ }\textbf {\bibinfo {volume} {40}},\ \bibinfo {pages} {153} (\bibinfo {year} {1968})}\BibitemShut {NoStop}%
\bibitem [{\citenamefont {Stanton}\ and\ \citenamefont {Bartlett}(1993)}]{stantonEquationMotionCoupledcluster1993}%
  \BibitemOpen
  \bibfield  {author} {\bibinfo {author} {\bibfnamefont {J.~F.}\ \bibnamefont {Stanton}}\ and\ \bibinfo {author} {\bibfnamefont {R.~J.}\ \bibnamefont {Bartlett}},\ }\href {\doibase 10.1063/1.464746} {\bibfield  {journal} {\bibinfo  {journal} {J. Chem. Phys.}\ }\textbf {\bibinfo {volume} {98}},\ \bibinfo {pages} {7029} (\bibinfo {year} {1993})}\BibitemShut {NoStop}%
\bibitem [{\citenamefont {Nooijen}\ and\ \citenamefont {Bartlett}(1995)}]{nooijenEquationMotionCoupled1995}%
  \BibitemOpen
  \bibfield  {author} {\bibinfo {author} {\bibfnamefont {M.}~\bibnamefont {Nooijen}}\ and\ \bibinfo {author} {\bibfnamefont {R.~J.}\ \bibnamefont {Bartlett}},\ }\href {\doibase 10.1063/1.468592} {\bibfield  {journal} {\bibinfo  {journal} {J. Chem. Phys.}\ }\textbf {\bibinfo {volume} {102}},\ \bibinfo {pages} {3629} (\bibinfo {year} {1995})}\BibitemShut {NoStop}%
\bibitem [{\citenamefont {Krylov}(2008)}]{krylovEquationofMotionCoupledClusterMethods2008}%
  \BibitemOpen
  \bibfield  {author} {\bibinfo {author} {\bibfnamefont {A.~I.}\ \bibnamefont {Krylov}},\ }\href {\doibase 10.1146/annurev.physchem.59.032607.093602} {\bibfield  {journal} {\bibinfo  {journal} {Annu. Rev. Phys. Chem.}\ }\textbf {\bibinfo {volume} {59}},\ \bibinfo {pages} {433} (\bibinfo {year} {2008})}\BibitemShut {NoStop}%
\bibitem [{\citenamefont {Kowalski}\ and\ \citenamefont {Piecuch}(2000)}]{kowalski2000active}%
  \BibitemOpen
  \bibfield  {author} {\bibinfo {author} {\bibfnamefont {K.}~\bibnamefont {Kowalski}}\ and\ \bibinfo {author} {\bibfnamefont {P.}~\bibnamefont {Piecuch}},\ }\href {\doibase https://doi.org/10.1063/1.1318757} {\bibfield  {journal} {\bibinfo  {journal} {J. Chem. Phys.}\ }\textbf {\bibinfo {volume} {113}},\ \bibinfo {pages} {8490} (\bibinfo {year} {2000})}\BibitemShut {NoStop}%
\bibitem [{\citenamefont {Koch}\ and\ \citenamefont {J{\o}rgensen}(1990)}]{kochCoupledClusterResponse1990}%
  \BibitemOpen
  \bibfield  {author} {\bibinfo {author} {\bibfnamefont {H.}~\bibnamefont {Koch}}\ and\ \bibinfo {author} {\bibfnamefont {P.}~\bibnamefont {J{\o}rgensen}},\ }\href {\doibase 10.1063/1.458814} {\bibfield  {journal} {\bibinfo  {journal} {J. Chem. Phys.}\ }\textbf {\bibinfo {volume} {93}},\ \bibinfo {pages} {3333} (\bibinfo {year} {1990})}\BibitemShut {NoStop}%
\bibitem [{\citenamefont {Monkhorst}(1977)}]{monkhorstCalculationPropertiesCoupledcluster1977}%
  \BibitemOpen
  \bibfield  {author} {\bibinfo {author} {\bibfnamefont {H.~J.}\ \bibnamefont {Monkhorst}},\ }\href {\doibase 10.1002/qua.560120850} {\bibfield  {journal} {\bibinfo  {journal} {Int. J. Quantum Chem.}\ }\textbf {\bibinfo {volume} {12}},\ \bibinfo {pages} {421} (\bibinfo {year} {1977})}\BibitemShut {NoStop}%
\bibitem [{\citenamefont {Mukherjee}\ and\ \citenamefont {Mukherjee}(1979)}]{mukherjeeResponsefunctionApproachDirect1979}%
  \BibitemOpen
  \bibfield  {author} {\bibinfo {author} {\bibfnamefont {D.}~\bibnamefont {Mukherjee}}\ and\ \bibinfo {author} {\bibfnamefont {P.~K.}\ \bibnamefont {Mukherjee}},\ }\href {\doibase 10.1016/0301-0104(79)80153-6} {\bibfield  {journal} {\bibinfo  {journal} {Chem. Phys.}\ }\textbf {\bibinfo {volume} {39}},\ \bibinfo {pages} {325} (\bibinfo {year} {1979})}\BibitemShut {NoStop}%
\bibitem [{\citenamefont {Bagus}\ and\ \citenamefont {Schaefer}(1971)}]{bagusDirectNearHartreeFockCalculations1971}%
  \BibitemOpen
  \bibfield  {author} {\bibinfo {author} {\bibfnamefont {P.~S.}\ \bibnamefont {Bagus}}\ and\ \bibinfo {author} {\bibfnamefont {H.~F.}\ \bibnamefont {Schaefer}, \bibfnamefont {III}},\ }\href {\doibase 10.1063/1.1676248} {\bibfield  {journal} {\bibinfo  {journal} {J. Chem. Phys.}\ }\textbf {\bibinfo {volume} {55}},\ \bibinfo {pages} {1474} (\bibinfo {year} {1971})}\BibitemShut {NoStop}%
\bibitem [{\citenamefont {Bagus}(1965)}]{bagusSelfConsistentFieldWaveFunctions1965}%
  \BibitemOpen
  \bibfield  {author} {\bibinfo {author} {\bibfnamefont {P.~S.}\ \bibnamefont {Bagus}},\ }\href {\doibase 10.1103/PhysRev.139.A619} {\bibfield  {journal} {\bibinfo  {journal} {Phys. Rev.}\ }\textbf {\bibinfo {volume} {139}},\ \bibinfo {pages} {A619} (\bibinfo {year} {1965})}\BibitemShut {NoStop}%
\bibitem [{\citenamefont {Lee}, \citenamefont {Small},\ and\ \citenamefont {{Head-Gordon}}(2019)}]{leeExcitedStatesCoupled2019}%
  \BibitemOpen
  \bibfield  {author} {\bibinfo {author} {\bibfnamefont {J.}~\bibnamefont {Lee}}, \bibinfo {author} {\bibfnamefont {D.~W.}\ \bibnamefont {Small}}, \ and\ \bibinfo {author} {\bibfnamefont {M.}~\bibnamefont {{Head-Gordon}}},\ }\href {\doibase 10.1063/1.5128795} {\bibfield  {journal} {\bibinfo  {journal} {J. Chem. Phys.}\ }\textbf {\bibinfo {volume} {151}},\ \bibinfo {pages} {214103} (\bibinfo {year} {2019})}\BibitemShut {NoStop}%
\bibitem [{\citenamefont {Zheng}\ and\ \citenamefont {Cheng}(2019)}]{zhengPerformanceDeltaCoupledClusterMethods2019}%
  \BibitemOpen
  \bibfield  {author} {\bibinfo {author} {\bibfnamefont {X.}~\bibnamefont {Zheng}}\ and\ \bibinfo {author} {\bibfnamefont {L.}~\bibnamefont {Cheng}},\ }\href {\doibase 10.1021/acs.jctc.9b00568} {\bibfield  {journal} {\bibinfo  {journal} {J. Chem. Theory Comput.}\ }\textbf {\bibinfo {volume} {15}},\ \bibinfo {pages} {4945} (\bibinfo {year} {2019})}\BibitemShut {NoStop}%
\bibitem [{\citenamefont {South}\ \emph {et~al.}(2016)\citenamefont {South}, \citenamefont {Shee}, \citenamefont {Mukherjee}, \citenamefont {Wilson},\ and\ \citenamefont {Saue}}]{south4ComponentRelativisticCalculations2016}%
  \BibitemOpen
  \bibfield  {author} {\bibinfo {author} {\bibfnamefont {C.}~\bibnamefont {South}}, \bibinfo {author} {\bibfnamefont {A.}~\bibnamefont {Shee}}, \bibinfo {author} {\bibfnamefont {D.}~\bibnamefont {Mukherjee}}, \bibinfo {author} {\bibfnamefont {A.~K.}\ \bibnamefont {Wilson}}, \ and\ \bibinfo {author} {\bibfnamefont {T.}~\bibnamefont {Saue}},\ }\href {\doibase 10.1039/C6CP00262E} {\bibfield  {journal} {\bibinfo  {journal} {Phys. Chem. Chem. Phys.}\ }\textbf {\bibinfo {volume} {18}},\ \bibinfo {pages} {21010} (\bibinfo {year} {2016})}\BibitemShut {NoStop}%
\bibitem [{\citenamefont {Watts}\ and\ \citenamefont {Bartlett}(1994)}]{wattsInclusionConnectedTriple1994}%
  \BibitemOpen
  \bibfield  {author} {\bibinfo {author} {\bibfnamefont {J.~D.}\ \bibnamefont {Watts}}\ and\ \bibinfo {author} {\bibfnamefont {R.~J.}\ \bibnamefont {Bartlett}},\ }\href {\doibase 10.1063/1.467620} {\bibfield  {journal} {\bibinfo  {journal} {J. Chem. Phys.}\ }\textbf {\bibinfo {volume} {101}},\ \bibinfo {pages} {3073} (\bibinfo {year} {1994})}\BibitemShut {NoStop}%
\bibitem [{\citenamefont {Kucharski}\ \emph {et~al.}(2001)\citenamefont {Kucharski}, \citenamefont {W{\l}och}, \citenamefont {Musia{\l}},\ and\ \citenamefont {Bartlett}}]{kucharski2001coupled}%
  \BibitemOpen
  \bibfield  {author} {\bibinfo {author} {\bibfnamefont {S.~A.}\ \bibnamefont {Kucharski}}, \bibinfo {author} {\bibfnamefont {M.}~\bibnamefont {W{\l}och}}, \bibinfo {author} {\bibfnamefont {M.}~\bibnamefont {Musia{\l}}}, \ and\ \bibinfo {author} {\bibfnamefont {R.~J.}\ \bibnamefont {Bartlett}},\ }\href {\doibase https://doi.org/10.1063/1.1416173} {\bibfield  {journal} {\bibinfo  {journal} {J. Chem. Phys.}\ }\textbf {\bibinfo {volume} {115}},\ \bibinfo {pages} {8263} (\bibinfo {year} {2001})}\BibitemShut {NoStop}%
\bibitem [{\citenamefont {Hirata}(2004)}]{hirataHigherorderEquationofmotionCoupledcluster2004}%
  \BibitemOpen
  \bibfield  {author} {\bibinfo {author} {\bibfnamefont {S.}~\bibnamefont {Hirata}},\ }\href {\doibase 10.1063/1.1753556} {\bibfield  {journal} {\bibinfo  {journal} {J. Chem. Phys.}\ }\textbf {\bibinfo {volume} {121}},\ \bibinfo {pages} {51} (\bibinfo {year} {2004})}\BibitemShut {NoStop}%
\bibitem [{\citenamefont {Dyall}\ and\ \citenamefont {F{\ae}gri~Jr.}(2007)}]{dyallIntroductionRelativisticQuantum2007}%
  \BibitemOpen
  \bibfield  {author} {\bibinfo {author} {\bibfnamefont {K.~G.}\ \bibnamefont {Dyall}}\ and\ \bibinfo {author} {\bibfnamefont {K.}~\bibnamefont {F{\ae}gri~Jr.}},\ }\href {\doibase 10.1093/oso/9780195140866.001.0001} {\emph {\bibinfo {title} {Introduction to {{Relativistic Quantum Chemistry}}}}}\ (\bibinfo  {publisher} {Oxford University Press},\ \bibinfo {address} {Oxford, New York},\ \bibinfo {year} {2007})\BibitemShut {NoStop}%
\bibitem [{\citenamefont {Shee}\ \emph {et~al.}(2018)\citenamefont {Shee}, \citenamefont {Saue}, \citenamefont {Visscher},\ and\ \citenamefont {Gomes}}]{sheeEquationofmotionCoupledclusterTheory2018}%
  \BibitemOpen
  \bibfield  {author} {\bibinfo {author} {\bibfnamefont {A.}~\bibnamefont {Shee}}, \bibinfo {author} {\bibfnamefont {T.}~\bibnamefont {Saue}}, \bibinfo {author} {\bibfnamefont {L.}~\bibnamefont {Visscher}}, \ and\ \bibinfo {author} {\bibfnamefont {A.~S.~P.}\ \bibnamefont {Gomes}},\ }\href {\doibase 10.1063/1.5053846} {\bibfield  {journal} {\bibinfo  {journal} {J. Chem. Phys.}\ }\textbf {\bibinfo {volume} {149}},\ \bibinfo {pages} {174113} (\bibinfo {year} {2018})}\BibitemShut {NoStop}%
\bibitem [{\citenamefont {Mukhopadhyay}\ \emph {et~al.}(2025)\citenamefont {Mukhopadhyay}, \citenamefont {Chakraborty}, \citenamefont {Chamoli}, \citenamefont {Nayak},\ and\ \citenamefont {Dutta}}]{mukhopadhyayAnalyticCalculationTransition2025}%
  \BibitemOpen
  \bibfield  {author} {\bibinfo {author} {\bibfnamefont {T.}~\bibnamefont {Mukhopadhyay}}, \bibinfo {author} {\bibfnamefont {S.}~\bibnamefont {Chakraborty}}, \bibinfo {author} {\bibfnamefont {S.}~\bibnamefont {Chamoli}}, \bibinfo {author} {\bibfnamefont {M.~K.}\ \bibnamefont {Nayak}}, \ and\ \bibinfo {author} {\bibfnamefont {A.~K.}\ \bibnamefont {Dutta}},\ }\href {\doibase 10.1063/5.0229955} {\bibfield  {journal} {\bibinfo  {journal} {J. Chem. Phys.}\ }\textbf {\bibinfo {volume} {162}},\ \bibinfo {pages} {054115} (\bibinfo {year} {2025})}\BibitemShut {NoStop}%
\bibitem [{\citenamefont {Eliav}\ \emph {et~al.}(2024)\citenamefont {Eliav}, \citenamefont {Borschevsky}, \citenamefont {Zaitsevskii}, \citenamefont {Oleynichenko},\ and\ \citenamefont {Kaldor}}]{EliavFockSpace2021}%
  \BibitemOpen
  \bibfield  {author} {\bibinfo {author} {\bibfnamefont {E.}~\bibnamefont {Eliav}}, \bibinfo {author} {\bibfnamefont {A.}~\bibnamefont {Borschevsky}}, \bibinfo {author} {\bibfnamefont {A.}~\bibnamefont {Zaitsevskii}}, \bibinfo {author} {\bibfnamefont {A.~V.}\ \bibnamefont {Oleynichenko}}, \ and\ \bibinfo {author} {\bibfnamefont {U.}~\bibnamefont {Kaldor}},\ }in\ \href {\doibase 10.1016/B978-0-12-821978-2.00042-8} {\emph {\bibinfo {booktitle} {Comprehensive Computational Chemistry}}}\ (\bibinfo  {publisher} {Elsevier},\ \bibinfo {address} {Oxford},\ \bibinfo {year} {2024})\ pp.\ \bibinfo {pages} {79--93}\BibitemShut {NoStop}%
\bibitem [{\citenamefont {Hess}(1986)}]{hessRelativisticElectronicstructureCalculations1986}%
  \BibitemOpen
  \bibfield  {author} {\bibinfo {author} {\bibfnamefont {B.~A.}\ \bibnamefont {Hess}},\ }\href {\doibase 10.1103/PhysRevA.33.3742} {\bibfield  {journal} {\bibinfo  {journal} {Phys. Rev. A}\ }\textbf {\bibinfo {volume} {33}},\ \bibinfo {pages} {3742} (\bibinfo {year} {1986})}\BibitemShut {NoStop}%
\bibitem [{\citenamefont {{van Lenthe}}\ \emph {et~al.}(1996)\citenamefont {{van Lenthe}}, \citenamefont {{van Leeuwen}}, \citenamefont {Baerends},\ and\ \citenamefont {Snijders}}]{vanlentheRelativisticRegularTwocomponent1996}%
  \BibitemOpen
  \bibfield  {author} {\bibinfo {author} {\bibfnamefont {E.}~\bibnamefont {{van Lenthe}}}, \bibinfo {author} {\bibfnamefont {R.}~\bibnamefont {{van Leeuwen}}}, \bibinfo {author} {\bibfnamefont {E.~J.}\ \bibnamefont {Baerends}}, \ and\ \bibinfo {author} {\bibfnamefont {J.~G.}\ \bibnamefont {Snijders}},\ }\href {\doibase 10.1002/(SICI)1097-461X(1996)57:3<281::AID-QUA2>3.0.CO;2-U} {\bibfield  {journal} {\bibinfo  {journal} {Int. J. Quantum Chem.}\ }\textbf {\bibinfo {volume} {57}},\ \bibinfo {pages} {281} (\bibinfo {year} {1996})}\BibitemShut {NoStop}%
\bibitem [{\citenamefont {Dyall}(1997)}]{dyallInterfacingRelativisticNonrelativistic1997}%
  \BibitemOpen
  \bibfield  {author} {\bibinfo {author} {\bibfnamefont {K.~G.}\ \bibnamefont {Dyall}},\ }\href {\doibase 10.1063/1.473860} {\bibfield  {journal} {\bibinfo  {journal} {J. Chem. Phys.}\ }\textbf {\bibinfo {volume} {106}},\ \bibinfo {pages} {9618} (\bibinfo {year} {1997})}\BibitemShut {NoStop}%
\bibitem [{\citenamefont {Nakajima}\ and\ \citenamefont {Hirao}(1999)}]{nakajimaNewRelativisticTheory1999}%
  \BibitemOpen
  \bibfield  {author} {\bibinfo {author} {\bibfnamefont {T.}~\bibnamefont {Nakajima}}\ and\ \bibinfo {author} {\bibfnamefont {K.}~\bibnamefont {Hirao}},\ }\href {\doibase 10.1016/S0009-2614(99)00150-5} {\bibfield  {journal} {\bibinfo  {journal} {Chem. Phys. Lett.}\ }\textbf {\bibinfo {volume} {302}},\ \bibinfo {pages} {383} (\bibinfo {year} {1999})}\BibitemShut {NoStop}%
\bibitem [{\citenamefont {Barysz}\ and\ \citenamefont {Sadlej}(2001)}]{baryszTwocomponentMethodsRelativistic2001}%
  \BibitemOpen
  \bibfield  {author} {\bibinfo {author} {\bibfnamefont {M.}~\bibnamefont {Barysz}}\ and\ \bibinfo {author} {\bibfnamefont {A.~J.}\ \bibnamefont {Sadlej}},\ }\href {\doibase 10.1016/S0166-1280(01)00542-5} {\bibfield  {journal} {\bibinfo  {journal} {J. Mol. Struct. THEOCHEM}\ }\textbf {\bibinfo {volume} {573}},\ \bibinfo {pages} {181} (\bibinfo {year} {2001})}\BibitemShut {NoStop}%
\bibitem [{\citenamefont {Liu}\ and\ \citenamefont {Peng}(2009)}]{liuExactTwocomponentHamiltonians2009}%
  \BibitemOpen
  \bibfield  {author} {\bibinfo {author} {\bibfnamefont {W.}~\bibnamefont {Liu}}\ and\ \bibinfo {author} {\bibfnamefont {D.}~\bibnamefont {Peng}},\ }\href {\doibase 10.1063/1.3159445} {\bibfield  {journal} {\bibinfo  {journal} {J. Chem. Phys.}\ }\textbf {\bibinfo {volume} {131}},\ \bibinfo {pages} {031104} (\bibinfo {year} {2009})}\BibitemShut {NoStop}%
\bibitem [{\citenamefont {Saue}(2011)}]{saueRelativisticHamiltoniansChemistry2011}%
  \BibitemOpen
  \bibfield  {author} {\bibinfo {author} {\bibfnamefont {T.}~\bibnamefont {Saue}},\ }\href {\doibase 10.1002/cphc.201100682} {\bibfield  {journal} {\bibinfo  {journal} {ChemPhysChem}\ }\textbf {\bibinfo {volume} {12}},\ \bibinfo {pages} {3077} (\bibinfo {year} {2011})}\BibitemShut {NoStop}%
\bibitem [{\citenamefont {He{\ss}}\ \emph {et~al.}(1996)\citenamefont {He{\ss}}, \citenamefont {Marian}, \citenamefont {Wahlgren},\ and\ \citenamefont {Gropen}}]{hessMeanfieldSpinorbitMethod1996}%
  \BibitemOpen
  \bibfield  {author} {\bibinfo {author} {\bibfnamefont {B.~A.}\ \bibnamefont {He{\ss}}}, \bibinfo {author} {\bibfnamefont {C.~M.}\ \bibnamefont {Marian}}, \bibinfo {author} {\bibfnamefont {U.}~\bibnamefont {Wahlgren}}, \ and\ \bibinfo {author} {\bibfnamefont {O.}~\bibnamefont {Gropen}},\ }\href {\doibase 10.1016/0009-2614(96)00119-4} {\bibfield  {journal} {\bibinfo  {journal} {Chem. Phys. Lett.}\ }\textbf {\bibinfo {volume} {251}},\ \bibinfo {pages} {365} (\bibinfo {year} {1996})}\BibitemShut {NoStop}%
\bibitem [{\citenamefont {Liu}\ and\ \citenamefont {Cheng}(2018)}]{liuAtomicMeanfieldSpinorbit2018}%
  \BibitemOpen
  \bibfield  {author} {\bibinfo {author} {\bibfnamefont {J.}~\bibnamefont {Liu}}\ and\ \bibinfo {author} {\bibfnamefont {L.}~\bibnamefont {Cheng}},\ }\href {\doibase 10.1063/1.5023750} {\bibfield  {journal} {\bibinfo  {journal} {J. Chem. Phys.}\ }\textbf {\bibinfo {volume} {148}},\ \bibinfo {pages} {144108} (\bibinfo {year} {2018})}\BibitemShut {NoStop}%
\bibitem [{\citenamefont {Zhang}\ and\ \citenamefont {Cheng}(2022)}]{zhangAtomicMeanFieldApproach2022}%
  \BibitemOpen
  \bibfield  {author} {\bibinfo {author} {\bibfnamefont {C.}~\bibnamefont {Zhang}}\ and\ \bibinfo {author} {\bibfnamefont {L.}~\bibnamefont {Cheng}},\ }\href {\doibase 10.1021/acs.jpca.2c02181} {\bibfield  {journal} {\bibinfo  {journal} {J. Phys. Chem. A}\ }\textbf {\bibinfo {volume} {126}},\ \bibinfo {pages} {4537} (\bibinfo {year} {2022})}\BibitemShut {NoStop}%
\bibitem [{\citenamefont {Knecht}\ \emph {et~al.}(2022)\citenamefont {Knecht}, \citenamefont {Repisky}, \citenamefont {Jensen},\ and\ \citenamefont {Saue}}]{knechtExactTwocomponentHamiltonians2022}%
  \BibitemOpen
  \bibfield  {author} {\bibinfo {author} {\bibfnamefont {S.}~\bibnamefont {Knecht}}, \bibinfo {author} {\bibfnamefont {M.}~\bibnamefont {Repisky}}, \bibinfo {author} {\bibfnamefont {H.~J.~A.}\ \bibnamefont {Jensen}}, \ and\ \bibinfo {author} {\bibfnamefont {T.}~\bibnamefont {Saue}},\ }\href {\doibase 10.1063/5.0095112} {\bibfield  {journal} {\bibinfo  {journal} {J. Chem. Phys.}\ }\textbf {\bibinfo {volume} {157}},\ \bibinfo {pages} {114106} (\bibinfo {year} {2022})}\BibitemShut {NoStop}%
\bibitem [{\citenamefont {Kelley}\ and\ \citenamefont {Shiozaki}(2013)}]{kelleyLargescaleDiracFock2013}%
  \BibitemOpen
  \bibfield  {author} {\bibinfo {author} {\bibfnamefont {M.~S.}\ \bibnamefont {Kelley}}\ and\ \bibinfo {author} {\bibfnamefont {T.}~\bibnamefont {Shiozaki}},\ }\href {\doibase 10.1063/1.4807612} {\bibfield  {journal} {\bibinfo  {journal} {J. Chem. Phys.}\ }\textbf {\bibinfo {volume} {138}},\ \bibinfo {pages} {204113} (\bibinfo {year} {2013})}\BibitemShut {NoStop}%
\bibitem [{\citenamefont {Bates}\ and\ \citenamefont {Shiozaki}(2015)}]{batesFullyRelativisticComplete2015}%
  \BibitemOpen
  \bibfield  {author} {\bibinfo {author} {\bibfnamefont {J.~E.}\ \bibnamefont {Bates}}\ and\ \bibinfo {author} {\bibfnamefont {T.}~\bibnamefont {Shiozaki}},\ }\href {\doibase 10.1063/1.4906344} {\bibfield  {journal} {\bibinfo  {journal} {J. Chem. Phys.}\ }\textbf {\bibinfo {volume} {142}},\ \bibinfo {pages} {044112} (\bibinfo {year} {2015})}\BibitemShut {NoStop}%
\bibitem [{\citenamefont {{Helmich-Paris}}, \citenamefont {Repisky},\ and\ \citenamefont {Visscher}(2019)}]{helmich-parisRelativisticCholeskydecomposedDensity2019}%
  \BibitemOpen
  \bibfield  {author} {\bibinfo {author} {\bibfnamefont {B.}~\bibnamefont {{Helmich-Paris}}}, \bibinfo {author} {\bibfnamefont {M.}~\bibnamefont {Repisky}}, \ and\ \bibinfo {author} {\bibfnamefont {L.}~\bibnamefont {Visscher}},\ }\href {\doibase 10.1016/j.chemphys.2018.11.009} {\bibfield  {journal} {\bibinfo  {journal} {Chem. Phys.}\ }\textbf {\bibinfo {volume} {518}},\ \bibinfo {pages} {38} (\bibinfo {year} {2019})}\BibitemShut {NoStop}%
\bibitem [{\citenamefont {Banerjee}\ \emph {et~al.}(2023)\citenamefont {Banerjee}, \citenamefont {Zhang}, \citenamefont {Dyall},\ and\ \citenamefont {Li}}]{banerjeeRelativisticResolutionoftheidentityCholesky2023}%
  \BibitemOpen
  \bibfield  {author} {\bibinfo {author} {\bibfnamefont {S.}~\bibnamefont {Banerjee}}, \bibinfo {author} {\bibfnamefont {T.}~\bibnamefont {Zhang}}, \bibinfo {author} {\bibfnamefont {K.~G.}\ \bibnamefont {Dyall}}, \ and\ \bibinfo {author} {\bibfnamefont {X.}~\bibnamefont {Li}},\ }\href {\doibase 10.1063/5.0161871} {\bibfield  {journal} {\bibinfo  {journal} {J. Chem. Phys.}\ }\textbf {\bibinfo {volume} {159}},\ \bibinfo {pages} {114119} (\bibinfo {year} {2023})}\BibitemShut {NoStop}%
\bibitem [{\citenamefont {Uhl{\'i}{\v r}ov{\'a}}\ \emph {et~al.}(2024)\citenamefont {Uhl{\'i}{\v r}ov{\'a}}, \citenamefont {Cianchino}, \citenamefont {Nottoli}, \citenamefont {Lipparini},\ and\ \citenamefont {Gauss}}]{uhlirovaCholeskyDecompositionSpinFree2024}%
  \BibitemOpen
  \bibfield  {author} {\bibinfo {author} {\bibfnamefont {T.}~\bibnamefont {Uhl{\'i}{\v r}ov{\'a}}}, \bibinfo {author} {\bibfnamefont {D.}~\bibnamefont {Cianchino}}, \bibinfo {author} {\bibfnamefont {T.}~\bibnamefont {Nottoli}}, \bibinfo {author} {\bibfnamefont {F.}~\bibnamefont {Lipparini}}, \ and\ \bibinfo {author} {\bibfnamefont {J.}~\bibnamefont {Gauss}},\ }\href {\doibase 10.1021/acs.jpca.4c04353} {\bibfield  {journal} {\bibinfo  {journal} {J. Phys. Chem. A}\ }\textbf {\bibinfo {volume} {128}},\ \bibinfo {pages} {8292} (\bibinfo {year} {2024})}\BibitemShut {NoStop}%
\bibitem [{\citenamefont {Zhang}\ \emph {et~al.}(2024)\citenamefont {Zhang}, \citenamefont {Lipparini}, \citenamefont {Stopkowicz}, \citenamefont {Gauss},\ and\ \citenamefont {Cheng}}]{zhangCholeskyDecompositionBasedImplementation2024}%
  \BibitemOpen
  \bibfield  {author} {\bibinfo {author} {\bibfnamefont {C.}~\bibnamefont {Zhang}}, \bibinfo {author} {\bibfnamefont {F.}~\bibnamefont {Lipparini}}, \bibinfo {author} {\bibfnamefont {S.}~\bibnamefont {Stopkowicz}}, \bibinfo {author} {\bibfnamefont {J.}~\bibnamefont {Gauss}}, \ and\ \bibinfo {author} {\bibfnamefont {L.}~\bibnamefont {Cheng}},\ }\href {\doibase 10.1021/acs.jctc.3c01236} {\bibfield  {journal} {\bibinfo  {journal} {J. Chem. Theory Comput.}\ }\textbf {\bibinfo {volume} {20}},\ \bibinfo {pages} {787} (\bibinfo {year} {2024})}\BibitemShut {NoStop}%
\bibitem [{\citenamefont {Pototschnig}\ \emph {et~al.}(2021)\citenamefont {Pototschnig}, \citenamefont {Papadopoulos}, \citenamefont {Lyakh}, \citenamefont {Repisky}, \citenamefont {Halbert}, \citenamefont {Gomes}, \citenamefont {Jensen},\ and\ \citenamefont {Visscher}}]{pototschnigImplementationRelativisticCoupled2021}%
  \BibitemOpen
  \bibfield  {author} {\bibinfo {author} {\bibfnamefont {J.~V.}\ \bibnamefont {Pototschnig}}, \bibinfo {author} {\bibfnamefont {A.}~\bibnamefont {Papadopoulos}}, \bibinfo {author} {\bibfnamefont {D.~I.}\ \bibnamefont {Lyakh}}, \bibinfo {author} {\bibfnamefont {M.}~\bibnamefont {Repisky}}, \bibinfo {author} {\bibfnamefont {L.}~\bibnamefont {Halbert}}, \bibinfo {author} {\bibfnamefont {A.~S.~P.}\ \bibnamefont {Gomes}}, \bibinfo {author} {\bibfnamefont {H.~J.~A.}\ \bibnamefont {Jensen}}, \ and\ \bibinfo {author} {\bibfnamefont {L.}~\bibnamefont {Visscher}},\ }\href {\doibase 10.1021/acs.jctc.1c00260} {\bibfield  {journal} {\bibinfo  {journal} {J. Chem. Theory Comput.}\ }\textbf {\bibinfo {volume} {17}},\ \bibinfo {pages} {5509} (\bibinfo {year} {2021})}\BibitemShut {NoStop}%
\bibitem [{\citenamefont {DePrince}\ and\ \citenamefont {Hammond}(2011)}]{deprinceCoupledClusterTheory2011}%
  \BibitemOpen
  \bibfield  {author} {\bibinfo {author} {\bibfnamefont {A.~E.~I.}\ \bibnamefont {DePrince}}\ and\ \bibinfo {author} {\bibfnamefont {J.~R.}\ \bibnamefont {Hammond}},\ }\href {\doibase 10.1021/ct100584w} {\bibfield  {journal} {\bibinfo  {journal} {J. Chem. Theory Comput.}\ }\textbf {\bibinfo {volume} {7}},\ \bibinfo {pages} {1287} (\bibinfo {year} {2011})}\BibitemShut {NoStop}%
\bibitem [{\citenamefont {Chamoli}\ \emph {et~al.}(2022)\citenamefont {Chamoli}, \citenamefont {Surjuse}, \citenamefont {Jangid}, \citenamefont {Nayak},\ and\ \citenamefont {Dutta}}]{chamoliReducedCostFourcomponent2022}%
  \BibitemOpen
  \bibfield  {author} {\bibinfo {author} {\bibfnamefont {S.}~\bibnamefont {Chamoli}}, \bibinfo {author} {\bibfnamefont {K.}~\bibnamefont {Surjuse}}, \bibinfo {author} {\bibfnamefont {B.}~\bibnamefont {Jangid}}, \bibinfo {author} {\bibfnamefont {M.~K.}\ \bibnamefont {Nayak}}, \ and\ \bibinfo {author} {\bibfnamefont {A.~K.}\ \bibnamefont {Dutta}},\ }\href {\doibase 10.1063/5.0085932} {\bibfield  {journal} {\bibinfo  {journal} {J. Chem. Phys.}\ }\textbf {\bibinfo {volume} {156}},\ \bibinfo {pages} {204120} (\bibinfo {year} {2022})}\BibitemShut {NoStop}%
\bibitem [{\citenamefont {L{\"o}wdin}(1955)}]{lowdinQuantumTheoryManyParticle1955}%
  \BibitemOpen
  \bibfield  {author} {\bibinfo {author} {\bibfnamefont {P.-O.}\ \bibnamefont {L{\"o}wdin}},\ }\href {\doibase 10.1103/PhysRev.97.1474} {\bibfield  {journal} {\bibinfo  {journal} {Phys. Rev.}\ }\textbf {\bibinfo {volume} {97}},\ \bibinfo {pages} {1474} (\bibinfo {year} {1955})}\BibitemShut {NoStop}%
\bibitem [{\citenamefont {Majee}\ \emph {et~al.}(2024)\citenamefont {Majee}, \citenamefont {Chakraborty}, \citenamefont {Mukhopadhyay}, \citenamefont {Nayak},\ and\ \citenamefont {Dutta}}]{majeeReducedCostFourcomponent2024}%
  \BibitemOpen
  \bibfield  {author} {\bibinfo {author} {\bibfnamefont {K.}~\bibnamefont {Majee}}, \bibinfo {author} {\bibfnamefont {S.}~\bibnamefont {Chakraborty}}, \bibinfo {author} {\bibfnamefont {T.}~\bibnamefont {Mukhopadhyay}}, \bibinfo {author} {\bibfnamefont {M.~K.}\ \bibnamefont {Nayak}}, \ and\ \bibinfo {author} {\bibfnamefont {A.~K.}\ \bibnamefont {Dutta}},\ }\href {\doibase 10.1063/5.0207091} {\bibfield  {journal} {\bibinfo  {journal} {J. Chem. Phys.}\ }\textbf {\bibinfo {volume} {161}},\ \bibinfo {pages} {034101} (\bibinfo {year} {2024})}\BibitemShut {NoStop}%
\bibitem [{\citenamefont {Surjuse}\ \emph {et~al.}(2022)\citenamefont {Surjuse}, \citenamefont {Chamoli}, \citenamefont {Nayak},\ and\ \citenamefont {Dutta}}]{surjuseLowcostFourcomponentRelativistic2022}%
  \BibitemOpen
  \bibfield  {author} {\bibinfo {author} {\bibfnamefont {K.}~\bibnamefont {Surjuse}}, \bibinfo {author} {\bibfnamefont {S.}~\bibnamefont {Chamoli}}, \bibinfo {author} {\bibfnamefont {M.~K.}\ \bibnamefont {Nayak}}, \ and\ \bibinfo {author} {\bibfnamefont {A.~K.}\ \bibnamefont {Dutta}},\ }\href {\doibase 10.1063/5.0125868} {\bibfield  {journal} {\bibinfo  {journal} {J. Chem. Phys.}\ }\textbf {\bibinfo {volume} {157}},\ \bibinfo {pages} {204106} (\bibinfo {year} {2022})}\BibitemShut {NoStop}%
\bibitem [{\citenamefont {Chamoli}\ \emph {et~al.}(2025)\citenamefont {Chamoli}, \citenamefont {Wang}, \citenamefont {Zhang}, \citenamefont {Nayak},\ and\ \citenamefont {Dutta}}]{chamoli2025frozen}%
  \BibitemOpen
  \bibfield  {author} {\bibinfo {author} {\bibfnamefont {S.}~\bibnamefont {Chamoli}}, \bibinfo {author} {\bibfnamefont {X.}~\bibnamefont {Wang}}, \bibinfo {author} {\bibfnamefont {C.}~\bibnamefont {Zhang}}, \bibinfo {author} {\bibfnamefont {M.~K.}\ \bibnamefont {Nayak}}, \ and\ \bibinfo {author} {\bibfnamefont {A.~K.}\ \bibnamefont {Dutta}},\ }\href {\doibase https://pubs.acs.org/doi/full/10.1021/acs.jctc.5c00199} {\bibfield  {journal} {\bibinfo  {journal} {J. Chem. Theory Comput.}\ }\textbf {\bibinfo {volume} {21}},\ \bibinfo {pages} {4532} (\bibinfo {year} {2025})}\BibitemShut {NoStop}%
\bibitem [{\citenamefont {Yuan}\ \emph {et~al.}(2023)\citenamefont {Yuan}, \citenamefont {Halbert}, \citenamefont {Visscher},\ and\ \citenamefont {Gomes}}]{yuanFrequencyDependentQuadraticResponse2023}%
  \BibitemOpen
  \bibfield  {author} {\bibinfo {author} {\bibfnamefont {X.}~\bibnamefont {Yuan}}, \bibinfo {author} {\bibfnamefont {L.}~\bibnamefont {Halbert}}, \bibinfo {author} {\bibfnamefont {L.}~\bibnamefont {Visscher}}, \ and\ \bibinfo {author} {\bibfnamefont {A.~S.~P.}\ \bibnamefont {Gomes}},\ }\href {\doibase 10.1021/acs.jctc.3c01011} {\bibfield  {journal} {\bibinfo  {journal} {J. Chem. Theory Comput.}\ }\textbf {\bibinfo {volume} {19}},\ \bibinfo {pages} {9248} (\bibinfo {year} {2023})}\BibitemShut {NoStop}%
\bibitem [{\citenamefont {Lindroth}(1988)}]{lindrothNumericalSolutionRelativistic1988}%
  \BibitemOpen
  \bibfield  {author} {\bibinfo {author} {\bibfnamefont {E.}~\bibnamefont {Lindroth}},\ }\href {\doibase 10.1103/PhysRevA.37.316} {\bibfield  {journal} {\bibinfo  {journal} {Phys. Rev. A}\ }\textbf {\bibinfo {volume} {37}},\ \bibinfo {pages} {316} (\bibinfo {year} {1988})}\BibitemShut {NoStop}%
\bibitem [{\citenamefont {Visscher}, \citenamefont {Lee},\ and\ \citenamefont {Dyall}(1996)}]{visscherFormulationImplementationRelativistic1996}%
  \BibitemOpen
  \bibfield  {author} {\bibinfo {author} {\bibfnamefont {L.}~\bibnamefont {Visscher}}, \bibinfo {author} {\bibfnamefont {T.~J.}\ \bibnamefont {Lee}}, \ and\ \bibinfo {author} {\bibfnamefont {K.~G.}\ \bibnamefont {Dyall}},\ }\href {\doibase 10.1063/1.472655} {\bibfield  {journal} {\bibinfo  {journal} {J. Chem. Phys.}\ }\textbf {\bibinfo {volume} {105}},\ \bibinfo {pages} {8769} (\bibinfo {year} {1996})}\BibitemShut {NoStop}%
\bibitem [{\citenamefont {Visscher}, \citenamefont {Eliav},\ and\ \citenamefont {Kaldor}(2001)}]{visscherFormulationImplementationRelativistic2001}%
  \BibitemOpen
  \bibfield  {author} {\bibinfo {author} {\bibfnamefont {L.}~\bibnamefont {Visscher}}, \bibinfo {author} {\bibfnamefont {E.}~\bibnamefont {Eliav}}, \ and\ \bibinfo {author} {\bibfnamefont {U.}~\bibnamefont {Kaldor}},\ }\href {\doibase 10.1063/1.1415746} {\bibfield  {journal} {\bibinfo  {journal} {J. Chem. Phys.}\ }\textbf {\bibinfo {volume} {115}},\ \bibinfo {pages} {9720} (\bibinfo {year} {2001})}\BibitemShut {NoStop}%
\bibitem [{\citenamefont {Sucher}(1980)}]{sucherFoundationsRelativisticTheory1980}%
  \BibitemOpen
  \bibfield  {author} {\bibinfo {author} {\bibfnamefont {J.}~\bibnamefont {Sucher}},\ }\href {\doibase 10.1103/PhysRevA.22.348} {\bibfield  {journal} {\bibinfo  {journal} {Phys. Rev. A}\ }\textbf {\bibinfo {volume} {22}},\ \bibinfo {pages} {348} (\bibinfo {year} {1980})}\BibitemShut {NoStop}%
\bibitem [{\citenamefont {Reiher}\ and\ \citenamefont {Wolf}(2015)}]{reiherRelativisticQuantumChemistry2015}%
  \BibitemOpen
  \bibfield  {author} {\bibinfo {author} {\bibfnamefont {M.}~\bibnamefont {Reiher}}\ and\ \bibinfo {author} {\bibfnamefont {A.}~\bibnamefont {Wolf}},\ } \href {https://onlinelibrary.wiley.com/doi/book/10.1002/9783527667550} {\emph {\bibinfo {title} {Relativistic {{Quantum Chemistry}}: {{The Fundamental Theory}} of {{Molecular Science}}}}}\ (\bibinfo  {publisher} {John Wiley \& Sons},\ \bibinfo {year} {2015})\BibitemShut {NoStop}%
\bibitem [{\citenamefont {Yuan}, \citenamefont {Visscher},\ and\ \citenamefont {Gomes}(2022)}]{yuanAssessingMP2Frozen2022}%
  \BibitemOpen
  \bibfield  {author} {\bibinfo {author} {\bibfnamefont {X.}~\bibnamefont {Yuan}}, \bibinfo {author} {\bibfnamefont {L.}~\bibnamefont {Visscher}}, \ and\ \bibinfo {author} {\bibfnamefont {A.~S.~P.}\ \bibnamefont {Gomes}},\ }\href {\doibase 10.1063/5.0087243} {\bibfield  {journal} {\bibinfo  {journal} {J. Chem. Phys.}\ }\textbf {\bibinfo {volume} {156}},\ \bibinfo {pages} {224108} (\bibinfo {year} {2022})}\BibitemShut {NoStop}%
\bibitem [{\citenamefont {Chamoli}, \citenamefont {Nayak},\ and\ \citenamefont {Dutta}(2024)}]{chamoliRelativisticReducedDensity2024}%
  \BibitemOpen
  \bibfield  {author} {\bibinfo {author} {\bibfnamefont {S.}~\bibnamefont {Chamoli}}, \bibinfo {author} {\bibfnamefont {M.~K.}\ \bibnamefont {Nayak}}, \ and\ \bibinfo {author} {\bibfnamefont {A.~K.}\ \bibnamefont {Dutta}},\ }in\ \href {\doibase 10.1002/9781394217656.ch5} {\emph {\bibinfo {booktitle} {Electron {{Density}}}}}\ (\bibinfo  {publisher} {John Wiley \& Sons, Ltd},\ \bibinfo {year} {2024})\ Chap.~\bibinfo {chapter} {5}, pp.\ \bibinfo {pages} {83--96}\BibitemShut {NoStop}%
\bibitem [{\citenamefont {Helmich}\ and\ \citenamefont {H{\"a}ttig}(2011)}]{helmich2011local}%
  \BibitemOpen
  \bibfield  {author} {\bibinfo {author} {\bibfnamefont {B.}~\bibnamefont {Helmich}}\ and\ \bibinfo {author} {\bibfnamefont {C.}~\bibnamefont {H{\"a}ttig}},\ }\href {\doibase 10.1063/1.3664902} {\bibfield  {journal} {\bibinfo  {journal} {J. Chem. Phys.}\ }\textbf {\bibinfo {volume} {135}},\ \bibinfo {pages} {214106} (\bibinfo {year} {2011})}\BibitemShut {NoStop}%
\bibitem [{\citenamefont {Mester}, \citenamefont {Nagy},\ and\ \citenamefont {K{\'a}llay}(2017)}]{mester2017reduced}%
  \BibitemOpen
  \bibfield  {author} {\bibinfo {author} {\bibfnamefont {D.}~\bibnamefont {Mester}}, \bibinfo {author} {\bibfnamefont {P.~R.}\ \bibnamefont {Nagy}}, \ and\ \bibinfo {author} {\bibfnamefont {M.}~\bibnamefont {K{\'a}llay}},\ }\href {\doibase 10.1063/1.4983277} {\bibfield  {journal} {\bibinfo  {journal} {J. Chem. Phys.}\ }\textbf {\bibinfo {volume} {146}},\ \bibinfo {pages} {194102} (\bibinfo {year} {2017})}\BibitemShut {NoStop}%
\bibitem [{\citenamefont {Peng}, \citenamefont {Clement},\ and\ \citenamefont {Valeev}(2018)}]{peng2018state}%
  \BibitemOpen
  \bibfield  {author} {\bibinfo {author} {\bibfnamefont {C.}~\bibnamefont {Peng}}, \bibinfo {author} {\bibfnamefont {M.~C.}\ \bibnamefont {Clement}}, \ and\ \bibinfo {author} {\bibfnamefont {E.~F.}\ \bibnamefont {Valeev}},\ }\href {\doibase 10.1021/acs.jctc.8b00171} {\bibfield  {journal} {\bibinfo  {journal} {J. Chem. Theory Comput.}\ }\textbf {\bibinfo {volume} {14}},\ \bibinfo {pages} {5597} (\bibinfo {year} {2018})}\BibitemShut {NoStop}%
\bibitem [{\citenamefont {Dutta}\ \emph {et~al.}(2018)\citenamefont {Dutta}, \citenamefont {Nooijen}, \citenamefont {Neese},\ and\ \citenamefont {Izs{\'a}k}}]{dutta2018exploring}%
  \BibitemOpen
  \bibfield  {author} {\bibinfo {author} {\bibfnamefont {A.~K.}\ \bibnamefont {Dutta}}, \bibinfo {author} {\bibfnamefont {M.}~\bibnamefont {Nooijen}}, \bibinfo {author} {\bibfnamefont {F.}~\bibnamefont {Neese}}, \ and\ \bibinfo {author} {\bibfnamefont {R.}~\bibnamefont {Izs{\'a}k}},\ }\href {\doibase 10.1021/acs.jctc.7b00802} {\bibfield  {journal} {\bibinfo  {journal} {J. Chem. Theory Comput.}\ }\textbf {\bibinfo {volume} {14}},\ \bibinfo {pages} {72} (\bibinfo {year} {2018})}\BibitemShut {NoStop}%
\bibitem [{\citenamefont {Manna}\ and\ \citenamefont {Dutta}(2025)}]{manna2025reducedcostequationmotion}%
  \BibitemOpen
  \bibfield  {author} {\bibinfo {author} {\bibfnamefont {A.}~\bibnamefont {Manna}}\ and\ \bibinfo {author} {\bibfnamefont {A.~K.}\ \bibnamefont {Dutta}},\ }\href {https://arxiv.org/abs/2506.16894} {} (\bibinfo {year} {2025}),\ \Eprint {http://arxiv.org/abs/2506.16894} {arXiv:2506.16894 [physics.chem-ph]} \BibitemShut {NoStop}%
\bibitem [{\citenamefont {Pernpointner}(2014)}]{pernpointnerRelativisticPolarizationPropagator2014}%
  \BibitemOpen
  \bibfield  {author} {\bibinfo {author} {\bibfnamefont {M.}~\bibnamefont {Pernpointner}},\ }\href {\doibase 10.1063/1.4865964} {\bibfield  {journal} {\bibinfo  {journal} {J. Chem. Phys.}\ }\textbf {\bibinfo {volume} {140}},\ \bibinfo {pages} {084108} (\bibinfo {year} {2014})}\BibitemShut {NoStop}%
\bibitem [{\citenamefont {Pernpointner}, \citenamefont {Visscher},\ and\ \citenamefont {Trofimov}(2018)}]{pernpointnerFourComponentPolarizationPropagator2018}%
  \BibitemOpen
  \bibfield  {author} {\bibinfo {author} {\bibfnamefont {M.}~\bibnamefont {Pernpointner}}, \bibinfo {author} {\bibfnamefont {L.}~\bibnamefont {Visscher}}, \ and\ \bibinfo {author} {\bibfnamefont {A.~B.}\ \bibnamefont {Trofimov}},\ }\href {\doibase 10.1021/acs.jctc.7b01056} {\bibfield  {journal} {\bibinfo  {journal} {J. Chem. Theory Comput.}\ }\textbf {\bibinfo {volume} {14}},\ \bibinfo {pages} {1510} (\bibinfo {year} {2018})}\BibitemShut {NoStop}%
\bibitem [{\citenamefont {Chakraborty}\ \emph {et~al.}(2025)\citenamefont {Chakraborty}, \citenamefont {Mukhopadhyay}, \citenamefont {Nayak},\ and\ \citenamefont {Dutta}}]{chakraborty2025relativistic}%
  \BibitemOpen
  \bibfield  {author} {\bibinfo {author} {\bibfnamefont {S.}~\bibnamefont {Chakraborty}}, \bibinfo {author} {\bibfnamefont {T.}~\bibnamefont {Mukhopadhyay}}, \bibinfo {author} {\bibfnamefont {M.~K.}\ \bibnamefont {Nayak}}, \ and\ \bibinfo {author} {\bibfnamefont {A.~K.}\ \bibnamefont {Dutta}},\ }\href {\doibase 10.1063/5.0246920} {\bibfield  {journal} {\bibinfo  {journal} {J. Chem. Phys.}\ }\textbf {\bibinfo {volume} {162}},\ \bibinfo {pages} {104106} (\bibinfo {year} {2025})}\BibitemShut {NoStop}%
\bibitem [{\citenamefont {Schirmer}\ and\ \citenamefont {Trofimov}(2004)}]{schirmer2004intermediate}%
  \BibitemOpen
  \bibfield  {author} {\bibinfo {author} {\bibfnamefont {J.}~\bibnamefont {Schirmer}}\ and\ \bibinfo {author} {\bibfnamefont {A.~B.}\ \bibnamefont {Trofimov}},\ }\href {\doibase 10.1063/1.1752875} {\bibfield  {journal} {\bibinfo  {journal} {J. Chem. Phys.}\ }\textbf {\bibinfo {volume} {120}},\ \bibinfo {pages} {11449} (\bibinfo {year} {2004})}\BibitemShut {NoStop}%
\bibitem [{\citenamefont {Hirao}\ and\ \citenamefont {Nakatsuji}(1982)}]{hiraoGeneralizationDavidsonsMethod1982}%
  \BibitemOpen
  \bibfield  {author} {\bibinfo {author} {\bibfnamefont {K.}~\bibnamefont {Hirao}}\ and\ \bibinfo {author} {\bibfnamefont {H.}~\bibnamefont {Nakatsuji}},\ }\href {\doibase 10.1016/0021-9991(82)90119-X} {\bibfield  {journal} {\bibinfo  {journal} {J. Comput. Phys.}\ }\textbf {\bibinfo {volume} {45}},\ \bibinfo {pages} {246} (\bibinfo {year} {1982})}\BibitemShut {NoStop}%
\bibitem [{\citenamefont {Dyall}(1994)}]{dyallExactSeparationSpinfree1994}%
  \BibitemOpen
  \bibfield  {author} {\bibinfo {author} {\bibfnamefont {K.~G.}\ \bibnamefont {Dyall}},\ }\href {\doibase 10.1063/1.466508} {\bibfield  {journal} {\bibinfo  {journal} {J. Chem. Phys.}\ }\textbf {\bibinfo {volume} {100}},\ \bibinfo {pages} {2118} (\bibinfo {year} {1994})}\BibitemShut {NoStop}%
\bibitem [{\citenamefont {Dutta}\ \emph {et~al.}(2025)\citenamefont {Dutta}, \citenamefont {Manna}, \citenamefont {Jangid}, \citenamefont {Majee}, \citenamefont {Surjuse}, \citenamefont {Mukherjee}, \citenamefont {Thapa}, \citenamefont {Arora}, \citenamefont {Chamoli}, \citenamefont {Haldar}, \citenamefont {Chakraborty}, \citenamefont {Mandal},\ and\ \citenamefont {Mukhopadhyay}}]{duttaBAGHQuantumChemistry2025}%
  \BibitemOpen
  \bibfield  {author} {\bibinfo {author} {\bibfnamefont {A.~K.}\ \bibnamefont {Dutta}}, \bibinfo {author} {\bibfnamefont {A.}~\bibnamefont {Manna}}, \bibinfo {author} {\bibfnamefont {B.}~\bibnamefont {Jangid}}, \bibinfo {author} {\bibfnamefont {K.}~\bibnamefont {Majee}}, \bibinfo {author} {\bibfnamefont {K.}~\bibnamefont {Surjuse}}, \bibinfo {author} {\bibfnamefont {M.}~\bibnamefont {Mukherjee}}, \bibinfo {author} {\bibfnamefont {M.}~\bibnamefont {Thapa}}, \bibinfo {author} {\bibfnamefont {S.}~\bibnamefont {Arora}}, \bibinfo {author} {\bibfnamefont {S.}~\bibnamefont {Chamoli}}, \bibinfo {author} {\bibfnamefont {S.}~\bibnamefont {Haldar}}, \bibinfo {author} {\bibfnamefont {S.}~\bibnamefont {Chakraborty}}, \bibinfo {author} {\bibfnamefont {S.}~\bibnamefont {Mandal}}, \ and\ \bibinfo {author} {\bibfnamefont {T.}~\bibnamefont {Mukhopadhyay}},\ }\href {https://bagh-doc.readthedocs.io/en/latest/} {\enquote {\bibinfo {title} {{{BAGH}}: {{A}} quantum chemistry software package},}\ } (\bibinfo {year}
  {2025})\BibitemShut {NoStop}%
\bibitem [{\citenamefont {Sun}(2015)}]{sunLibcintEfficientGeneral2015}%
  \BibitemOpen
  \bibfield  {author} {\bibinfo {author} {\bibfnamefont {Q.}~\bibnamefont {Sun}},\ }\href {\doibase https://doi.org/10.1002/jcc.23981} {\bibfield  {journal} {\bibinfo  {journal} {J. Comput. Chem.}\ }\textbf {\bibinfo {volume} {36}},\ \bibinfo {pages} {1664} (\bibinfo {year} {2015})}\BibitemShut {NoStop}%
\bibitem [{\citenamefont {Sun}\ \emph {et~al.}(2018)\citenamefont {Sun}, \citenamefont {Berkelbach}, \citenamefont {Blunt}, \citenamefont {Booth}, \citenamefont {Guo}, \citenamefont {Li}, \citenamefont {Liu}, \citenamefont {McClain}, \citenamefont {Sayfutyarova}, \citenamefont {Sharma}, \citenamefont {Wouters},\ and\ \citenamefont {Chan}}]{sunPySCFPythonbasedSimulations2018}%
  \BibitemOpen
  \bibfield  {author} {\bibinfo {author} {\bibfnamefont {Q.}~\bibnamefont {Sun}}, \bibinfo {author} {\bibfnamefont {T.~C.}\ \bibnamefont {Berkelbach}}, \bibinfo {author} {\bibfnamefont {N.~S.}\ \bibnamefont {Blunt}}, \bibinfo {author} {\bibfnamefont {G.~H.}\ \bibnamefont {Booth}}, \bibinfo {author} {\bibfnamefont {S.}~\bibnamefont {Guo}}, \bibinfo {author} {\bibfnamefont {Z.}~\bibnamefont {Li}}, \bibinfo {author} {\bibfnamefont {J.}~\bibnamefont {Liu}}, \bibinfo {author} {\bibfnamefont {J.~D.}\ \bibnamefont {McClain}}, \bibinfo {author} {\bibfnamefont {E.~R.}\ \bibnamefont {Sayfutyarova}}, \bibinfo {author} {\bibfnamefont {S.}~\bibnamefont {Sharma}}, \bibinfo {author} {\bibfnamefont {S.}~\bibnamefont {Wouters}}, \ and\ \bibinfo {author} {\bibfnamefont {G.~K.-L.}\ \bibnamefont {Chan}},\ }\href {\doibase 10.1002/wcms.1340} {\enquote {\bibinfo {title} {{{PySCF}}: The {{Python}}-based simulations of chemistry framework},}\ } (\bibinfo {year} {2018})\BibitemShut {NoStop}%
\bibitem [{\citenamefont {Sun}\ \emph {et~al.}(2020)\citenamefont {Sun}, \citenamefont {Zhang}, \citenamefont {Banerjee}, \citenamefont {Bao}, \citenamefont {Barbry}, \citenamefont {Blunt}, \citenamefont {Bogdanov}, \citenamefont {Booth}, \citenamefont {Chen}, \citenamefont {Cui}, \citenamefont {Eriksen}, \citenamefont {Gao}, \citenamefont {Guo}, \citenamefont {Hermann}, \citenamefont {Hermes}, \citenamefont {Koh}, \citenamefont {Koval}, \citenamefont {Lehtola}, \citenamefont {Li}, \citenamefont {Liu}, \citenamefont {Mardirossian}, \citenamefont {McClain}, \citenamefont {Motta}, \citenamefont {Mussard}, \citenamefont {Pham}, \citenamefont {Pulkin}, \citenamefont {Purwanto}, \citenamefont {Robinson}, \citenamefont {Ronca}, \citenamefont {Sayfutyarova}, \citenamefont {Scheurer}, \citenamefont {Schurkus}, \citenamefont {Smith}, \citenamefont {Sun}, \citenamefont {Sun}, \citenamefont {Upadhyay}, \citenamefont {Wagner}, \citenamefont {Wang}, \citenamefont {White}, \citenamefont {Whitfield}, \citenamefont
  {Williamson}, \citenamefont {Wouters}, \citenamefont {Yang}, \citenamefont {Yu}, \citenamefont {Zhu}, \citenamefont {Berkelbach}, \citenamefont {Sharma}, \citenamefont {Sokolov},\ and\ \citenamefont {Chan}}]{sunRecentDevelopmentsSCF2020}%
  \BibitemOpen
  \bibfield  {author} {\bibinfo {author} {\bibfnamefont {Q.}~\bibnamefont {Sun}}, \bibinfo {author} {\bibfnamefont {X.}~\bibnamefont {Zhang}}, \bibinfo {author} {\bibfnamefont {S.}~\bibnamefont {Banerjee}}, \bibinfo {author} {\bibfnamefont {P.}~\bibnamefont {Bao}}, \bibinfo {author} {\bibfnamefont {M.}~\bibnamefont {Barbry}}, \bibinfo {author} {\bibfnamefont {N.~S.}\ \bibnamefont {Blunt}}, \bibinfo {author} {\bibfnamefont {N.~A.}\ \bibnamefont {Bogdanov}}, \bibinfo {author} {\bibfnamefont {G.~H.}\ \bibnamefont {Booth}}, \bibinfo {author} {\bibfnamefont {J.}~\bibnamefont {Chen}}, \bibinfo {author} {\bibfnamefont {Z.-H.}\ \bibnamefont {Cui}}, \bibinfo {author} {\bibfnamefont {J.~J.}\ \bibnamefont {Eriksen}}, \bibinfo {author} {\bibfnamefont {Y.}~\bibnamefont {Gao}}, \bibinfo {author} {\bibfnamefont {S.}~\bibnamefont {Guo}}, \bibinfo {author} {\bibfnamefont {J.}~\bibnamefont {Hermann}}, \bibinfo {author} {\bibfnamefont {M.~R.}\ \bibnamefont {Hermes}}, \bibinfo {author} {\bibfnamefont {K.}~\bibnamefont {Koh}},
  \bibinfo {author} {\bibfnamefont {P.}~\bibnamefont {Koval}}, \bibinfo {author} {\bibfnamefont {S.}~\bibnamefont {Lehtola}}, \bibinfo {author} {\bibfnamefont {Z.}~\bibnamefont {Li}}, \bibinfo {author} {\bibfnamefont {J.}~\bibnamefont {Liu}}, \bibinfo {author} {\bibfnamefont {N.}~\bibnamefont {Mardirossian}}, \bibinfo {author} {\bibfnamefont {J.~D.}\ \bibnamefont {McClain}}, \bibinfo {author} {\bibfnamefont {M.}~\bibnamefont {Motta}}, \bibinfo {author} {\bibfnamefont {B.}~\bibnamefont {Mussard}}, \bibinfo {author} {\bibfnamefont {H.~Q.}\ \bibnamefont {Pham}}, \bibinfo {author} {\bibfnamefont {A.}~\bibnamefont {Pulkin}}, \bibinfo {author} {\bibfnamefont {W.}~\bibnamefont {Purwanto}}, \bibinfo {author} {\bibfnamefont {P.~J.}\ \bibnamefont {Robinson}}, \bibinfo {author} {\bibfnamefont {E.}~\bibnamefont {Ronca}}, \bibinfo {author} {\bibfnamefont {E.~R.}\ \bibnamefont {Sayfutyarova}}, \bibinfo {author} {\bibfnamefont {M.}~\bibnamefont {Scheurer}}, \bibinfo {author} {\bibfnamefont {H.~F.}\ \bibnamefont {Schurkus}},
  \bibinfo {author} {\bibfnamefont {J.~E.~T.}\ \bibnamefont {Smith}}, \bibinfo {author} {\bibfnamefont {C.}~\bibnamefont {Sun}}, \bibinfo {author} {\bibfnamefont {S.-N.}\ \bibnamefont {Sun}}, \bibinfo {author} {\bibfnamefont {S.}~\bibnamefont {Upadhyay}}, \bibinfo {author} {\bibfnamefont {L.~K.}\ \bibnamefont {Wagner}}, \bibinfo {author} {\bibfnamefont {X.}~\bibnamefont {Wang}}, \bibinfo {author} {\bibfnamefont {A.}~\bibnamefont {White}}, \bibinfo {author} {\bibfnamefont {J.~D.}\ \bibnamefont {Whitfield}}, \bibinfo {author} {\bibfnamefont {M.~J.}\ \bibnamefont {Williamson}}, \bibinfo {author} {\bibfnamefont {S.}~\bibnamefont {Wouters}}, \bibinfo {author} {\bibfnamefont {J.}~\bibnamefont {Yang}}, \bibinfo {author} {\bibfnamefont {J.~M.}\ \bibnamefont {Yu}}, \bibinfo {author} {\bibfnamefont {T.}~\bibnamefont {Zhu}}, \bibinfo {author} {\bibfnamefont {T.~C.}\ \bibnamefont {Berkelbach}}, \bibinfo {author} {\bibfnamefont {S.}~\bibnamefont {Sharma}}, \bibinfo {author} {\bibfnamefont {A.~Y.}\ \bibnamefont {Sokolov}},
  \ and\ \bibinfo {author} {\bibfnamefont {G.~K.-L.}\ \bibnamefont {Chan}},\ }\href {\doibase 10.1063/5.0006074} {\enquote {\bibinfo {title} {Recent developments in the {{P}} {\textsc{y}} {{SCF}} program package},}\ } (\bibinfo {year} {2020})\BibitemShut {NoStop}%
\bibitem [{\citenamefont {Wang}(2025)}]{wangXubwaSocutils2025}%
  \BibitemOpen
  \bibfield  {author} {\bibinfo {author} {\bibfnamefont {X.}~\bibnamefont {Wang}},\ }\href {https://github.com/xubwa/socutils} {\enquote {\bibinfo {title} {Xubwa/socutils},}\ } (\bibinfo {year} {2025})\BibitemShut {NoStop}%
\bibitem [{\citenamefont {Bast}\ \emph {et~al.}(2023)\citenamefont {Bast}, \citenamefont {Gomes}, \citenamefont {Saue}, \citenamefont {Visscher}, \citenamefont {Jensen}, \citenamefont {Aucar}, \citenamefont {Bakken}, \citenamefont {Chibueze}, \citenamefont {Creutzberg}, \citenamefont {Dyall}, \citenamefont {Dubillard}, \citenamefont {Ekstr{\"o}m}, \citenamefont {Eliav}, \citenamefont {Enevoldsen}, \citenamefont {Fa{\ss}hauer}, \citenamefont {Fleig}, \citenamefont {Fossgaard}, \citenamefont {Halbert}, \citenamefont {Hedeg{\aa}rd}, \citenamefont {Helgaker}, \citenamefont {{Helmich-Paris}}, \citenamefont {Henriksson}, \citenamefont {{van Horn}}, \citenamefont {Ilia{\v s}}, \citenamefont {Jacob}, \citenamefont {Knecht}, \citenamefont {Komorovsk{\'y}}, \citenamefont {Kullie}, \citenamefont {L{\ae}rdahl}, \citenamefont {Larsen}, \citenamefont {Lee}, \citenamefont {List}, \citenamefont {Nataraj}, \citenamefont {Nayak}, \citenamefont {Norman}, \citenamefont {Nyvang}, \citenamefont {Olejniczak}, \citenamefont {Olsen},
  \citenamefont {Olsen}, \citenamefont {Papadopoulos}, \citenamefont {Park}, \citenamefont {Pedersen}, \citenamefont {Pernpointner}, \citenamefont {Pototschnig}, \citenamefont {Di~Remigio~Eik{\aa}s}, \citenamefont {Repisk{\'y}}, \citenamefont {Ruud}, \citenamefont {Sa{\l}ek}, \citenamefont {Schimmelpfennig}, \citenamefont {Senjean}, \citenamefont {Shee}, \citenamefont {Sikkema}, \citenamefont {Sunaga}, \citenamefont {Thyssen}, \citenamefont {{van Stralen}}, \citenamefont {Vidal}, \citenamefont {Villaume}, \citenamefont {Visser}, \citenamefont {Winther}, \citenamefont {Yamamoto},\ and\ \citenamefont {Yuan}}]{bastDIRAC232023}%
  \BibitemOpen
  \bibfield  {author} {\bibinfo {author} {\bibfnamefont {R.}~\bibnamefont {Bast}}, \bibinfo {author} {\bibfnamefont {A.~S.~P.}\ \bibnamefont {Gomes}}, \bibinfo {author} {\bibfnamefont {T.}~\bibnamefont {Saue}}, \bibinfo {author} {\bibfnamefont {L.}~\bibnamefont {Visscher}}, \bibinfo {author} {\bibfnamefont {H.~J.~{\relax Aa}.}\ \bibnamefont {Jensen}}, \bibinfo {author} {\bibfnamefont {I.~A.}\ \bibnamefont {Aucar}}, \bibinfo {author} {\bibfnamefont {V.}~\bibnamefont {Bakken}}, \bibinfo {author} {\bibfnamefont {C.}~\bibnamefont {Chibueze}}, \bibinfo {author} {\bibfnamefont {J.}~\bibnamefont {Creutzberg}}, \bibinfo {author} {\bibfnamefont {K.~G.}\ \bibnamefont {Dyall}}, \bibinfo {author} {\bibfnamefont {S.}~\bibnamefont {Dubillard}}, \bibinfo {author} {\bibfnamefont {U.}~\bibnamefont {Ekstr{\"o}m}}, \bibinfo {author} {\bibfnamefont {E.}~\bibnamefont {Eliav}}, \bibinfo {author} {\bibfnamefont {T.}~\bibnamefont {Enevoldsen}}, \bibinfo {author} {\bibfnamefont {E.}~\bibnamefont {Fa{\ss}hauer}}, \bibinfo {author}
  {\bibfnamefont {T.}~\bibnamefont {Fleig}}, \bibinfo {author} {\bibfnamefont {O.}~\bibnamefont {Fossgaard}}, \bibinfo {author} {\bibfnamefont {L.}~\bibnamefont {Halbert}}, \bibinfo {author} {\bibfnamefont {E.~D.}\ \bibnamefont {Hedeg{\aa}rd}}, \bibinfo {author} {\bibfnamefont {T.}~\bibnamefont {Helgaker}}, \bibinfo {author} {\bibfnamefont {B.}~\bibnamefont {{Helmich-Paris}}}, \bibinfo {author} {\bibfnamefont {J.}~\bibnamefont {Henriksson}}, \bibinfo {author} {\bibfnamefont {M.}~\bibnamefont {{van Horn}}}, \bibinfo {author} {\bibfnamefont {M.}~\bibnamefont {Ilia{\v s}}}, \bibinfo {author} {\bibfnamefont {{\relax Ch}.~R.}\ \bibnamefont {Jacob}}, \bibinfo {author} {\bibfnamefont {S.}~\bibnamefont {Knecht}}, \bibinfo {author} {\bibfnamefont {S.}~\bibnamefont {Komorovsk{\'y}}}, \bibinfo {author} {\bibfnamefont {O.}~\bibnamefont {Kullie}}, \bibinfo {author} {\bibfnamefont {J.~K.}\ \bibnamefont {L{\ae}rdahl}}, \bibinfo {author} {\bibfnamefont {C.~V.}\ \bibnamefont {Larsen}}, \bibinfo {author} {\bibfnamefont
  {Y.~S.}\ \bibnamefont {Lee}}, \bibinfo {author} {\bibfnamefont {N.~H.}\ \bibnamefont {List}}, \bibinfo {author} {\bibfnamefont {H.~S.}\ \bibnamefont {Nataraj}}, \bibinfo {author} {\bibfnamefont {M.~K.}\ \bibnamefont {Nayak}}, \bibinfo {author} {\bibfnamefont {P.}~\bibnamefont {Norman}}, \bibinfo {author} {\bibfnamefont {A.}~\bibnamefont {Nyvang}}, \bibinfo {author} {\bibfnamefont {G.}~\bibnamefont {Olejniczak}}, \bibinfo {author} {\bibfnamefont {J.}~\bibnamefont {Olsen}}, \bibinfo {author} {\bibfnamefont {J.~M.~H.}\ \bibnamefont {Olsen}}, \bibinfo {author} {\bibfnamefont {A.}~\bibnamefont {Papadopoulos}}, \bibinfo {author} {\bibfnamefont {Y.~C.}\ \bibnamefont {Park}}, \bibinfo {author} {\bibfnamefont {J.~K.}\ \bibnamefont {Pedersen}}, \bibinfo {author} {\bibfnamefont {M.}~\bibnamefont {Pernpointner}}, \bibinfo {author} {\bibfnamefont {J.~V.}\ \bibnamefont {Pototschnig}}, \bibinfo {author} {\bibfnamefont {R.}~\bibnamefont {Di~Remigio~Eik{\aa}s}}, \bibinfo {author} {\bibfnamefont {M.}~\bibnamefont
  {Repisk{\'y}}}, \bibinfo {author} {\bibfnamefont {K.}~\bibnamefont {Ruud}}, \bibinfo {author} {\bibfnamefont {P.}~\bibnamefont {Sa{\l}ek}}, \bibinfo {author} {\bibfnamefont {B.}~\bibnamefont {Schimmelpfennig}}, \bibinfo {author} {\bibfnamefont {B.}~\bibnamefont {Senjean}}, \bibinfo {author} {\bibfnamefont {A.}~\bibnamefont {Shee}}, \bibinfo {author} {\bibfnamefont {J.}~\bibnamefont {Sikkema}}, \bibinfo {author} {\bibfnamefont {A.}~\bibnamefont {Sunaga}}, \bibinfo {author} {\bibfnamefont {J.}~\bibnamefont {Thyssen}}, \bibinfo {author} {\bibfnamefont {J.}~\bibnamefont {{van Stralen}}}, \bibinfo {author} {\bibfnamefont {M.~L.}\ \bibnamefont {Vidal}}, \bibinfo {author} {\bibfnamefont {S.}~\bibnamefont {Villaume}}, \bibinfo {author} {\bibfnamefont {O.}~\bibnamefont {Visser}}, \bibinfo {author} {\bibfnamefont {T.}~\bibnamefont {Winther}}, \bibinfo {author} {\bibfnamefont {S.}~\bibnamefont {Yamamoto}}, \ and\ \bibinfo {author} {\bibfnamefont {X.}~\bibnamefont {Yuan}},\ }\href {\doibase 10.5281/zenodo.7670749}
  {\enquote {\bibinfo {title} {{{DIRAC: Program for Atomic and Molecular Direct Iterative Relativistic All-electron Calculations.}}}}\ } (\bibinfo {year} {2023})\BibitemShut {NoStop}%
\bibitem [{\citenamefont {Barca}\ \emph {et~al.}(2020)\citenamefont {Barca}, \citenamefont {Bertoni}, \citenamefont {Carrington}, \citenamefont {Datta}, \citenamefont {De~Silva}, \citenamefont {Deustua}, \citenamefont {Fedorov}, \citenamefont {Gour}, \citenamefont {Gunina}, \citenamefont {Guidez}, \citenamefont {Harville}, \citenamefont {Irle}, \citenamefont {Ivanic}, \citenamefont {Kowalski}, \citenamefont {Leang}, \citenamefont {Li}, \citenamefont {Li}, \citenamefont {Lutz}, \citenamefont {Magoulas}, \citenamefont {Mato}, \citenamefont {Mironov}, \citenamefont {Nakata}, \citenamefont {Pham}, \citenamefont {Piecuch}, \citenamefont {Poole}, \citenamefont {Pruitt}, \citenamefont {Rendell}, \citenamefont {Roskop}, \citenamefont {Ruedenberg}, \citenamefont {Sattasathuchana}, \citenamefont {Schmidt}, \citenamefont {Shen}, \citenamefont {Slipchenko}, \citenamefont {Sosonkina}, \citenamefont {Sundriyal}, \citenamefont {Tiwari}, \citenamefont {Galvez~Vallejo}, \citenamefont {Westheimer}, \citenamefont {W{\l}och},
  \citenamefont {Xu}, \citenamefont {Zahariev},\ and\ \citenamefont {Gordon}}]{barcaRecentDevelopmentsGeneral2020}%
  \BibitemOpen
  \bibfield  {author} {\bibinfo {author} {\bibfnamefont {G.~M.~J.}\ \bibnamefont {Barca}}, \bibinfo {author} {\bibfnamefont {C.}~\bibnamefont {Bertoni}}, \bibinfo {author} {\bibfnamefont {L.}~\bibnamefont {Carrington}}, \bibinfo {author} {\bibfnamefont {D.}~\bibnamefont {Datta}}, \bibinfo {author} {\bibfnamefont {N.}~\bibnamefont {De~Silva}}, \bibinfo {author} {\bibfnamefont {J.~E.}\ \bibnamefont {Deustua}}, \bibinfo {author} {\bibfnamefont {D.~G.}\ \bibnamefont {Fedorov}}, \bibinfo {author} {\bibfnamefont {J.~R.}\ \bibnamefont {Gour}}, \bibinfo {author} {\bibfnamefont {A.~O.}\ \bibnamefont {Gunina}}, \bibinfo {author} {\bibfnamefont {E.}~\bibnamefont {Guidez}}, \bibinfo {author} {\bibfnamefont {T.}~\bibnamefont {Harville}}, \bibinfo {author} {\bibfnamefont {S.}~\bibnamefont {Irle}}, \bibinfo {author} {\bibfnamefont {J.}~\bibnamefont {Ivanic}}, \bibinfo {author} {\bibfnamefont {K.}~\bibnamefont {Kowalski}}, \bibinfo {author} {\bibfnamefont {S.~S.}\ \bibnamefont {Leang}}, \bibinfo {author} {\bibfnamefont
  {H.}~\bibnamefont {Li}}, \bibinfo {author} {\bibfnamefont {W.}~\bibnamefont {Li}}, \bibinfo {author} {\bibfnamefont {J.~J.}\ \bibnamefont {Lutz}}, \bibinfo {author} {\bibfnamefont {I.}~\bibnamefont {Magoulas}}, \bibinfo {author} {\bibfnamefont {J.}~\bibnamefont {Mato}}, \bibinfo {author} {\bibfnamefont {V.}~\bibnamefont {Mironov}}, \bibinfo {author} {\bibfnamefont {H.}~\bibnamefont {Nakata}}, \bibinfo {author} {\bibfnamefont {B.~Q.}\ \bibnamefont {Pham}}, \bibinfo {author} {\bibfnamefont {P.}~\bibnamefont {Piecuch}}, \bibinfo {author} {\bibfnamefont {D.}~\bibnamefont {Poole}}, \bibinfo {author} {\bibfnamefont {S.~R.}\ \bibnamefont {Pruitt}}, \bibinfo {author} {\bibfnamefont {A.~P.}\ \bibnamefont {Rendell}}, \bibinfo {author} {\bibfnamefont {L.~B.}\ \bibnamefont {Roskop}}, \bibinfo {author} {\bibfnamefont {K.}~\bibnamefont {Ruedenberg}}, \bibinfo {author} {\bibfnamefont {T.}~\bibnamefont {Sattasathuchana}}, \bibinfo {author} {\bibfnamefont {M.~W.}\ \bibnamefont {Schmidt}}, \bibinfo {author} {\bibfnamefont
  {J.}~\bibnamefont {Shen}}, \bibinfo {author} {\bibfnamefont {L.}~\bibnamefont {Slipchenko}}, \bibinfo {author} {\bibfnamefont {M.}~\bibnamefont {Sosonkina}}, \bibinfo {author} {\bibfnamefont {V.}~\bibnamefont {Sundriyal}}, \bibinfo {author} {\bibfnamefont {A.}~\bibnamefont {Tiwari}}, \bibinfo {author} {\bibfnamefont {J.~L.}\ \bibnamefont {Galvez~Vallejo}}, \bibinfo {author} {\bibfnamefont {B.}~\bibnamefont {Westheimer}}, \bibinfo {author} {\bibfnamefont {M.}~\bibnamefont {W{\l}och}}, \bibinfo {author} {\bibfnamefont {P.}~\bibnamefont {Xu}}, \bibinfo {author} {\bibfnamefont {F.}~\bibnamefont {Zahariev}}, \ and\ \bibinfo {author} {\bibfnamefont {M.~S.}\ \bibnamefont {Gordon}},\ }\href {\doibase 10.1063/5.0005188} {\enquote {\bibinfo {title} {Recent developments in the general atomic and molecular electronic structure system},}\ } (\bibinfo {year} {2020})\BibitemShut {NoStop}%
\bibitem [{\citenamefont {Chamoli}\ \emph {et~al.}(2024)\citenamefont {Chamoli}, \citenamefont {Mishra}, \citenamefont {Dutta}, \citenamefont {Kesarkar},\ and\ \citenamefont {Sahoo}}]{chamoli2024relativistic}%
  \BibitemOpen
  \bibfield  {author} {\bibinfo {author} {\bibfnamefont {S.}~\bibnamefont {Chamoli}}, \bibinfo {author} {\bibfnamefont {A.}~\bibnamefont {Mishra}}, \bibinfo {author} {\bibfnamefont {A.~K.}\ \bibnamefont {Dutta}}, \bibinfo {author} {\bibfnamefont {R.~S.}\ \bibnamefont {Kesarkar}}, \ and\ \bibinfo {author} {\bibfnamefont {B.}~\bibnamefont {Sahoo}},\ }\href {\doibase 10.1103/PhysRevA.109.063111} {\bibfield  {journal} {\bibinfo  {journal} {Phys. Rev. A}\ }\textbf {\bibinfo {volume} {109}},\ \bibinfo {pages} {063111} (\bibinfo {year} {2024})}\BibitemShut {NoStop}%
\bibitem [{\citenamefont {Kramida}\ \emph {et~al.}(2024)\citenamefont {Kramida}, \citenamefont {Ralchenko}, \citenamefont {Reader} \emph {et~al.}}]{AtomicSpectraDatabase2009}%
  \BibitemOpen
  \bibfield  {author} {\bibinfo {author} {\bibfnamefont {A.}~\bibnamefont {Kramida}}, \bibinfo {author} {\bibfnamefont {Y.}~\bibnamefont {Ralchenko}}, \bibinfo {author} {\bibfnamefont {J.}~\bibnamefont {Reader}},  \emph {et~al.},\ }\href {\doibase https://dx.doi.org/10.18434/T4W30F} {\enquote {\bibinfo {title} {{NIST} {Atomic Spectra Database} (version 5.12)},}\ } (\bibinfo {year} {2024})\BibitemShut {NoStop}%
\bibitem [{\citenamefont {Huber}\ and\ \citenamefont {Herzberg}(1979)}]{huberConstantsDiatomicMolecules1979}%
  \BibitemOpen
  \bibfield  {author} {\bibinfo {author} {\bibfnamefont {K.~P.}\ \bibnamefont {Huber}}\ and\ \bibinfo {author} {\bibfnamefont {G.}~\bibnamefont {Herzberg}},\ }in\ \href {\doibase 10.1007/978-1-4757-0961-2_2} {\emph {\bibinfo {booktitle} {Molecular {{Spectra}} and {{Molecular Structure}}: {{IV}}. {{Constants}} of {{Diatomic Molecules}}}}}\ (\bibinfo  {publisher} {Springer US},\ \bibinfo {address} {Boston, MA},\ \bibinfo {year} {1979})\ pp.\ \bibinfo {pages} {8--689}\BibitemShut {NoStop}%
\bibitem [{\citenamefont {Borschevsky}\ \emph {et~al.}(2006)\citenamefont {Borschevsky}, \citenamefont {Eliav}, \citenamefont {Ishikawa},\ and\ \citenamefont {Kaldor}}]{BorschevskyAtomicTransitionEnergies2006}%
  \BibitemOpen
  \bibfield  {author} {\bibinfo {author} {\bibfnamefont {A.}~\bibnamefont {Borschevsky}}, \bibinfo {author} {\bibfnamefont {E.}~\bibnamefont {Eliav}}, \bibinfo {author} {\bibfnamefont {Y.}~\bibnamefont {Ishikawa}}, \ and\ \bibinfo {author} {\bibfnamefont {U.}~\bibnamefont {Kaldor}},\ }\href {\doibase 10.1103/PhysRevA.74.062505} {\bibfield  {journal} {\bibinfo  {journal} {Phys. Rev. A}\ }\textbf {\bibinfo {volume} {74}},\ \bibinfo {pages} {062505} (\bibinfo {year} {2006})}\BibitemShut {NoStop}%
\bibitem [{\citenamefont {Schreiber}\ \emph {et~al.}(2008)\citenamefont {Schreiber}, \citenamefont {Silva-Junior}, \citenamefont {Sauer},\ and\ \citenamefont {Thiel}}]{schreiber2008benchmarks}%
  \BibitemOpen
  \bibfield  {author} {\bibinfo {author} {\bibfnamefont {M.}~\bibnamefont {Schreiber}}, \bibinfo {author} {\bibfnamefont {M.~R.}\ \bibnamefont {Silva-Junior}}, \bibinfo {author} {\bibfnamefont {S.}~\bibnamefont {Sauer}}, \ and\ \bibinfo {author} {\bibfnamefont {W.}~\bibnamefont {Thiel}},\ }\href {\doibase 10.1063/1.2889385} {\bibfield  {journal} {\bibinfo  {journal} {J. Chem. Phys.}\ }\textbf {\bibinfo {volume} {128}},\ \bibinfo {pages} {134110} (\bibinfo {year} {2008})}\BibitemShut {NoStop}%
\bibitem [{\citenamefont {Landau}\ \emph {et~al.}(2004)\citenamefont {Landau}, \citenamefont {Eliav}, \citenamefont {Ishikawa},\ and\ \citenamefont {Kaldor}}]{landauMixedsectorIntermediateHamiltonian2004}%
  \BibitemOpen
  \bibfield  {author} {\bibinfo {author} {\bibfnamefont {A.}~\bibnamefont {Landau}}, \bibinfo {author} {\bibfnamefont {E.}~\bibnamefont {Eliav}}, \bibinfo {author} {\bibfnamefont {Y.}~\bibnamefont {Ishikawa}}, \ and\ \bibinfo {author} {\bibfnamefont {U.}~\bibnamefont {Kaldor}},\ }\href {\doibase 10.1063/1.1788652} {\bibfield  {journal} {\bibinfo  {journal} {J. Chem. Phys.}\ }\textbf {\bibinfo {volume} {121}},\ \bibinfo {pages} {6634} (\bibinfo {year} {2004})}\BibitemShut {NoStop}%
\bibitem [{\citenamefont {Andersson}\ \emph {et~al.}(1990)\citenamefont {Andersson}, \citenamefont {Malmqvist}, \citenamefont {Roos}, \citenamefont {Sadlej},\ and\ \citenamefont {Wolinski}}]{anderssonSecondorderPerturbationTheory1990}%
  \BibitemOpen
  \bibfield  {author} {\bibinfo {author} {\bibfnamefont {{\relax Kerstin}.}~\bibnamefont {Andersson}}, \bibinfo {author} {\bibfnamefont {P.~A.}\ \bibnamefont {Malmqvist}}, \bibinfo {author} {\bibfnamefont {B.~O.}\ \bibnamefont {Roos}}, \bibinfo {author} {\bibfnamefont {A.~J.}\ \bibnamefont {Sadlej}}, \ and\ \bibinfo {author} {\bibfnamefont {{\relax Krzysztof}.}~\bibnamefont {Wolinski}},\ }\href {\doibase 10.1021/j100377a012} {\bibfield  {journal} {\bibinfo  {journal} {J. Phys. Chem.}\ }\textbf {\bibinfo {volume} {94}},\ \bibinfo {pages} {5483} (\bibinfo {year} {1990})}\BibitemShut {NoStop}%
\bibitem [{\citenamefont {Roos}, \citenamefont {Taylor},\ and\ \citenamefont {Sigbahn}(1980)}]{roosCompleteActiveSpace1980}%
  \BibitemOpen
  \bibfield  {author} {\bibinfo {author} {\bibfnamefont {B.~O.}\ \bibnamefont {Roos}}, \bibinfo {author} {\bibfnamefont {P.~R.}\ \bibnamefont {Taylor}}, \ and\ \bibinfo {author} {\bibfnamefont {P.~E.~M.}\ \bibnamefont {Sigbahn}},\ }\href {\doibase 10.1016/0301-0104(80)80045-0} {\bibfield  {journal} {\bibinfo  {journal} {Chem. Phys.}\ }\textbf {\bibinfo {volume} {48}},\ \bibinfo {pages} {157} (\bibinfo {year} {1980})}\BibitemShut {NoStop}%
\bibitem [{\citenamefont {Vala}, \citenamefont {Kosloff},\ and\ \citenamefont {Harvey}(2001)}]{valaInitioDiatomicsMolecule2001}%
  \BibitemOpen
  \bibfield  {author} {\bibinfo {author} {\bibfnamefont {J.}~\bibnamefont {Vala}}, \bibinfo {author} {\bibfnamefont {R.}~\bibnamefont {Kosloff}}, \ and\ \bibinfo {author} {\bibfnamefont {J.~N.}\ \bibnamefont {Harvey}},\ }\href {\doibase 10.1063/1.1361248} {\bibfield  {journal} {\bibinfo  {journal} {J. Chem. Phys.}\ }\textbf {\bibinfo {volume} {114}},\ \bibinfo {pages} {7413} (\bibinfo {year} {2001})}\BibitemShut {NoStop}%
\bibitem [{\citenamefont {Fleig}, \citenamefont {Olsen},\ and\ \citenamefont {Visscher}(2003)}]{fleigGeneralizedActiveSpace2003}%
  \BibitemOpen
  \bibfield  {author} {\bibinfo {author} {\bibfnamefont {T.}~\bibnamefont {Fleig}}, \bibinfo {author} {\bibfnamefont {J.}~\bibnamefont {Olsen}}, \ and\ \bibinfo {author} {\bibfnamefont {L.}~\bibnamefont {Visscher}},\ }\href {\doibase 10.1063/1.1590636} {\bibfield  {journal} {\bibinfo  {journal} {J. Chem. Phys.}\ }\textbf {\bibinfo {volume} {119}},\ \bibinfo {pages} {2963} (\bibinfo {year} {2003})}\BibitemShut {NoStop}%
\bibitem [{\citenamefont {Fleig}\ \emph {et~al.}(2006)\citenamefont {Fleig}, \citenamefont {Jensen}, \citenamefont {Olsen},\ and\ \citenamefont {Visscher}}]{fleigGeneralizedActiveSpace2006}%
  \BibitemOpen
  \bibfield  {author} {\bibinfo {author} {\bibfnamefont {T.}~\bibnamefont {Fleig}}, \bibinfo {author} {\bibfnamefont {H.~J.~A.}\ \bibnamefont {Jensen}}, \bibinfo {author} {\bibfnamefont {J.}~\bibnamefont {Olsen}}, \ and\ \bibinfo {author} {\bibfnamefont {L.}~\bibnamefont {Visscher}},\ }\href {\doibase 10.1063/1.2176609} {\bibfield  {journal} {\bibinfo  {journal} {J. Chem. Phys.}\ }\textbf {\bibinfo {volume} {124}},\ \bibinfo {pages} {104106} (\bibinfo {year} {2006})}\BibitemShut {NoStop}%
\bibitem [{\citenamefont {Gross}\ and\ \citenamefont {Kohn}(1990)}]{grossTimeDependentDensityFunctionalTheory1990}%
  \BibitemOpen
  \bibfield  {author} {\bibinfo {author} {\bibfnamefont {E.~K.~U.}\ \bibnamefont {Gross}}\ and\ \bibinfo {author} {\bibfnamefont {W.}~\bibnamefont {Kohn}},\ }in\ \href {\doibase 10.1016/S0065-3276(08)60600-0} {\emph {\bibinfo {booktitle} {Advances in {{Quantum Chemistry}}}}},\ \bibinfo {series} {Density {{Functional Theory}} of {{Many-Fermion Systems}}}, Vol.~\bibinfo {volume} {21},\ \bibinfo {editor} {edited by\ \bibinfo {editor} {\bibfnamefont {P.-O.}\ \bibnamefont {L{\"o}wdin}}}\ (\bibinfo  {publisher} {Academic Press},\ \bibinfo {year} {1990})\ pp.\ \bibinfo {pages} {255--291}\BibitemShut {NoStop}%
\bibitem [{\citenamefont {Wang}, \citenamefont {Tu},\ and\ \citenamefont {Wang}(2014)}]{wangEquationofMotionCoupledClusterTheory2014}%
  \BibitemOpen
  \bibfield  {author} {\bibinfo {author} {\bibfnamefont {Z.}~\bibnamefont {Wang}}, \bibinfo {author} {\bibfnamefont {Z.}~\bibnamefont {Tu}}, \ and\ \bibinfo {author} {\bibfnamefont {F.}~\bibnamefont {Wang}},\ }\href {\doibase 10.1021/ct500854m} {\bibfield  {journal} {\bibinfo  {journal} {J. Chem. Theory Comput.}\ }\textbf {\bibinfo {volume} {10}},\ \bibinfo {pages} {5567} (\bibinfo {year} {2014})}\BibitemShut {NoStop}%
\bibitem [{\citenamefont {Zaitsevskii}, \citenamefont {Oleynichenko},\ and\ \citenamefont {Eliav}(2020)}]{zaitsevskiiFiniteFieldCalculationsTransition2020}%
  \BibitemOpen
  \bibfield  {author} {\bibinfo {author} {\bibfnamefont {A.}~\bibnamefont {Zaitsevskii}}, \bibinfo {author} {\bibfnamefont {A.~V.}\ \bibnamefont {Oleynichenko}}, \ and\ \bibinfo {author} {\bibfnamefont {E.}~\bibnamefont {Eliav}},\ }\href {\doibase 10.3390/sym12111845} {\bibfield  {journal} {\bibinfo  {journal} {Symmetry}\ }\textbf {\bibinfo {volume} {12}},\ \bibinfo {pages} {1845} (\bibinfo {year} {2020})}\BibitemShut {NoStop}%
\bibitem [{\citenamefont {Sansonetti}\ and\ \citenamefont {Martin}(2005)}]{sansonettiHandbookBasicAtomic2005}%
  \BibitemOpen
  \bibfield  {author} {\bibinfo {author} {\bibfnamefont {J.~E.}\ \bibnamefont {Sansonetti}}\ and\ \bibinfo {author} {\bibfnamefont {W.~C.}\ \bibnamefont {Martin}},\ }\href {\doibase 10.1063/1.1800011} {\bibfield  {journal} {\bibinfo  {journal} {J. Phys. Chem. Ref. Data}\ }\textbf {\bibinfo {volume} {34}},\ \bibinfo {pages} {1559} (\bibinfo {year} {2005})}\BibitemShut {NoStop}%
\bibitem [{\citenamefont {Margulis}, \citenamefont {Coker},\ and\ \citenamefont {Lynden-Bell}(2001)}]{margulis2001monte}%
  \BibitemOpen
  \bibfield  {author} {\bibinfo {author} {\bibfnamefont {C.}~\bibnamefont {Margulis}}, \bibinfo {author} {\bibfnamefont {D.}~\bibnamefont {Coker}}, \ and\ \bibinfo {author} {\bibfnamefont {R.}~\bibnamefont {Lynden-Bell}},\ }\href {\doibase https://doi.org/10.1063/1.1328757} {\bibfield  {journal} {\bibinfo  {journal} {J. Chem. Phys.}\ }\textbf {\bibinfo {volume} {114}},\ \bibinfo {pages} {367} (\bibinfo {year} {2001})}\BibitemShut {NoStop}%
\bibitem [{\citenamefont {Neese}(2022)}]{ORCA5}%
  \BibitemOpen
  \bibfield  {author} {\bibinfo {author} {\bibfnamefont {F.}~\bibnamefont {Neese}},\ }\href {\doibase 10.1002/wcms.1606} {\bibfield  {journal} {\bibinfo  {journal} {WIREs Comput. Mol. Sci.}\ }\textbf {\bibinfo {volume} {12}},\ \bibinfo {pages} {e1606} (\bibinfo {year} {2022})}\BibitemShut {NoStop}%
\end{thebibliography}
%
\end{document}